\documentstyle[preprint,aps]{revtex}
\tighten
\draft
\begin{document}
\widetext
\input epsf
\preprint{CLNS 97/1498, HUTP-97/A032, NUB 3162}
\bigskip
\bigskip
\title{A Review of Three-Family Grand Unified String Models}
\medskip
\author{Zurab Kakushadze$^{1,2}$\footnote{E-mail: 
zurab@string.harvard.edu}, 
Gary Shiu$^3$\footnote{E-mail: shiu@mail.lns.cornell.edu}, S.-H. Henry Tye$^3$\footnote{E-mail: tye@mail.lns.cornell.edu}, 
Yan Vtorov-Karevsky$^3$\footnote{E-mail: gilburd@mail.lns.cornell.edu}}

\bigskip

\address{$^1$Lyman Laboratory of Physics, Harvard University, Cambridge, 
MA 02138\\
$^2$Department of Physics, Northeastern University, Boston, MA 02115\\
$^3$Newman Laboratory of Nuclear Studies, Cornell University,
Ithaca, NY 14853}
\date{\today}
\bigskip
\medskip
\maketitle

\begin{abstract}
{}We review the construction and classification of three-family grand 
unified models within the framework of asymmetric orbifolds in
 perturbative heterotic superstring. We give a detailed survey of all 
such models which is organized to aid analysis of their phenomenological
 properties. We compute tree-level superpotentials for these models. 
These superpotentials are used to analyze the issues of proton stability 
(doublet-triplet splitting and $R$-parity violating terms) and Yukawa 
mass matrices. To have agreement with phenomenological data all these 
models seem to require certain degree of fine-tuning. We also analyze 
the possible patterns of supersymmetry breaking in these models.
 We find that the supersymmetry breaking scale comes out either too high 
to explain the electroweak hierarchy problem, or below the electroweak 
scale unless some degree of fine-tuning is involved. Thus, none of the 
models at hand seem to be phenomenologically flawless. 
\end{abstract}
\narrowtext

\section{Introduction}

{}If string theory is relevant to nature, it must possess a vacuum
whose low energy effective field theory describes the Standard Model of
strong and electroweak interactions. The question whether such a string 
vacuum exists
is difficult to answer as the space of classical string vacua has a very 
large degeneracy, and there lack objective criteria that would select a 
particular string vacuum among the numerous possibilities. One might expect 
non-perturbative string dynamics to lift, partially or completely, this 
degeneracy in the moduli space. If this lifting is complete, then a thorough
understanding of string dynamics may be sufficient to find a complete
description of our universe. {\em A priori}, however, it is reasonable to 
suspect that non-perturbative dynamics may {\em not} select the unique 
vacuum, but rather pick out 
a number of consistent vacua, some in four dimensions with completely 
broken supersymmetry; 
and our universe would be realized as one of the consistent vacua in 
this (probably large) set. 
If so, we would need to impose some additional, namely, 
phenomenological constraints to find the string vacuum in which we live. 
This approach has been known as ``superstring phenomenology''. The latter
 must still be augmented with an understanding of non-perturbative dynamics 
as superstring is believed not to break supersymmetry perturbatively.   

{}It might seem, at least naively, that there is more than enough 
phenomenological data to identify the superstring vacuum corresponding to 
our universe. It is, however, not known how to fully embed the Standard 
Model into string theory with all of its complexity, so one is bound to 
try to incorporate only a few phenomenologically desirable features at a 
time (such as, say, the gauge group, number of families, {\em etc.}). Since 
such constraints are not very stringent, 
this ultimately leads to numerous possibilities for embedding the Standard 
Model in superstring that {\em a priori} seem reasonable \cite{rev}. Thus, 
to make progress in superstring phenomenology, one needs as restrictive 
phenomenological constraints as possible.  It might be advantageous to impose
 such phenomenological constraints taking into account 
specifics of a partcular framework for superstring model building.   

{} In the past decade, the main arena for model building within the context
 of superstring phenomenology has been perturbative heterotic superstring. 
The reason is that such model building is greatly facilitated by existence
 of relatively simple rules (such as free-fermionic \cite{FFC} and orbifold
 \cite{Orb,NonAbe} constructions). Moreover, many calculational tools 
(such as, say, scattering amplitudes and rules for computing 
superpotentials \cite{scatt})
are either readily available, or can be developed for certain cases of 
interest. Despite enormous progress made in the past few years in 
understanding non-perturbative superstring vacua, the state of the art there
 is still far from being competitive with perturbative heterotic superstring. 
The tools available in the latter framework  must be first generalized to 
include the non-perturbative vacua before superstring phenomenology 
can step into this 
new terrain.

{}To be specific, let us concentrate on perturbative heterotic superstring. Within this framework the total rank of the gauge group (for $N=1$ space-time supersymmetric models) is 22 or less. After accommodating the Standard Model of strong and electroweak interactions (with gauge group $SU(3)_c \otimes SU(2)_w \otimes U(1)_Y$ whose rank is 4), the left-over rank for the hidden and/or horizontal gauge symmetry is 18 or less. The possible choices here are myriad \cite{standard} and largely unexplored. The situation is similar for embedding unification within a {\em semi-simple} \cite{semi} gauge group $G\supset SU(3)_c \otimes SU(2)_w \otimes U(1)_Y$ ({\em e.g.}, $SU(5)\otimes U(1)$).

{}The state of affairs is quite different if one tries to embed grand unification within a {\em simple} gauge group  $G\supset SU(3)_c \otimes SU(2)_w \otimes U(1)_Y$. Thus, an adjoint 
(or some other appropriate higher dimensional) Higgs field must be present 
among the light degrees of freedom in effective field theory to break the 
grand unified gauge group $G$ down to that of the Standard Model. 
In perturbative heterotic superstring such states in the massless 
spectrum are compatible with $N=1$ supersymmetry and chiral fermions 
only if the grand unified gauge group is realized via a current algebra
 at level $k>1$ (see, {\em e.g.}, Ref \cite{Lew}). This ultimately leads
 to reduction of total rank of the gauge group, and, therefore, to 
smaller room for hidden/horizontal symmetry, which greatly limits 
the number of possible models. 

{}The limited number of possibilities is not the only distinguishing feature 
of 
grand unified models in superstring theory. Grand unified theories (GUTs) 
possess a number of  
properties not shared by superstring models with either the Standard Model or 
a semi-simple gauge group. One such property concerns the gauge coupling
 unification problem in superstring theory \cite{dienu}. Thus, the strong 
and electroweak couplings 
$\alpha_3$, $\alpha_2$ and $\alpha_1$ of 
$SU(3)_c\otimes SU(2)_w\otimes U(1)_Y$ in 
the minimal supersymmetric Standard Model (MSSM) unify at the GUT scale 
$M_{GUT}\sim 10^{16}~{\mbox{GeV}}$ \cite{coupling} at the value of 
$\alpha_{GUT}\sim 1/24$. 
Running of these couplings is schematically shown in Fig.1a as a function of 
the energy scale $E$. For comparative purposes, a dimensionless gravitational 
coupling $\alpha_{G} =G_N E^2$ is introduced, where $G_N$ is the Newton's 
constant.
In string theory 
the unification demands that all couplings meet at a single scale. 
Note that the gravitational coupling becomes equal $\alpha_{GUT}$ at a scale
roughly two orders of magnitude higher than $M_{GUT}$ (Fig.1a).
Several possible approaches have been proposed to reconcile this apparent 
discrepancy \cite{dienu}, some of which are listed below:\\ 
$\bullet$ The subgroups of $SU(3)_c\otimes SU(2)_w\otimes U(1)_Y$ unify 
into a single GUT gauge group $G$, and the gauge coupling of $G$ meets 
with $\alpha_G$ as shown in Fig.1b.\\ 
$\bullet$ The gauge group remains that of the MSSM all the way up, but  
some extra (compared with the MSSM spectrum) matter multiplets 
are present which 
change the
running of the couplings so that their unification scale is pushed up to meet
$\alpha_G$. Such a  scenario requires a judicious choice of the extra fields, 
as well as their masses \cite{dienf}. The situation is similar for the 
models with 
other semi-simple gauge groups.\\ 
$\bullet$ Another possibility is suggested by $M$-theory \cite{witten}, 
where 
the graviton can propagate in a 5-dimensional spacetime (the bulk), while 
the gauge and 
matter fields live on 4-dimensional ``walls''. Below the energy scale of 
the order of the inverse
``thickness'' of the bulk, all fields propagate in 4 space-time 
dimensions \cite{witten}. Above this scale, however, the gauge and matter
fields propagate in 4 dimensions, while the graviton propagates in
5 dimensions. As a result, $\alpha_G$ runs faster and catches up with the
strong and electroweak couplings as shown in Fig.1c.

{}The simplest way to obtain higher-level current algebras (required for 
GUT embeddings) in perturbative heterotic superstring is via the 
following construction. Start from a 
$k$-fold product $G\otimes G\otimes \cdots \otimes G$ of the grand 
unified gauge group $G$ (of rank $r$) realized via a level-1 
current algebra. The diagonal subgroup 
$G_{diag}\subset  G\otimes G\otimes \cdots \otimes G$ is then 
realized via level $k$ current algebra. (Note that in 
carrying out this procedure the rank of the gauge group is reduced 
from $kr$ to $r$.) As far as the 
Hilbert space is concerned, here we are identifying the states 
under the ${\bf Z}_k$ cyclic symmetry of the $k$-fold product 
$G\otimes G\otimes \cdots \otimes G$. This is nothing but 
${\bf Z}_k$ orbifold action, namely, modding out by the outer automorphism.

\begin{figure}[h]
\epsfysize=6in
\centerline{\epsfbox{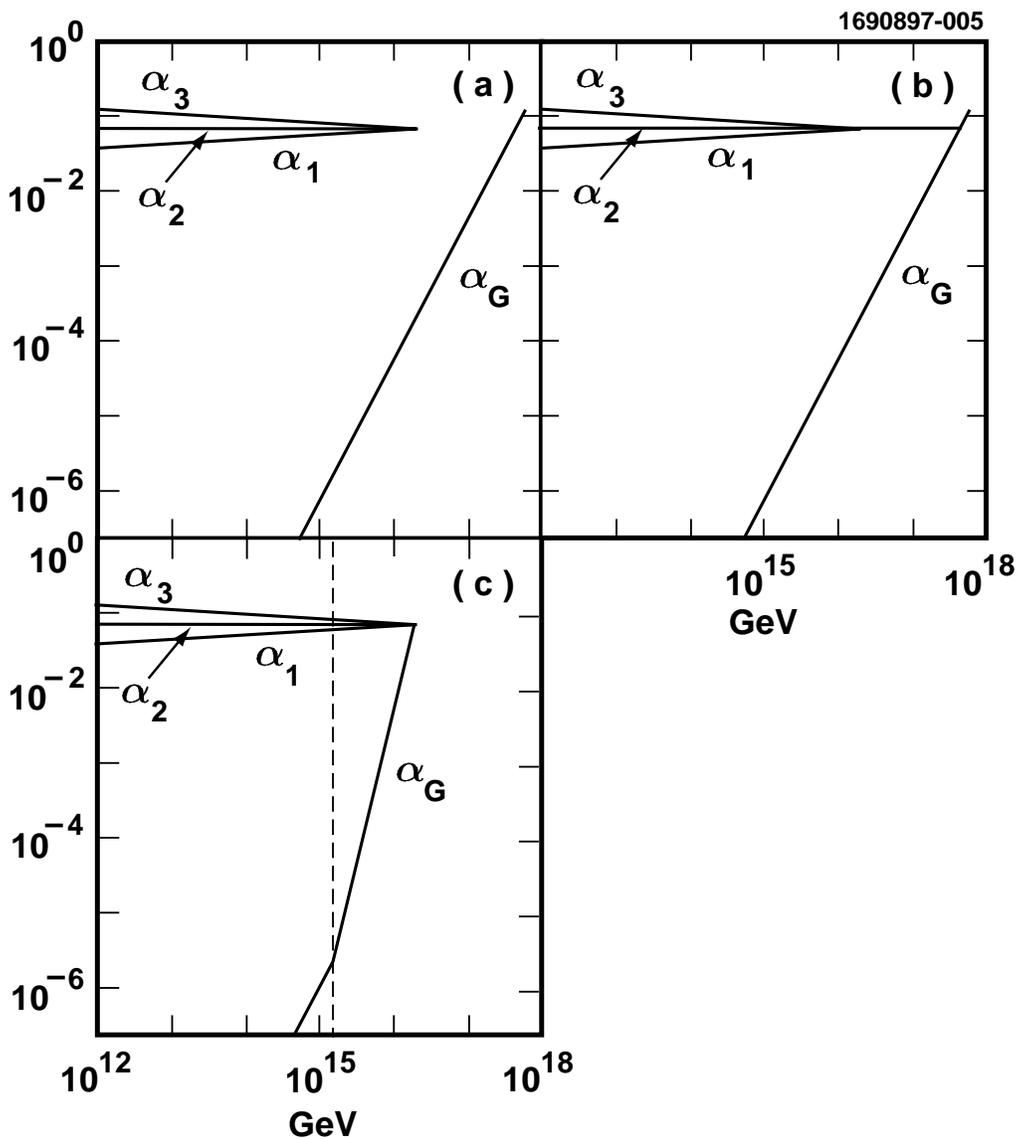}} 
\caption{Running of the MSSM couplings vs. dimensionless gravitational 
coupling}
\end{figure}

{}An immediate implication of the above construction is a rather 
limited number of possibilities. For example, for a grand unified gauge group 
$G=SO(10)$ with, say, $k=3$, the left-over rank (for the hidden and/or 
horizontal gauge symmetry) is at most 7 ($=22-3\times 5$). This is to be 
compared with the left-over rank 18 in the case of the Standard Model 
embedding. Taking into account that the number of models grows rapidly as 
a function of the left-over rank, it becomes clear that grand 
unified model building is much more restricted than other embeddings.

{}Since desired adjoint (or higher dimensional) Higgs fields are allowed 
already at level $k=2$, multiple attempts have been made in the past 
several years to construct level-2 grand unified string models. 
None of them, however, have yielded 3-family models. The first $SO(10)$ 
string GUT realized via a level-2 current algebra was obtained by 
Lewellen \cite{Lew} within the framework of free-fermionic construction 
\cite{FFC}. Soon after Schwartz \cite{Sch} obtained an $SU(5)$ 
level-2 string GUT 
within the same framework. Both of these models have four chiral 
families. Multiple attempts have been made ever since to construct
 three-family string 
GUTs realized via level-two current algebras within free-fermionic
 construction \cite{CCHL} and within the framework of 
symmetric \cite{AFIU} as well as asymmetric \cite{Erler}
Abelian orbifolds \cite{Orb}. Finally, three of us tried 
non-Abelian orbifolds \cite{kst0} within 
both the free-fermionic and the bosonic formulations \cite{NonAbe}. 
There is no formal proof that 3-family models cannot be 
obtained from level-2 constructions, but one can intuitively 
understand why attempts to find such models have failed. In the 
$k=2$ construction the orbifold group is ${\bf Z}_2$. So the numbers 
of fixed points in the twisted sectors, which are related to 
the number of chiral families, are always even in this case. 

{}Thus, it is natural to consider $k=3$ models. The orbifold action in 
this case is ${\bf Z}_3$, and one might hope to obtain models with 3 
families as the numbers of fixed points in the twisted sectors are 
some powers of 3. The level-3 model building appears to be more involved 
than that for level-2 constructions. The latter are facilitated by 
existence of the $E_8\otimes E_8$ heterotic superstring in 10 
dimensions which explicitly possesses a ${\bf Z}_2$ outer automorphism 
symmetry of the two $E_8$'s. Constructing a level-2 model then can 
be carried out in two steps: first one embeds the grand unified gauge 
group $G$ in each of the $E_8$'s, and then performs the outer 
automorphism ${\bf Z}_2$ twist. In contrast to the $k=2$ 
construction, $k=3$ model building requires explicitly 
realizing ${\bf Z}_3$ outer automorphism symmetry which 
is not present in 10 dimensions. The implication of the 
above discussion is that one needs relatively simple rules to 
facilitate model building. Such rules have been
 derived \cite{kt} within the framework of 
{\em asymmetric orbifolds} \cite{vafa}.

{}With the appropriate model building tools available, it became possible 
to construct \cite{kt,kt10,kt5} and classify \cite{class10,class5} 
3-family grand unified string models within the framework of 
asymmetric orbifolds in perturbative heterotic string theory. 
Here we briefly discuss the results of this classification. 
(A detailed survey of three-family grand unified string models 
is given in section II.) For each model we list here, there are 
additional models connected to it via classically flat 
directions \cite{class10,class5}:\\
$\bullet$ One $E_6$ model with 5 left-handed and 2 
right-handed families, and asymptotically free $SU(2)$ hidden sector 
with 1 ``flavor''.\\ 
$\bullet$ One $SO(10)$ model with 4 left-handed and 1 right-handed 
families, and $SU(2)\otimes SU(2)\otimes SU(2)$ hidden sector which 
is {\em not} asymptotically free at the string scale.\\
$\bullet$ Three $SU(6)$ models:\\
({\em i}) The first model has 6 left-handed and 3 right-handed families, 
and asymptotically free $SU(3)$ hidden sector with 3 ``flavors''.\\
({\em ii}) The second model has 3 left-handed and no right-handed families, 
and asymptotically free $SU(2)\otimes SU(2)$ hidden sector with matter 
content consisting of doublets of each $SU(2)$ subgroup as well as 
bi-fundamentals.\\
({\em iii}) The third model has 3 left-handed and no right-handed 
families, and asymptotically free $SU(4)$ hidden sector with 
3 ``flavors''. (This model has not been explicitly given in 
Ref \cite{class5}, and will be presented in Appendix \ref{models}.)\\
$\bullet$ Finally, there are some additional $SU(5)$ models which 
do not seem to be phenomenologically appealing (see section \ref{gut} 
for details).

{}All of the above models share some common phenomenological features. 
Thus, there is only one adjoint and no other higher dimensional Higgs 
fields in all of these models. The $E_6$ and $SO(10)$ models 
(and other related models) do {\em not} possess anomalous $U(1)$. All 
three $SU(6)$ models listed above {\em do} have anomalous $U(1)$  
(which in string theory is broken via the Green-Schwarz mechanism 
\cite{GSDSW}). The above models all possess non-Abelian hidden sector. 
There, however, exist models
where the hidden sector is completely broken.

{}To study phenomenological properties of these models it is first 
necessary to deduce their tree-level superpotentials. This turns 
out to be a rather non-trivial task as it involves understanding 
scattering in asymmetric orbifolds. There, however, are certain 
simplifying circumstances here due to the fact that asymmetric 
orbifold models possess enhanced discrete and continuous gauge 
symmetries. Making use of these symmetries,
 the tools for computing tree-level superpotentials for a class of 
asymmetric orbifold models (which includes the models of interest for us 
here) have been developed in Ref \cite{kst}. The perturbative 
superpotentials for some of the three-family grand unified 
string models were computed in Ref \cite{kst}. We compute 
perturbative superpotentials for other relevant models in 
this paper following the rules of Ref \cite{kst}.

{}The knowledge of tree-level superpotentials allows one to analyze 
certain phenomenological issues such as proton stability 
(doublet-triplet splitting and
$R$-parity violating terms) and Yukawa mass matrices. The question 
of supersymmetry breaking can also be addressed by augmenting the 
tree-level superpotentials with non-perturbative contributions which 
are under control in $N=1$ supersymmetric field theories \cite{Seiberg}.  

{}Thus, in Ref \cite{kstv} doublet-triplet splitting problem and Yukawa
mass matrices were studied for the $SO(10)$ models.
It was found that certain degree of
fine-tuning is required to solve the doublet-triplet splitting
problem, suppress dangerous $R$-parity violating terms and achieve
realistic Yukawa mass matrices.
All $SU(5)$ models suffer from the severe fine-tuning problem steming from
 the doublet-triplet splitting as there are no ``exotic'' higher
dimensional Higgs fields among the light degrees of freedom.
The latter are required by all known field theory solutions to the
problem.

{}In this paper we study similar issues for the $SU(6)$ models. The 
results of our analysis indicate that the doublet-triplet 
splitting does not seem to be as big of a problem for the $SU(6)$ 
models as it is for their $SO(10)$ and $SU(5)$ counterparts. However, 
the troubles 
with $R$-parity violating terms and Yukawa mass matrices still persist for 
these models.

{}In this paper we also analyze the possible patterns of 
supersymmetry breaking in the three-family grand unified string models. 
We find that the supersymmetry breaking scale in these 
models comes out either too high to explain the electroweak 
hierarchy problem, or below the 
electroweak scale unless some degree of fine-tuning is involved.

{}Since none of the three-family grand unified string models 
constructed to date appear to be phenomenologically flawless, 
one naturally wonders whether there may exist (even within 
perturbative heterotic superstring vacua) other such models 
with improved phenomenological characteristics.
 Thus, all {\em a priori} 
possible free-field embeddings of higher-level current algebras 
within perturbative heterotic superstring framework have been 
classified \cite{dien}. This, however, does not guarantee that 
any given embedding can be incorporated in a consistent string model,
 and even if this is indeed possible, there need not exist 
three-family models within such an embeding. The three-family 
grand unified string models descussed in this review are 
concrete realizations of the {\em diagonal} level-3 embedding 
$G_{diag}\supset G\otimes G\otimes G$. However, even if our 
models do exhaust all three-family grand unified string models 
within free-field realized perturbative heterotic superstring, 
there may exist non-free-field 
grand unified string models with three families. Tools for 
constructing such models are not available at this moment, so that 
for years to come the asymmetric orbifold models we discuss in this 
review might be the only ones available. Regardless of their 
phenomenological viability they provide the {\em proof of existence} 
for three-family grand unified string models; these models could also serve as 
a stringy paradigm for such model building in general, and also 
give insight to the ``bottom-up'' approach.   

{}This review is organized as follows. In section \ref{survey} we give a detailed survey of three-family grand unified string models classified in Refs \cite{class10,class5}. In section \ref{hidden} we discuss supersymmetry (SUSY) breaking in these models. In section \ref{gut} we discuss some aspects of SUSY GUT phenomenology in these models. In section \ref{conc} we give our conclusions and outlook. All the details are relegated to appendices.

\section{Survey of Three-Family Grand Unified String Models}\label{survey}

{}In this section we give a systematic survey of all the (relevant) 
three-family grand unified string models. This survey is organized by the 
grand unified gauge group, number of generations and anti-generations, and 
hidden sector gauge group and matter. We will not list all the models 
(as most of them can be either found in Refs \cite{class10,class5}, or 
deduced from those listed there), but give only the models that are 
representative of classes of models with similar features, or stand by their 
own (for they possess special features). Some of the latter models, namely, 
with $SU(6)$ grand unified gauge group, have not been presented elsewhere, 
so we briefly outline their construction in Appendix \ref{models}.

\subsection{The $E_6$ Model}

{}There is one $E_6$ model (referred to as the $E1$ or $E2$ model in 
Ref \cite{class10}; $E1$ and $E2$ stand for two different ways of 
constructing the same model \cite{class10,kst}). It has $5$ left-handed and 
$2$ right-handed families, and $SU(2)_H$ hidden sector with $2$ chiral 
doublets (corresponding to one ``flavor"). The massless spectrum of the 
$E_6$ model is summarized in Table I.

{}The $E_6$ model is a special (enhanced gauge symmetry) point in a certain 
moduli space, which we will refer to as ${\cal M}$, that also contains other 
three-family models with grand unified gauge groups $SO(10)$, $SU(5)$ and 
$SU(6)$ \cite{class10,class5}. All of these models are connected to each 
other (and to the $E_6$ model) via a web of (classically) flat directions.  
They all have $5$ left-handed and $2$ right-handed families, and $SU(2)_H$ 
hidden sector. There are two features that distinguish them: the hidden 
sector matter content (number of ``flavors" of $SU(2)_H$) and the horizontal 
gauge symmetry. These two features are not completely uncorrelated. Thus, 
breaking the horizontal symmetry (via Higgsing) sometimes leads to some of 
the hidden sector doublets acquiring mass (via tree-level superpotential). 
On the other hand, sometimes hidden sector doublets can get heavy without 
breaking the horizontal symmetry, {\em i.e.}, via coupling to gauge singlets 
(present in some models) acquiring vevs. Depending on the number of hidden 
sector doublets, the $SU(2)_H$ group may or may not be asymptotically free. 
Here we are interested only in models with asymptotically free hidden sector 
(for the phenomenological reasons discussed earlier). If the number of hidden 
sector ``flavors" is too large in a given model so that $SU(2)_H$ is not 
asymptotically free, it can always be connected to another model with 
smaller matter content of $SU(2)_H$ such that the latter is asymptotically 
free. Thus, we can confine our attention to the models with asymptotically 
free $SU(2)_H$ without loss of generality.

{}The moduli space ${\cal M}$ is multiply connected, {\em i.e.}, it has 
different branches. Thus, there are points where more than one branch 
meet, yet these branches have no other intersection points. Some of the 
branches lead to models with no hidden sector. We will not consider them 
here.

{}Let us summarize the possibilities for the $SU(2)_H$ matter content. 
There are models with one and three ``flavors" (these are some of the 
$SO(10)$, $SU(5)$ and $SU(6)$ models \cite{class10,class5}), and this 
feature is generic in these models, {\em i.e.}, there are branches in  
the moduli space with this matter content. Note that there are no models 
with $4$ chiral doublets (two ``flavors") of $SU(2)_H$. (There is also 
one $SU(5)$ model with five ``flavors"
\cite{class5}; this model exists only at a special point in the moduli 
space, and generically, the number 
of ``flavors" is one on the branch where it lives 
because the other four become heavy via a coupling 
to a gauge singlet whose vev is a flat direction. The model with 
one ``flavor" is the $F11(1,0)$ model given in Table IV; see 
subsection D of this section and subsection 4 of 
Appendix \ref{models} for details.) 

{}The $SO(10)$, $SU(5)$ and $SU(6)$ models with one ``flavor" can be 
obtained from the $E_6$ model by adjoint breaking (the vev of the adjoint 
is a flat direction). There are four such breakings: 
$E_6\supset SO(10)\otimes U(1)$, $E_6\supset SU(6)\otimes U(1)$, 
$E_6\supset SU(5)\otimes SU(2) \otimes U(1)$ and 
$E_6\supset SU(5)\otimes U(1)^2$. The models with three ``flavors" are 
connected to the models with one ``flavor" via points with enhanced 
horizontal gauge symmetry where the branches with three ``flavors" 
and one ``flavor" intersect.

{}We note that the generation and anti-generation structure of the 
models with different numbers of the $SU(2)_H$ ``flavors" is the same. 
In particular, the couplings involving the corresponding (grand unified) 
matter fields are similar. Thus, to study the SUSY GUT phenomenological 
aspects of these models it suffices to consider the models with 
one $SU(2)_H$ ``flavor". The conclusions generically hold for all the 
other models. This implies that the $E_6$ model does not require a 
separate study as long as we analyze the $SO(10)$, $SU(5)$ and $SU(6)$ 
models obtained from the former by the adjoint breakings listed above. 
This is because with just one adjoint (${\bf 78}$) and pairs of 
fundamentals (${\bf 27}$) and anti-fundamentals 
(${\overline {\bf 27}}$) of $E_6$, the breaking to the standard model 
gauge group is bound to be via $SO(10)$, $SU(5)$ or $SU(6)$ branchings. 
(Thus, say, $E_6 \supset SU(3)_c \otimes SU(3)_w \otimes SU(3)$ breaking 
would require at least one ${\bf 650}$ of  $E_6$, which is not present
 amongst the massless fields of the $E_6$ model.)

{}Finally, we note that all the models in the moduli space ${\cal M}$ 
are $U(1)$ anomaly free. 

\subsection{The $SO(10)$ Models}

{}One class of $SO(10)$ models are those living in the moduli 
space ${\cal M}$ discussed in the previous subsection. (All of 
these models have been explicitly constructed \cite{class10}.) In the 
context of our discussion there, we need only consider the model that is 
obtained from the $E_6$ model via the adjoint breaking 
$E_6\supset SO(10)\otimes U(1)$. The massless spectrum of this model 
(referred to as the $T1(1,1)$ or $T2(1,1)$ model in 
Ref \cite{class10,kst}) is given in Table II.

{}There is one other $SO(10)$ model that does not live in the moduli space 
${\cal M}$ (referred to as the $T5(1,0)$ or $T6(1,0)$ model in 
Ref \cite{class10}; $T5(1,0)$ and $T6(1,0)$ stand for two different 
ways of constructing the same model \cite{class10}). The massless 
spectrum of this model is summarized in Table III. It has $4$ 
left-handed and $1$ right-handed families, and 
$SU(2)\otimes SU(2)\otimes SU(2)$ hidden sector. The latter is not 
asymptotically free at the string scale as each $SU(2)$ subgroup has 
$16$ chiral doublets (eight ``flavors"). It is possible that some of 
these doublets acquire masses once the horizontal gauge symmetry is 
(partially) Higgsed. Examining the model reveals that the resulting 
model would have six ``flavors" in each $SU(2)$ subgroup. The one-loop 
$\beta$-function coefficient for this matter content is zero, and although 
the group is asymptotically free in the two-loop order, the scale of 
non-perturbative dynamics comes out too low (see section \ref{hidden} for 
details). We note that this model is $U(1)$ anomaly free.

\subsection{The $SU(6)$ Models}

{}Just as in the case of $SO(10)$ models, there are $SU(6)$ models that 
live in the moduli space ${\cal M}$. We need only consider the model that 
is obtained from the $E_6$ model via the adjoint breaking 
$E_6\supset SU(6)\otimes U(1)$. The massless spectrum of this model 
(referred to as the $S1(1,1)$ or $S2(1,1)$ model in Ref \cite{class5})
 is given in Table IV. 

{}Note that the spectrum of the $S1(1,1)$ model is similar to that of the 
$T1(1,1)$ model. In particular, the former can be obtained from the latter 
via replacing $SO(10)\otimes U(1)$ (the last $U(1)$ in the first column of 
Table II) by $SU(6) \otimes U(1)$ (the last $U(1)$ in the first column of 
Table IV). Under this substitution, ${\bf 45}(0)$ of $SO(10)\otimes U(1)$ 
is replaced by ${\bf 35}(0)$ of $SU(6)\otimes U(1)$, and 
${\bf 16}(-1)+{\bf 10}(+2)+{\bf 1}(-4)$ is replaced by 
${\bf 15}(0)+{\overline {\bf 6}}(+1)+{\overline {\bf 6}}(-1)$. 
(The complex conjugates of these states are substituted similarly.) 
This replacement procedure can be applied to all the $SO(10)$ models in 
the moduli space ${\cal M}$ to obtain the corresponding $SU(6)$ models. 
(Some of these models have been explicitly constructed previously 
\cite{class5}; others should also exist, albeit their explicit constructions 
have not been presented elsewhere.) In any case, for our purposes it suffices 
to consider the $S1(1,1)$ model. (Here we should mention that starting 
from the $T5(1,0)$ model that does not live in ${\cal M}$, we could obtain 
the corresponding $SU(6)$ model $S5(1,0)$. This model has not been explicitly 
constructed, but it should exist just as other $SU(6)$ models that do live in 
the moduli space ${\cal M}$. The hidden sector of this model is the same as 
that of the $T5(1,0)$ model, so that the discussion presented for the latter 
in the previous subsection also applies to the former.) 

{}There are five additional $SU(6)$ models that do not live in the moduli 
space ${\cal M}$. Two of them have been previously constructed, whereas the 
explicit construction of the other three models appears in 
Appendix \ref{models} for the first time. Next, we turn to describing 
these five models.

{}One model, referred to as $S1$ (or $S3$) in Ref \cite{class5},
 has 6 left-handed and 3 right-handed families, and 
$SU(3)_H$ hidden sector with 3 chiral triplets and antitriplets 
(corresponding to three ``flavors"). The model has anomalous $U(1)$. 
Examining the superpotential of the model reveals that generically one 
of the ``flavors" of $SU(3)_H$ is heavy, and the other two are light 
(see section \ref{hidden} for details). The massless spectrum of this 
model is given in Table V. 

{}Another model referred to as $S2$ (or $S4$) in Ref \cite{class5},
 has 3 left-handed and no right-handed families (the number 
of families in the $SU(6)$ case is determined by the number of ${\bf 15}$'s), 
plus additional six ${\bf 6}+{\overline {\bf 6}}$ pairs. The hidden sector 
of the model is $SU(2)\otimes SU(2)$, where the first subgroup has $6$ 
chiral doublets (3 ``flavors"), whereas the second one has $10$ chiral 
doublets (5 ``flavors"). Some of these come in the bi-fundamental 
representation of $SU(2)\otimes SU(2)$. The model has anomalous $U(1)$. 
Examining the superpotential of the model reveals that generically two 
of the ``flavors" of the second $SU(2)$ subgroup are heavy, and one is 
left with 3 ``flavors" in each of the two subgroups (see section 
\ref{hidden} for details). The massless spectrum of this model is 
given in Table V.

{}One of the models constructed in Appendix \ref{models}, referred 
to as $\Sigma 1$, has 5 left-handed and 2 right-handed families, and 
$SU(3)_H$ hidden sector  with no matter. The massless spectrum of this 
model is given in Table VI.
Its part charged under the grand unified gauge group is very similar to 
that of the model $S1(1,1)$. The only difference is in the Higgs sector of 
${\bf 6}+{\overline {\bf 6}}$'s. Note that this model does not have 
anomalous $U(1)$.

{}Another model, referred to as $\Sigma 2$, has 3 left-handed and no 
right-handed
families, and the $SU(2)_H$ hidden sector with two chiral doublets 
(one ``flavor"). It is also $U(1)$ anomaly free. Its massless spectrum 
is given in Table VI. In Appendix \ref{S2} we show that this model is 
connected to the $S2$ model via a classically flat direction. 

{}Finally, the third model of Appendix \ref{models}, referred to as 
$\Sigma 3$, has
3 left-handed and no right-handed families, and $SU(4)_H$ hidden 
sector with 3 chiral quartets and antiquartets (corresponding to 
three ``flavors"). The model has anomalous $U(1)$. Examining the 
superpotential of the model reveals that generically two of 
the ``flavors" of $SU(4)_H$ are heavy, and the other one is 
light (see section \ref{hidden} for details).
The massless spectrum of this model is given in Table VII. Note that 
unlike the other two $SU(6)$ models ($S2$ and $\Sigma 2$) with 3 
left-handed and no right-handed families, the $\Sigma 3$ model is 
minimal in the sense that it is a minimal $SU(6)$ extension of the 
minimal $SU(5)$ model \cite{Georgi}. That is, the $\Sigma 3$ model 
possesses no ${\bf 6}+{\overline {\bf 6}}$ Higgs sector. 
Phenomenologically this is unappealing since the doublet-triplet 
splitting cannot be achieved without fine-tuning (just as in the 
minimal $SU(5)$ model).

{}In Appendix \ref{models} we also construct four other models referred 
to as $\Sigma 1A$, $\Sigma 1B$, $\Sigma 2A$ and $\Sigma 2B$. The 
$\Sigma 1A$, $\Sigma 1B$ models are connected to the $\Sigma 1$ model 
by classically flat directions.
Similarly, $\Sigma 2A$ and $\Sigma 2B$ live in the same moduli space as 
the $\Sigma 2$ model. The massless spectra of these models are given in 
Table VIII and Table IX. Note that these models have enhanced horizontal 
gauge symmetry ($U(1)$ is enhanced to $SU(2)_3$), and at the enhanced 
symmetry point they possess 4 chiral supermultiplets in ${\bf 20}$ of 
$SU(6)$. (Also note that the $\Sigma 2A$ and $\Sigma 2B$ models have 
3 ``flavors" of $SU(2)_H$ at the enhanced symmetry point.)

{}The appearance of ${\bf 20}$'s of $SU(6)$ is amusing as they have been 
considered in certain scenarios where the doublet-triplet splitting is 
achieved via the pseudo-Goldstone mechanism \cite{pseudo}. That is, 
electroweak Higgs doublets arise as pseudo-Goldstones in breaking 
$SU(6)$ grand unified gauge symmetry down to that of the Standard 
Model. These doublets then do not couple to matter that easily 
(to produce, say, top Yukawa coupling) unless some special 
arrangements are made. One of the solutions proposed in the 
literature \cite{pseudo} involves (an odd number of) 
${\bf 20}$'s of $SU(6)$ (so that they do not pair up and get Dirac mass). 
However, in the models $\Sigma 1A$, $\Sigma 1B$, $\Sigma 2A$ and $\Sigma 2B$ 
the ${\bf 20}$'s (whose number is even) are vector-like and are massless 
only at the special point (of enhanced horizontal gauge symmetry) in the 
moduli space discussed above. Generically they are heavy and thus 
decouple from the massless sector. 
We will, therefore, regard the $\Sigma 1A$, $\Sigma 1B$, 
$\Sigma 2A$ and $\Sigma 2B$ as some special points in the respective 
moduli spaces of models $\Sigma 1$ and $\Sigma 2$, and consider only 
the latter for phenomenological purposes.

\subsection{The $SU(5)$ models}                      

{}As we already mentioned, there are various $SU(5)$ models most of 
which are connected (via flat directions) to the $E_6$, $SO(10)$ and 
$SU(6)$ models discussed above, and some of which are isolated points 
\cite{kt,kt5,class5}. 
All of these models have only one adjoint but no higher dimensional Higgs 
fields in their massless spectra. Thus, they do not possess ingredients 
necessary for solving the doublet-triplet splitting without fine-tuning 
(as all the known solutions are based on existence of massless 
higher-dimensional Higgs fields). We will, therefore, not consider 
any of these 
models in detail. For illustrative purposes 
we give one of these models (referred to as $F11(1,0)$ in Ref \cite{class5}) 
in Table IV. (Note that this model lives in the moduli space ${\cal M}$, and 
can be obtained from the $E_6$ model via the adjoint breaking 
$E_6\supset SU(5)\otimes SU(2) \otimes U(1)$.) 

{}We note that the $SO(10)$ and $SU(6)$ models {\em a priori} do not look 
as hopeless (from the doublet-triplet splitting point of view) as the 
$SU(5)$ models, and this is why we investigate them in more detail.   

\subsection{Summary}

{}Let us summarize the above discussion, and list the models we should 
focus upon. As far as $SO(10)$ models are concerned, we will concentrate 
on the $T1(1,1)$ model. We will also show in section \ref{hidden} 
that the hidden sector in the $T5(1,0)$ model does not get strong at 
desired energy scale. The $SU(6)$ models that we need to consider are 
$S1(1,1)$, $S1$, $S2$, $\Sigma1$ and $\Sigma2$. 
We will also point out some of the
phenomenological problems in the $\Sigma3$ model. Finally, as we 
explained before, we need not consider the $E_6$ model 
separately.     

{}Here we would like to emphasize one feature that is common to all of 
the three-family grand unified models: they have one adjoint but no higher 
dimensional Higgs fields in their massless spectra.

\section{Hidden Sector Dynamics}\label{hidden}

{}In this section we discuss some issues relevant for strong coupling 
dynamics in the hidden sector. The latter is important for possible 
supersymmetry breaking and stabilization of moduli. Although the 
current state of the art in this subject does not always allow one to 
carry out detailed quantitative analysis, in models with relatively 
simple hidden sectors (and all of the models we are studying in this 
paper are of this type) it is oftentimes possible to see whether 
supersymmetry is broken at all, and to estimate the scale of supersymmetry 
breaking. If in a given model this scale comes out too high compared 
with the electroweak scale $M_{EW}\sim 100~{\mbox {GeV}}$, then it is 
difficult to imagine how the electroweak gauge hierarchy would be 
explained in such a model without fine-tuning. On the other hand, 
if this scale comes out way below $M_{EW}$, then supersymmetry 
breaking in such a model is not compatible with experiment.   

\subsection{Hidden Sector Scale}

{}Thus, consider a hidden sector with a 
simple gauge group $G_H$ and some matter content (some number of ``flavors") 
with the $\beta$-function coefficient $b_0$ at the string scale $M_{st}$. In 
some cases there could be a scale $M_X$ below the string scale at which some 
of the matter charged under $G_H$ becomes heavy and decouples. Such a scale, 
for instance, could be an anomalous $U(1)_A$ breaking scale, and if some of 
the matter couples 
(say, via three-point Yukawa couplings)
to singlets (charged under $U(1)_A$) that acquire vevs to break $U(1)_A$, 
then it could develop Dirac mass. Let the $\beta$-function coefficient below 
the scale $M_X$ be ${\tilde b}_0$. Let ${\tilde b}_0>0$, so that the hidden 
sector is asymptotically free below the scale $M_X$. Above the scale $M_X$ 
it may (in which case $b_0\geq 0$) or may not ({\em i.e.}, $b_0<0$ be 
asymptotically free. It is simple to see that the gauge coupling of $G_H$ 
blows up at the scale $M_H$ given by
\begin{equation}\label{M_H}
 M_H=\left({M_X\over{M_{st}}}\right)^{1-b_0/{\tilde b}_0} 
             M_{st}\exp(-{2\pi\over
 {\alpha_{st} {\tilde b}_0}})~.
\end{equation}    
Here $\alpha_{st}$ is the gauge coupling of $G_H$ at the string scale. Note 
that in all of our models the hidden sector gauge group is realized via a 
level-1 current algebra, while the grand unified gauge group is realized 
via a level-3 current algebra. Thus, the relation between the gauge 
couplings of $G_H$ and the observable sector (let the latter gauge 
coupling be $\alpha^\prime_{st}$) at the string scale is given by 
$\alpha_{st}=3\alpha^\prime_{st}$. Also note that if any of these models 
were to match the experimental data, we would ultimately have to assume 
that $\alpha^\prime_{st}\geq \alpha_{GUT}$, where $\alpha_{GUT}\sim 1/24$ 
is the unification coupling obtained by extrapolating the LEP data with 
the assumption of minimal matter content and superpartner thresholds at 
$\sim 1~{\mbox {TeV}}$. Thus, we necessarily have $\alpha_{st}\geq 1/8$. 
For the value $\alpha_{st}=1/8$, we can estimate the string scale \cite{Kap} 
$M_{st}\sim 5g_{st}\times 10^{17}~{\mbox {GeV}}=6\times 10^{17}~{\mbox {GeV}}$
(note 
that $\alpha_{st}=g_{st}^2 /4\pi$).   

{}In models with no anomalous $U(1)$ the natural value for the scale $M_X$ 
is $M_X\sim M_{Pl}$. In the models with anomalous $U(1)$ (precisely which 
are of interest for us), this scale (provided that $X$ is, say, a singlet 
under $G_H$ but carries $Q_X$ anomalous $U(1)_A$ charge whose sign is 
opposite to that of the total $U(1)_A$ trace anomaly ${\mbox{Tr}}\,(Q_A)$) 
can be estimated from the vanishing of the corresponding Fayet-Iliopoulos 
$D$-term. 

\subsection{Supersymmetry Breaking}

{}Next, we turn to the supersymmetry breaking in the models we are studying. 
In Appendix \ref{SUSYBR} we provide all the necessary details, 
so our discussion here will be brief.

{}First consider the $\Sigma3$ model. Its hidden sector is $SU(3)$ with 
no matter. 
There is a possibility that local supersymmetry is broken dynamically 
if suitable 
non-perturbative corrections are present in the 
K{\"a}hler potential \cite{Wu}, but the question remains open since computing 
these corrections is out of reach of our present technology. In any case, 
even if supersymmetry was broken dynamically in the hidden sector of the 
$\Sigma3$ model, it would be transmitted (via the standard supergravity 
mediated scenario \cite{SUGRA}) to the observable sector at a scale 
much higher than the one favored by phenomenology (see Appendix \ref{SUSYBR}).

{}Another case we need to consider is $SU(N_c)$ with $0<N_f<N_c$, where $N_f$ 
is the number of ``flavors". Within the framework of global supersymmetry, 
this case 
is known to have no quantum vacuum as the meson fields 
(which are quark bi-linears) 
have a runaway behavior. That is, the there is a non-perturbative 
superpotential such that 
the meson $F$-terms (as well as the superpotential itself) are 
non-vanishing at any finite values of the meson vevs. Once we 
incorporate this theory within the local supergravity framework, 
however, the tree-level K{\"a}hler potential is enough to stabilize 
the meson vev \cite{Gia},
and local supersymmetry is broken. All the details can be found in Appendix \ref{SUSYBR}.
There we also give the corresponding supersymmetry breaking scales for the models $S1$
($SU(3)$ with two ``flavors''), $T1(1,1)$, $S1(1,1)$ and $\Sigma2$ ($SU(2)$ with one ``flavor''), and $\Sigma3$ ($SU(4)$ with one ``flavor''). We find that the supersymmetry breaking scales in the models $S1$ and $\Sigma3$ come out too high compared with $M_{EW}$, whereas for the models $T1(1,1)$, $S1(1,1)$ and $\Sigma2$ they come out way below $M_{EW}$ unless it is assumed that the supersymmetry breaking messenger scale is $\sim10^{10}-10^{11}~{\mbox {GeV}}$. In these models obtaining such low scales for the messenger $U(1)$'s seems to require some fine-tuning. 

{}We are left with only one model to consider, namely, the $S2$ model. 
Below the anomalous $U(1)$ breaking scale this model has the following 
matter (chiral supermultiplets) in the hidden sector (the hidden sector 
gauge group is $SU(2)\otimes SU(2)$): $2(\Box,\Box)$, $2(\Box,{\bf 1})$ 
and $2({\bf 1},\Box)$ ($\Box$ stands for the fundamental, {\em i.e.}, 
doublet, of $SU(2)$) with different dynamically generated scales 
$\Lambda_1$ and $\Lambda_2$ for the two $SU(2)$'s, and the tree-level 
superpotential. In Appendix \ref{S2} we investigate the 
non-perturbative dynamics in this model. We show that there is no 
dynamically generated superpotential, and the quantum moduli space 
is the same as the classical one, unless one of the $SU(2)$ subgroups is completely Higgsed.
In the latter case we simply flow into the $\Sigma2$ model (which has $SU(2)$ with one ``flavor'' in its hidden sector), which we already discussed above.  

\subsection{Summary}

{}Thus, the analysis in Appendix \ref{SUSYBR} seems to indicate that 
none of the models at hand are flawless as far as supersymmetry 
breaking is concerned. To make more definitive 
conclusions about phenomenological viability of these models, in the 
subsequent sections 
we will discuss issues that should be resolved at tree-level once the 
non-perturbative dynamics is assumed to work in favor of the models we 
are studying. If we can find some 
unsatisfactory features even at this level, then phenomenological 
viability of the models will most certainly be (at least) difficult 
to render.

\section{Aspects of SUSY GUT Phenomenology}\label{gut}

{}In this section we address the following three 
SUSY GUT phenomenological issues of the 
models we are studying: ({\em i}) doublet-triplet 
splitting, ({\em ii}) $R$-parity violating terms, and ({\em iii}) 
Yukawa mass matrices. The first two issues are related to proton 
stability. In the third one we will confine our attention to figuring 
out if one can naturally get one of the three up-quarks to be heavy 
(top-like) while having the other two light. We note that all of these 
questions in a phenomenologically viable model should be resolved at the 
level of perturbative superpotential. Thus, we will assume that the 
non-perturbative dynamics in the hidden sector discussed in the previous 
section works out in favor of phenomenological viability of these models. 
Then we can see whether the three aspects mentioned above can work out 
without some unnatural fine-tuning of vevs, which in this approach are 
free parameters.

\subsection{The $T1(1,1)$ Model}

{}In Ref \cite{kstv} we studied the $T1(1,1)$ model in detail.
Let us first summarize the important features of this model (see 
Table II for details). There is one adjoint Higgs field (${\bf 45}$), 5 
generations (${\bf 16}$'s), 2 antigenerations (${\overline {\bf 16}}$'s), 
7 Higgs fields (${\bf 10}$'s), and various singlets of $SO(10)$ grand 
unified gauge group. Note that these fields descend from the  adjoint 
Higgs field (${\bf 78}$), 5 generations (${\bf 27}$'s), 2 antigenerations 
(${\overline {\bf 27}}$'s), and singlets of $E_6$ in the $E_6$ model. 
(Recall the branching ${\bf 27}={\bf 16}(-1)+{\bf 10}(+2)+{\bf 1}(-4)$ 
under the breaking $E_6 \supset SO(10)\otimes U(1)$. Note that 
${\bf 78}={\bf 45}(0)+{\bf 1}(0)+{\bf 16}(+3)+{\overline {\bf 16}}(-3)$. 
The ${\bf 16}(+3)+{\overline {\bf 16}}(-3)$ fields from the adjoint of 
$E_6$ are missing in the $SO(10)$ model as they are eaten in the 
super-Higgs mechanism upon breaking $E_6$ down to $SO(10)\otimes U(1)$.)

{}We note that due to the $F$-flatness conditions in the 
superpotential of the $T1(1,1)$ model (derived in \cite{kst}) one 
of the ${\bf 10}$'s does not couple to any of the other fields and 
can be dropped out of the analysis. Thus, we have only 6 ${\bf 10}$'s 
to consider. For the same reason of $F$-flatness conditions, only one 
${\bf 16}+{\overline {\bf 16}}$ pair becomes heavy at the GUT scale. 
In particular, this implies that another ${\bf 16}+{\overline {\bf 16}}$ 
can only become heavy at a much lower scale (of the order of $M_{SUSY}$ 
due to relaxation of the $F$-flatness conditions). Thus, we have 4 ${\bf 16}$'s and 1 
${\overline {\bf 16}}$ to consider below the GUT scale.

{}Next, let us consider the possible scenarios for breaking $SO(10)$ down 
to $SU(3)_c\otimes SU(2)_w\otimes U(1)_Y$. There is only one adjoint Higgs 
field in the massless spectrum of the model. The adjoint must break $SO(10)$ 
down to its subgroup that contains $SU(3)_c\otimes SU(2)_w\otimes U(1)_Y$. 
This implies that the adjoint $\Phi$ must acquire a vev in the following 
direction: $\Phi=\epsilon\otimes {\mbox{diag}}(a,a,a,b,b)$, where $\epsilon$ 
is a $2\times 2$ antisymmetric matrix.

{}Now, let us discuss the Higgs sector (${\bf 10}$'s). We will use the 
notation ${\bf 10}_i$, $i=1,\dots,6$, to refer to these Higgs fields. 
The mass matrices for the fields ({\em i.e.}, Higgs doublets and triplets) 
coming from ${\bf 10}_i$ have the following form: 
$\lambda_{ij} {\bf 10}_i {\bf 10}_j+\lambda^\prime_{ij} {\bf 10}_i 
{\bf 10}_j {\bf 45}$. Here $\lambda_{ij}$ are symmetric couplings 
(coming from the couplings of ${\bf 10}_i$ to singlets), whereas 
$\lambda^\prime_{ij}$ are antisymmetric couplings (coming from the 
couplings of ${\bf 10}_i$ to singlets and the adjoint). The latter 
couplings are antisymmetric since the adjoint ${\bf 45}$ of $SO(10)$ 
is antisymmetric. In Ref \cite{kstv} we showed that generically all six 
pairs of doublets and triplets coming from ${\bf 10}_i$ are heavy due to 
the above couplings. To have a pair of light doublets one then needs one 
fine-tuning (of vevs that determine the above couplings). 

{}Suppose now we have one pair of light doublets and all the other fields 
coming from ${\bf 10}_i$ are heavy (in doing this we have to impose one 
fine-tuning, but let us do this as we are trying to make another point). 
There are terms in the superpotential that have the following form: 
$\lambda_{ABi} {\bf 16}_A
{\bf 16}_B {\bf 10}_i$. Here ${\bf 16}_A$, $A=1,\dots,5$, are the five 
generations. Note that in order to break $SO(10)$ all the way down to 
$SU(3)_c\otimes SU(2)_w\otimes U(1)_Y$, one ${\bf 16}+{\overline {\bf 16}}$ 
pair has to acquire a vev (along with the adjoint). This pair is a linear 
combination of the five generations and two antigenerations. Thus, once this 
pair has a non-zero vev we ultimately get mixing between the fields in 
${\bf 10}_i$, {\em i.e.}, the Higgs sector, and ${\bf 16}_A$, {\em i.e.}, 
the matter sector. Here one should be careful as {\em a priori} the mixing 
could be between states that become heavy and decouple. Thus, what one 
really needs to check is if there is mixing between the light pair of 
doublets from ${\bf 10}_i$ and the three light generations that would 
have to correspond to the three families of quarks and leptons below 
the $M_{SUSY}$ scale. Generically, such a mixing is present, and to avoid 
it one fine-tuning is required. 

{}The reason why the above mixing should be avoided is the following: it 
generically generates $R$-parity violating terms which 
would lead to too rapid proton decay. Thus to avoid these $R$-parity 
violating terms we need another fine-tuning as described above. Note 
that the $R$-parity violating terms in this case are ``vev induced", 
{\em i.e.}, they appear after certain fields acquire vevs. The reason 
is that there is no clear separation 
between the matter sector and the Higgs sector in this model at the 
string scale, {\em i.e.}, no well defined 
$R$-parity. This turns out to be a generic feature of all of the models we 
are studying.

{}Finally, we would like to address the issue of Yukawa mass matrices. 
Suppose we take the pair of light Higgs doublets coming from ${\bf 10}_i$ 
(this requires one fine-tuning) and impose the requirement that the 
$R$-parity violating terms be absent (this requires another fine-tuning). 
Then we may ask a question of what the Yukawa coupling matrix of these 
Higgs fields to the three generations looks like. In Ref \cite{kstv} we 
showed that generically this $3\times 3$ Yukawa (mass) matrix has rank 3, 
and, therefore, two additional fine-tunings are required to get rank 1 
{\em i.e.} to have only one heavy (top-like) quark with electric 
charge $+2/3$. 

{}In conclusion we see that ( according to the analysis of Ref \cite{kstv}) 
the $T1(1,1)$, and all the other three-family $SO(10)$ grand unified string 
models of Ref \cite{class10}, do not meet the above three phenomenological 
requirements at tree-level. To achieve desired phenomenological features 
they generically require multiple fine-tunings. The only chance for them 
to arise 
is via some very contrived non-perturbative dynamics (which by some 
``string miracle" would have to fix all the vevs at their fine-tuned values).
This seems unnatural and improbable.       

\subsection{The $SU(6)$ Models}

{}In this subsection we discuss some of the issues in $SU(6)$ SUSY GUT 
phenomenology. In Appendix \ref{GUT6} the reader can find more details 
of our analysis for particular models we are considering here.

{}As far as the doublet-triplet splitting is concerned, there is a crucial 
difference between $SO(10)$ and $SU(6)$ GUTs. It can be easily understood 
from the following point of view. In $SO(10)$ GUTs quarks and leptons 
come from ${\bf 16}$ irreps, while doublet and triplet Higgses come 
from ${\bf 10}$ irreps. Note that ${\bf 10}$ of $SO(10)$ does {\em not} 
contain singlets of $SU(3)_c\otimes SU(2)_w\otimes U(1)_Y$. Thus, they 
cannot participate in breaking the $SO(10)$ grand unified gauge group 
down to that of the Standard Model. In our models, therefore, the only 
fields that can take part in reducing $SO(10)$ to 
$SU(3)_c\otimes SU(2)_w\otimes U(1)_Y$ are the ${\bf 45}$ and a 
${\bf 16}+{\overline{\bf 16}}$ pair. Then the doublets and triplets
 from the ${\bf 10}$'s generically couple to these fields in much 
the same way, and one needs fine-tuning of vevs (as we discussed in
 the previous subsection) to achieve doublet-triplet splitting.

{}In $SU(6)$ GUTs the situations is quite different. Each ``generation"
 arises from one ${\bf 15}$ and a pair of ${\overline{\bf 6}}$'s. There
 are some states within these irreps that are extra compared with the 
Standard Model quarks and leptons. They can play the role of Higgs 
doublets and triplets. In this case, however, the adjoint must acquire 
vev in the direction that breaks the $SU(6)$ to $SU(5)\otimes U(1)$. 
This can be easily seen from examining the quantum numbers of the states 
in ${\bf 15}$ and ${\overline{\bf 6}}$'s in terms of 
$SU(3)_c\otimes SU(2)_w\otimes U(1)_Y$. The situation here is the same 
as in  $SU(5)$ GUTs, and doublet-triplet splitting cannot be achieved 
without fine-tuning. (This is true taking into account that in our 
models there are no
``exotic" massless higher-dimensional Higgs fields.) There is an 
alternative scenario, namely, to have extra ${\bf 6}+{\overline{\bf 6}}$ 
pairs. Then Higgs doublets and triplets could come from these additional 
fields. Naturally, the extra states in ${\bf 15}$ and 
${\overline{\bf 6}}$'s must become heavy which can be 
achieved by having Yukawa couplings of 
the type ${\bf 15}\cdot{\overline{\bf 6}}\cdot{\overline{\bf 6}}$ 
and giving vev to a ${\bf 6}+{\overline{\bf 6}}$. This is allowed
 since these states do contain a singlet of 
$SU(3)_c\otimes SU(2)_w\otimes U(1)_Y$. Note that in this case the adjoint
 (${\bf 35}$) of $SU(6)$ must acquire vev in a direction different from 
the one that breaks $SU(6)$ to $SU(5)\otimes U(1)$ (or else a 
${\bf 15} +{\overline {\bf 15}}$ pair would have to acquire vev 
to break $SU(5)\otimes U(1)$ further down to 
$SU(3)_c\otimes SU(2)_w\otimes U(1)_Y$, which is not possible
 since a ${\bf 6}+{\overline{\bf 6}}$ pair already has 
acquired a vev). There are two possibilities
 here for the adjoint breaking: $SU(6)\supset SU(4)_c\otimes SU(2)_w\otimes U(1)$ and $SU(6)\supset SU(3)_c\otimes SU(3)_w\otimes U(1)$. In the first case ${\bf 6}+{\overline{\bf 6}}$ vev breaks $SU(4)_c\otimes U(1)$ to  
$SU(3)_c\otimes U(1)_Y$, whereas in the second case $SU(3)_w\otimes U(1)$ is broken to $SU(2)_w\otimes U(1)_Y$. It is not difficult to see that doublet and triplet Higgses in these scenarios are not treated on the equal footing, and, therefore, doublet-triplet splitting may be achieved without fine-tuning \cite{pseudo}.

{}One of the conclusions we can immediately draw from the above discussion is that the minimal three-family $SU(6)$ model, {\em i.e.}, the $\Sigma3$ model, does not seem to be phenomenologically viable, and we will not consider it any further.

{}Although the doublet-triplet splitting problem can be relatively 
easily solved in $SU(6)$ GUTs, there are some additional difficulties. 
The most pressing of those is generating the top-quark mass. Note that 
to generate masses for up-quarks in general (assuming that massless 
${\bf 20}$'s of $SU(6)$ are absent), one needs couplings of the form 
${\bf 15}\cdot{\bf 15}\cdot{ {\bf 6}}\cdot{{\bf 6}}\cdot f({\bf 35})$.
 Here one of the ${\overline {\bf 6}}$'s acquires vev at a high scale 
($\sim\!\!M_{st}$), whereas the other one contains the electroweak doublet 
responsible for generating the up-quark masses. The form of the 
polynomial function $f({\bf 35})$ depends upon details of the scenario. 
In any case, if couplings of this type are absent in a given model, it 
can be rendered phenomenologically inviable. In Appendix \ref{SupSU(6)} 
we give superpotentilas for the $SU(6)$ models $S1(1,1)$, $S1$, $S2$, 
$\Sigma1$ and $\Sigma2$. From these superpotentials one can see that the 
$S1(1,1)$, $S1$, and $\Sigma1$ models do
possess the above type of couplings, and we will investigate them further. The models $S2$ and $\Sigma2$, however, lack such couplings altogether, and are, therefore, phenomenologically inviable. We will not discuss them any further.  

{}Thus, we are left with the $S1(1,1)$, $S1$ and $\Sigma1$ models to consider.
First, we would like to address the issue of Yukawa mass matrices. 
The $S1(1,1)$ and $\Sigma1$ models are very similar. In particular, the 
terms relevant for generating Yukawa mass matrices are of the same form. 
In Appendix \ref{GUT6} we present the details, so here we simply state 
the results: there are two top-like quark generations in these models, 
and unlike, say, the $T1(1,1)$ model, even fine-tuning does not seem to 
help in obtaining rank-1 up-quark Yukawa mass matrix.  

{}Note that since in the $S1(1,1)$ and $\Sigma1$ models there is no well-defined $R$-parity at the string scale ({\em i.e.}, the Higgs sector is not clearly separated from the matter sector), there will generically be $R$-parity violating terms once a ${\bf 6}+{\overline{\bf 6}}$ pair acquires vev. These terms come from the Yukawa couplings ${\bf 15}\cdot{\bf 6}\cdot{\bf 6}$. Some degree of fine-tuning will be needed to suppress these $R$-parity violating terms.

{}Finally, we comment on the $S1$ model without going into any detail. The problems with Yukawa mass matrices and vev-induced $R$-parity violating terms that we have encountered in the $T1(1,1)$, $S1(1,1)$ and $\Sigma1$ models are present in the $S1$ model as well. (In this model, however, it appears to be possible to get rank-1 up-quark Yukawa mass matrix in the leading approximation at the cost of fine-tuning.)
This can be verified by analyzing the superpotential for this model given in Appendix \ref{SupSU(6)}.
   
\subsection{Summary}

{}In this section we saw that fine-tuning seems to be required 
to achieve doublet-triplet splitting in the three-family $SO(10)$ grand 
unified string models we are considering. 
In our $SU(6)$ models the doublet-triplet splitting problem can be solved 
via mechanisms already developed in the field theory context \cite{pseudo}. 
All of the models seem to require fine-tuning to suppress vev generated 
$R$-parity violating terms. In some of the models it is possible to 
obtain rank-1 up-quark mass matrix (in the leading order) at the cost 
of fine-tuning, whereas in other models even this seems to be problematic.  

\section{Conclusions and Outlook}\label{conc}

{}None of the three-family grand unified string models 
considered in this review agree with 
phenomenology unless some fine-tuning is involved. The 
possibility of such fine-tuning is 
due to the presence of moduli, or flat directions, in the 
classical supersymmetric vacuum. 
These flat directions could be lifted after dynamical SUSY breaking. 
Then such fine-tuning could {\em a priori} arise via some very 
contrived non-perturbative dynamics (which by some ``string miracle'' 
would have to fix all the vevs at their fine-tuned values). 
The latter possibility seems unlikely taking into account that 
multiple vevs have to be fine-tuned (in {\em a priori}
 uncorrelated fashion).

{}Here we would like to make a comparison between the field theory and 
string GUTs. Naively, one might expect a ``string miracle'' (such as a stringy 
discrete symmetry) that could solve some of the problems present 
in field theory GUTs. This, however, does not seem to be the case in the
 three-family grand unified string models considered in this review. Thus,
 the doublet-triplet splitting problem in the $SO(10)$ and $SU(5)$ string 
models was not any less severe than in their field theory counterparts. 
On the contrary, absence of ``exotic'' higher dimensional Higgs fields 
in the massless spectrum has led us to the conclusion that the
doublet-triplet splitting would require unnatural fine-tuning in these 
models. On the other hand, the doublet-triplet splitting problem in 
the $SU(6)$ string models would have to be achieved via the already 
well-known field theory mechanism \cite{pseudo}. Even in this case 
the field theory approach turns out to have an advantage: the successful 
solution of the doublet-triplet splitting problem does {\em not} guarantee 
proton stability as $R$-parity violating terms might still mediate
 too--rapid proton decay. Within field theory one can simply require 
$R$-parity violating terms to be absent already at the string scale. 
However, in our models there is no well-defined $R$-parity: 
there is simply no clear separation between the matter and Higgs 
sectors at the string scale. This leads to ``vev induced'' $R$-parity 
violating terms. Such $R$-parity violating terms are a common feature 
for all of our models, and could be present even in more generic classes of 
models (which are not necessarily string GUTs). 

{}Finally, we would like to comment on one of the ``original 
motivations'' to consider string GUTs. As discussed in 
Introduction, in string GUTs the gauge and gravitational 
coupling unification is automatic. Here one subtlety arises. 
In order to have agreement with the experimental data (such as 
$\sin^2\theta_W$), one must assume that there are {\em not} 
too many  extra light states
(compared with the MSSM spectrum) below the GUT scale. Such 
states (if in incomplete $SU(5)$ multiplets) generally tend 
to raise the unification scale (for the fixed $\sin^2\theta_W$ 
prediction) at one loop order, and sometimes could also spoil 
the unification of couplings. If they are too light, then the 
GUT scale can rise beyond the string scale which would not be 
compatible with the phenomenological data. Perhaps, some of the most
troublesome of such light states are those in the adjoint of 
$SU(3)_c\otimes SU(2)_w\otimes U(1)_Y$. They are the left-over states 
from the GUT gauge group $G$ breaking down to that of the Standard Model. 
Since the adjoint Higgs in $G$ is a flat direction in perturbative 
heterotic superstring, it is massless perturbatively. Non-perturbative 
effects responsible for supersymmetry breaking can typically generate 
masses of the order of $1~{\mbox{TeV}}$ for these fields. This, however, 
is not acceptable since the GUT scale is pushed up to the string scale 
already for the adjoints as heavy as 
$\sim 10^{12} -10^{13}~{\mbox{GeV}}$ \cite{adj}. There 
exist model-dependent mechanisms that can generate such high 
masses for the adjoints. However, unless these masses  are 
actually much higher (of the order of $10^{16} -10^{17}~{\mbox{GeV}}$), 
the GUT scale is basically the same as the string scale. In this case 
the original motivation for the automatic unification of the gauge and 
gravitational couplings disappears. One could turn this point around and
 argue that this could be another mechanism for solving the gauge 
coupling unification problem in string theory \cite{adj}. However, 
the price one has to pay (doublet-triplet splitting problem) seems 
to be too high.

{}In conclusion we see that the idea of string GUTs has its advantages,
 but it also faces many problems. Our particular models do not seem to 
provide satisfactory solutions to these problems, yet they give 
us a flavor of what a stringy paradigm for GUT model building might be.

\acknowledgments

{}We would like to thank Damiano Anselmi, Ignatios Antoniadis, Philip 
Argyres, Nima Arkani-Hamed, Richard Arnowitt, Paul Aspinwall,
Costas Bachas, Alexander Bais, 
Riccardo Barbieri, Steven Barr, Zurab Berezhiani, Michael Bershadsky, 
Stanley Brodsky, Gerry Cleaver, Mirjam Cveti\v{c}, Keith Dienes, 
Robbert Dijkgraaf, Michael Dine, Lance Dixon, Gia Dvali, John Ellis, 
Alon Faraggi, Glennys Farrar, Matthias Gaberdiel, Larry Hall, 
Martin Halpern, Luis Ib{\'a}{\~n}ez, Andrei Johansen, Elias Kiritsis, 
Costas Kounnas, Finn Larsen, Albion Lawrence, Joe Lykken, 
Marcos Mari{\~n}o, Pran Nath, Peter Nilles, Roberto Peccei, Michael Peskin, 
Joe Polchinski, Erich Poppitz, Philippe Pouliot, Fernando Quevedo, Stuart Raby, 
Lisa Randall, John Schwarz, Yuri Shirman, Michael Spalinski, 
Alessandro Strumia, Zurab Tavartkiladze, Tom Taylor, Angel Uranga, 
Cumrun Vafa, Erik Verlinde, Herman Verlinde, Mikhail Vysotsky, 
Carlos Wagner, Frederic Zamora, Barton Zwiebach, and many others 
for discussions at various times while completing this program of 
constructing 
and studying three-family grand unified string models.
The research of G.S., S.-H.H.T.  and Y.V.-K. was partially supported by 
National Science Foundation. G.S. would also like to thank Joyce M. Kuok 
Foundation for financial support. The work of Z.K. was supported in part 
by the grant NSF PHY-96-02074, and the DOE 1994 OJI award. Z.K. would like to thank CERN Theory Division for their kind hospitality while parts of this work were completed. Z.K. would also like to thank Albert and Ribena Yu for financial support.

\appendix

\section{Supersymmetry Breaking}\label{SUSYBR}

{}In this section we discuss supersymmetry breaking in the models $T1(1,1)$, $S1(1,1)$, $S1$, $\Sigma1$, $\Sigma2$ and $\Sigma3$. (The $S2$ model with be considered separately in Appendix \ref{S2}.) Note that all of these models have one simple non-Abelian gauge group in the hidden sector. All the models except for the $\Sigma1$ model have matter.  

{}Let us be general here and consider $SU(N_c)$ hidden sector with $0<N_f<N_c$ ``flavors''.
The non-perturbative superpotential for this theory is given by \cite{Seiberg}:
\begin{equation}
 W_{non-pert}=(N_c-N_f)\left( {{\Lambda^{3N_c-N_f }}\over {\det(M)} }\right)^{1/(N_c-N_f)}~.
\end{equation}
Here $M^i_{\bar j}\equiv Q^i {\tilde Q}_{\bar j}$ are the meson fields ($Q^i$ and ${\tilde Q}_{\bar j}$ are the fundamental quarks and anti-quarks), and $\Lambda$ is the dynamically generated scale. This superpotential has a ``runaway'' behavior. That is, for any finite values of the meson vevs $M^i_{\bar j}$, the $F$-flatness conditions cannot be simultaneously satisfied and the superpotential does not vanish. This implies local supersymmetry breaking due to K{\"a}hler potential contributions \cite{Gia}. Intuitively this can be understood by noting that once such a theory is coupled to supergravity there is a natural shut-down scale for all the runaway directions, namely, the Plank scale. Generically, this results in local supersymmetry breakdown. 
The supersymmetry breaking scale $M_{SUSY}$ in the observable sector is given by 
\begin{equation}\label{susybr0}
 M_{SUSY}\sim {\vert F \vert \over {M_M}}~.
\end{equation}
Here $M_M$ is the messenger scale, and $F$ is the supersymmetry breaking $F$-term. (Note that for a field $z$ the $F$-term $F_z$ is given by $F_z=\partial W/\partial z+M^{-2}_{Pl} W
\partial K/\partial z$, where $K$ is the K{\"a}hler potential.) Assuming that the meson vevs are stabilized at $\sim M_{Pl}$, and gravity is the messenger of supersymmetry breaking ({\em i.e.}, $M_M\sim M_{Pl}$), the supersymmetry breaking scale can be estimated to be $M_{SUSY}\sim \vert W_{non-pert}\vert /M_{Pl}^2$, where all the vevs of the canonically normalized fields in $W_{non-pert}$ are taken to be $\sim M_{Pl}$. This gives
\begin{equation}\label{susybr}
 M_{SUSY}\sim~{1\over{M_{Pl}^2}} \left( {{M_H^{3N_c-N_f}} 
 \over{M_{Pl}^{2N_f}} }\right)^{1\over {N_c-N_f}}=
 M_{st}\left( {M_{st} \over M_{Pl}}\right)^{{2N_c}\over {N_c-N_f}} \exp(-{2\pi\over
 {\alpha_{st}(N_c-N_f)}})~.
\end{equation}
This formula is in fact also correct for $N_f=0$. In this case we have gaugino condensate $\langle \lambda\lambda\rangle\sim M_H^3$. Then the standard gravity mediated  \cite{SUGRA} scenario predicts the following supersymmetry breaking scale in the observable sector:
\begin{equation}\label{gauginocond}
 M_{SUSY}\sim {{M_H^3}\over{M^2_{Pl}}}={{M^3_{st}}\over{M_{Pl}^2}}\exp(-{2\pi\over
 {\alpha_{st}N_c}})~.
\end{equation}

\subsection{The $\Sigma1$ Model}

{}Since there is no matter in the $SU(3)_H$ hidden sector in this model, supersymmetry breaking would have to occur via gaugino condensation. 
In our case of $SU(3)_H$ with no matter we have $b_0=9$, and according to Eq (\ref{M_H}) we have $M_H\sim 2\times 10^{15}~{\mbox{GeV}}$. Then from Eq (\ref{gauginocond}) the scale 
$M_{SUSY}{\ \lower-1.2pt\vbox{\hbox{\rlap{$>$}\lower5pt\vbox{\hbox{$\sim$}}}}\ }
10^5~{\mbox{TeV}}$. This is too high to be compatible with electroweak gauge hierarchy.

\subsection{The $S1$ Model}

{}The tree-level superpotential relevant for the hidden sector dynamics in this model reads (see Appendix \ref{SupSU(6)} for details and Table V for notation): 
\begin{eqnarray}
 W_{tree} = &\phantom{+}& \lambda_{1}  U_0
               ( U_{++}U_{--} +U_{+-}U_{-+} )
      +  \lambda_{2}  T_0
               ( {\tilde T}_{+} U_{-+} +{\tilde T}_{-} U_{++}) \nonumber\\
     &+& \lambda_{3}  {\tilde T}_0 T_- U_0 +
           \lambda_{4}  {\tilde T}_0 T_0 {\tilde U}_-~.
\end{eqnarray}
Note that $T_0,T_+,T_-$ are chiral triplets of $SU(3)_H$, whereas ${\tilde T}_0,{\tilde T}_+,{\tilde T}_-$ are chiral anti-triplets. $U$'s are $SU(3)_H$ singlets charged under various $U(1)$'s of the model. Thus, the mass matrix $m_{ij}T_i {\tilde T}_j$ ($i,j=0,+,-$) reads:
\begin{equation}
	(m_{ij})=\left( \begin{array}{ccc}
               \lambda_4 {\tilde U}_- & \lambda_2 U_{-+}  &  \lambda_2 U_{++}\\
               0& 0 & 0\\
               \lambda_3 U_0& 0& 0
               \end{array}
        \right)~.
\end{equation}
Next, we have the following $F$-flatness conditions: 
\begin{equation}
 0=\partial W/\partial U_{--}=U_0 U_{++}~,~~~0=\partial W/\partial U_{+-}=U_0 U_{-+}~.
\end{equation}
{}From this we conclude that generically one of the ``flavors'' of $SU(3)_H$ 
is heavy with the mass of the order of $\lambda U \sim M_X$. Therefore, 
below the anomalous $U(1)$ breaking scale $M_X$ we can treat the hidden 
sector as $SU(3)_H$ with two ``flavors''. For the $S1$ model ($N_c=3$ and 
$N_f=2$) $M_{SUSY}$ then comes out way too low. Note that if we assume 
that all four vevs $U_0,U_{++},U_{-+},{\tilde U}_-$ are zero, then 
supersymmetry is not broken at all.

\subsection{The $T1(1,1)$, $S1(1,1)$ and $\Sigma2$ Models}

{}The hidden sector gauge group and matter content are the same for the 
models $T1(1,1)$, $S1(1,1)$ and $\Sigma2$. Here we have $SU(2)_H$ with 
one ``flavor''. Using Eq (\ref{susybr}) one can easily verify that the 
scale $M_{SUSY}$ also comes out way too low.

\subsection{The  $\Sigma3$ Model}

{}The massless spectrum of this model is given in Table VIII. Following
 the rules for computing superpotentials given in Ref \cite{kst}, 
one can easily 
deduce the tree-level superpotential relevant for hidden sector dynamics:
\begin{equation}
 W_{tree}=U_0 (Q_{+} {\tilde Q}_{-}+Q_{-} {\tilde Q}_{+})~.
\end{equation}
Here $U_0\equiv ({\bf 1}, {\bf 1})(+6,0)$, 
$Q_{\pm}\equiv 2({\bf 4},{\bf 1})(-3,-3)$, and
${\tilde Q}_{\pm}\equiv 2({\overline {\bf 4}},{\bf 1})(-3,+3)$. Note 
that the first $U(1)$ is anomalous (with total trace anomaly equal $-72$), 
so that the field $U_0$ acquires vev to break it. Thus, the fields 
$Q_{\pm},{\tilde Q}_{\pm}$ get mass of the order of the anomalous $U(1)$ 
breaking scale $M_X$, and there is only one ``flavor'' 
(namely, $Q\equiv ({\bf 4},{\bf 1})(+3,+3)$ and 
${\tilde Q}_{\pm}\equiv ({\overline {\bf 4}},{\bf 1})(+3,-3)$) below 
this scale. The supersymmetry breaking scale in this model can be 
estimated to be 
$M_{SUSY}{\ \lower-1.2pt\vbox{\hbox{\rlap{$>$}\lower5pt\vbox{\hbox{$\sim$}}}}\ }10^{4}~{\mbox{TeV}}$, which is too high.

\subsection{Comments}

{}In section \ref{survey} we pointed out that there are models 
connected to the $E_6$ model (and, therefore, also to the models 
$T1(1,1)$ and $S1(1,1)$) with $SU(2)_H$ hidden sector with numbers of 
flavors $N_f>2$. In general, for $SU(N_c)$ with $N_f>N_c$ non-perturbative 
superpotential is not generated, and supersymmetry is not broken. So to 
break supersymmetry in these models all but one of the ``flavors'' 
(there are no models with two ``flavors'') would have to acquire masses. 
Then we are back to the $T1(1,1)$ and $S1(1,1)$ models. As we see, in 
none of the above models does the supersymmetry  breaking scale come 
out right. It is either too high (as in the $\Sigma1$ and 
$\Sigma3$ models), or way too low (as in the models $S1$, 
$T1(1,1)$, $S1(1,1)$ and $\Sigma2$). In the latter models one could try 
to replace $M_{Pl}$ by $M_X$ (which could be, say, anomalous $U(1)$ 
breaking scale) and argue that all the runaway vevs are stabilized 
at $M_X$, and also that the mediation of supersymmetry breaking is due to a 
messenger (different from gravity) whose scale $M_M$ is lower than 
$M_{Pl}$ (and is, perhaps, related to $M_X$). This does not really 
help since (as can readily be  
verified from Eq (\ref{susybr})) the main suppression comes from the 
exponential factor $\exp(-{2\pi/{\alpha_{st}(N_c-N_f)}})$, unless 
$M_M$ is taken to be rather low, namely, 
$M_M\sim 10^{10}-10^{11}~{\mbox{GeV}}$. The latter possibility seems 
to be very unlikely
since it is unclear what would generate such a low scale for messenger 
$U(1)$'s in the models 
$S1$, $T1(1,1)$, $S1(1,1)$ and $\Sigma2$, unless some fine-tuning is 
involved. (Note that the messengers in these models could be either 
gravity or $U(1)$'s, and the natural scale for the latter is $M_{Pl}$ 
in the models $T1(1,1)$, $S1(1,1)$ and $\Sigma2$, and $M_{Pl}$ or the 
anomalous $U(1)$ breaking scale $M_X$ in the 
$S1$ model.) To avoid such a large suppression, one could try to decrease 
the number of flavors $N_f$, which {\em a priori} is not impossible if one 
involves higher dimensional operators (such as, say, $\lambda_5$ and 
$\lambda_6$ couplings in the $E_6$ superpotential given in 
Appendix \ref{SupE_6}). A more careful examination of the corresponding 
superpotentials, however, reveals that in the $SO(10)$ models this would 
require giving ${\bf 10}$'s of $SO(10)$ large vevs which would break the 
Standard Model gauge group at a high scale. In the $SU(6)$ models one 
would ultimately have to turn on large vevs for ${\bf 15}$'s of $SU(6)$. 
This necessarily leads to breaking $SU(6)$ via $SU(5)\otimes U(1)$, and 
doublet-triplet splitting then cannot be achieved without fine-tuning 
(see section \ref{gut} and Appendix \ref{GUT6} for details). We, 
therefore, conclude that the models at hand do not seem to give 
rise to realistic supersymmetry breaking patterns.

\section{Hidden Sector Dynamics in the $S2$ Model}\label{S2}

{}In this section we study the hidden sector dynamics in the $S2$ model. 
The tree-level superpotential relevant for the hidden sector dynamics in 
this model reads (see Appendix \ref{SupSU(6)} for details and Table V for 
notation):
\begin{eqnarray}
 W_{tree}=\lambda_1 U_0(d_{++} d_{--}+d_{+-} d_{-+})+
 \lambda_2 (D_+ {\tilde d}_- \Delta_- +D_- {\tilde d}_+ \Delta_+)~.
\end{eqnarray} 

{}Note that the third $U(1)$ in this model is anomalous. In order to 
break the anomalous $U(1)$, the field $U_0$ must acquire a vev. Let the 
corresponding scale be $M_X$. Then the doublets $d_{+\pm}$ and $d_{-\pm}$ of 
the second $SU(2)$ subgroup decouple at this scale. At the end ({\em i.e.}, 
below the scale $M_X$) we are left with the $SU(2)\otimes SU(2)$ hidden 
sector with the following matter content: $2(\Box,\Box)$ 
(from $\Delta_\pm$), $2(\Box,{\bf 1})$ (from $D_\pm$), and $2({\bf 1},\Box)$ 
(from ${\tilde d}_{\pm}$). Here $\Box$ stands for the fundamental, 
{\em i.e.}, doublet, of $SU(2)$. The couplings of the two $SU(2)$'s 
are the same at the string scale, but they are {\em a priori} different at the $M_X$ 
scale for the their runnings between $M_{st}$ and $M_X$ are different. 
Thus, we have different 
dynamically generated scales $\Lambda_1$ and $\Lambda_2$ for the two $SU(2)$'s.
The tree-level superpotential below the scale $M_X$ then reads: 
\begin{eqnarray}\label{tree}
 W_{tree}=
 \lambda_2 (D_+ {\tilde d}_- \Delta_- +D_- {\tilde d}_+ \Delta_+)~.
\end{eqnarray}

\subsection{Exact Superpotential}

{}Next, we would like to understand the non-perturbative dynamics in this model. For later convenience, let us introduce some more appropriate notations. 
Consider $SU(2)_1\otimes SU(2)_2$ theory with $l$ fields 
$Q_{a\alpha\dot{\alpha}}$ transforming as fundamentals under both groups, 
$a = 1,\dots,l$; $2n$ $SU(2)_1$
fundamentals
$L_{i\alpha}$, $i = 1,\dots,2n$; and $2m$ $SU(2)_2$ fundamentals
$R_{p\dot{\alpha}}$, $p = 1,\dots,2m$. Here $\alpha = 1,2$ and 
${\dot{\alpha}} = 1,2$ are the gauge indices of $SU(2)_1$ and $SU(2)_2$, 
respectively. We refer to this theory as $[l,n,m]$ theory.   
This theory is non-chiral thus mass terms are allowed for the matter fields.
(In the $S2$ model we are interested in, we have the $[2,1,1]$ theory with 
$\Delta_\pm$ being $Q_{a\alpha\dot{\alpha}}$, $D_\pm$ being $L_{i\alpha}$ 
and ${\tilde d}_\pm$ being $R_{p\dot{\alpha}}$.)

{}We refer to a pair of fundamentals as one ``flavor''. The field content of 
the
$[l,n,m]$ theory is summarized in the following table. 
The $[l,n,m]$ theory has global symmetry $SU(2l)\otimes SU(2n)\otimes 
SU(2m)\otimes U(1)_A\otimes U(1)_R$. Note that the $U(1)_R$ charge assignment 
is {\em not} unique.

\begin{center}
\begin{tabular}{|c|c|c|c|c|c|c|c|}
\hline
&&&&&&&\\[-2mm]
&$\,SU(2)_1\,$&$\,SU(2)_2\,$&$\,SU(2l)\,$&$\,SU(2n)\,$&$\,SU(2m)\,$
&$U(1)_A$&$U(1)_R$\\[2mm]
\hline
&&&&&&&\\[-2mm]
$Q_{a\alpha\dot{\alpha}}$&$\Box$&$\Box$&$\Box$&1&1&1&0\\[1mm]
$L_{i\alpha}$&$\Box$&1&1&$\Box$&1&1&0\\[1mm]
$R_{p\dot{\alpha}}$&1&$\Box$&1&1&$\Box$&1&0\\[1mm]
$\,\Lambda_1^{6-l-n}\,$&1&1&1&1&1&$2l + 2n$&$\,4-2l-2n\,$\\[1mm]
$\,\Lambda_2^{6-l-m}\,$&1&1&1&1&1&$2l + 2m$&$\,4-2l-2m\,$\\[1mm]
\hline
\end{tabular}
\end{center}

{}Let us concentrate on the $[2,1,1]$ theory. The classical $D$-flat 
directions are given by the following operators:
\begin{eqnarray}
\label{xlr}
X_{ab}&\equiv& {1\over 2} \epsilon^{\alpha \beta}
\epsilon^{\dot{\alpha} \dot{\beta}}
Q_{a\alpha\dot{\alpha}} Q_{b\beta\dot{\beta}}~,\\
Y_{aip}&\equiv& {1\over 2}
 \epsilon^{\alpha \beta} \epsilon^{\dot{\alpha} \dot{\beta}} 
 Q_{a\alpha\dot{\alpha}} L_{i\beta} R_{p\dot{\beta}}~,\\
 {\cal L} &\equiv& {1\over 2}\epsilon^{ij}\epsilon^{\alpha \beta}
 L_{i\alpha} L_{j\beta}~,\\
 {\cal R} &\equiv& {1\over 2}\epsilon^{pq}\epsilon^{\dot{\alpha}\dot{\beta}}
 R_{p\dot{\alpha}} R_{q\dot{\beta}}~. 
\end{eqnarray}
There is the following classical constrain that the above flat directions satisfy: 
\begin{eqnarray}
\label{constraint0}
 {\cal L}{\cal R}X_{ab} = {\cal Y}_{ab}~,
\end{eqnarray}
where
\begin{eqnarray}
 {\cal Y}_{ab}\equiv {1\over 2}
 \epsilon^{ij}\epsilon^{pq} Y_{aip}\,Y_{bjq}~.
\end{eqnarray}

{}We would like to understand whether the classical moduli space is modified 
quantum mechanically, {\em i.e.}, to determine the exact superpotential. Our 
strategy will be to start from a known case and ``integrate in" \cite{int}. 
Then we can check our result in appropriate limits. 
The most convenient known case where we can ``integrate in" was considered in 
Ref \cite{int}. This is the $[1,1,1]$ theory. To get from the $[1,1,1]$ 
theory to the $[2,1,1]$ theory, we have to ``integrate in" one bi-fundamental 
({\em i.e.}, one $Q$ field).

{}The exact non-perturbative superpotential for the $[1,1,1]$ theory 
reads \cite{int}:
\begin{eqnarray}
\label{111}
 {\tilde {\cal W}} = A(X_{11}{\cal L}{\cal R} - {\tilde{\Lambda}}_1^4 
{\cal R} - 
 {\tilde{\Lambda}}_2^4 {\cal L} - {\cal Y}_{11})~.
\end{eqnarray}
Here $A$ is a Lagrange multiplier, and ${\tilde{\Lambda}}_1$ and 
${\tilde{\Lambda}}_2$ are the dynamical scales of the two $SU(2)$'s. 
Next, consider the following superpotential (which is obtained by adding 
Yukawa perturbations and mass terms to Eq (\ref{111})): 
\begin{eqnarray}
\label{in-in}
 {\tilde {\cal W}}^\prime
  &=& A(X_{11}{\cal L}{\cal R} - m_0\Lambda_1^3{\cal R} - 
 m_0\Lambda_2^3{\cal L} - {\cal Y}_{11})\nonumber\\
 && - (m_0 X_{22} + 2m_1 X_{12} +{m_1^2\over
  {m_0}}X_{11}) \nonumber\\
  &&-{1\over 2}\epsilon^{ij}\epsilon^{pq}\lambda_{ip}(Y_{2jq}+   
    {m_1\over{m_0}}Y_{1jq})\nonumber\\
  && -{1\over{8m_0}}\epsilon^{ij}\epsilon^{pq}\lambda_{ip}\lambda_{jq}
   {\cal L}{\cal R}~. 
\end{eqnarray}
Here $\Lambda_1$ and $\Lambda_2$ are the dynamical scales of the two 
$SU(2)$'s in the $[2,1,1]$ theory. The scale matching is given by:
\begin{eqnarray}
\label{scale}
 {\tilde {\Lambda}}_1^4 = m_0 \Lambda_1^3~,~~~
 {\tilde {\Lambda}}_2^4 = m_0 \Lambda_2^3~.
\end{eqnarray}

{}Next, we integrate out the chiral superfields 
$m_0$, $m_1$ and $\lambda_{ip}$ from the superpotential (\ref{in-in}). 
The result reads:
\begin{eqnarray}
\label{211}
{\cal W} = - {{\det({\cal L}{\cal R}X_{ab}-{\cal Y}_{ab})}\over{
{{\cal L}{\cal R}(\Lambda_1^3{\cal R} + \Lambda_2^3{\cal L})}}}~.
\end{eqnarray}
Here the determinant in the numerator is understood for the $2\times 2$ 
matrix with elements labeled by the indices $a,b$.

{}From the superpotential (\ref{211}) we see that the classical moduli 
space is not modified quantum mechanically. The non-perturbative 
superpotential (\ref{211}) vanishes due to the classical constraints 
(\ref{constraint0}). The exact superpotential then is given by the 
tree-level expression (\ref{tree})
\begin{eqnarray}\label{exact}
 W_{exact}=W_{tree}=\lambda_2(Y_{121}+Y_{212})=\lambda_2(Q_1 L_2 R_1+Q_2 
 L_1 R_2)~.
\end{eqnarray}
The theory has Higgs and Abelian Coulomb phases. (The Abelian Coulomb phase of this theory was recently discussed in Ref \cite{Coulomb}). The latter occurs when both 
$Q_1$ and $Q_2$ acquire vevs to break 
$SU(2)\otimes SU(2)\supset SU(2)_{diag}\supset U(1)$. Note that 
if we give a vev to $Q_1$ in the direction that breaks 
$SU(2)\otimes SU(2)$ to $SU(2)_{diag}$, and also to the 
$SU(2)_{diag}$ singlet in $Q_2$, we will get the $N=1$ 
theory with $SU(2)_{diag}$ gauge group and one chiral matter 
supermultiplet transforming in the adjoint of $SU(2)_{diag}$ 
(plus two singlets that are not charged under $SU(2)_{diag}$ and 
therefore decouple). This is the $N=2$ theory discussed in 
Ref \cite{SW} ($N=2$ $SU(2)$ Yang-Mills theory with no matter). 
The Abelian Coulomb branch mentioned above is then that of this $N=2$ theory. 

\subsection{The $\Lambda_1>>\Lambda_2$ Limit}

{}The above results can be obtained in a different way, namely, by 
considering the limit where the two scales $\Lambda_1$ and $\Lambda_2$ 
are very different. For definiteness let us take $\Lambda_1>>\Lambda_2$. 
In this limit we can deduce the superpotential as a sum of contributions 
generated by the two groups \cite{poppitz}. 

{}In fact, let us be more general and consider the $SU(N)_1\otimes SU(N)_2$ 
theory with the following chiral matter content: $({\bf N}_c,{\overline 
{\bf N}}_c)$, $({\overline {\bf N}}_c,{\bf N}_c)$, $({\bf N}_c,{{\bf 1}})$, 
$({\overline {\bf N}}_c,{\bf 1})$, $({{\bf 1}},{\bf N}_c)$, 
$({\bf 1},{\overline {\bf N}}_c)$. Note that for $N_c=2$ we have 
the $[2,1,1]$ theory considered in the previous subsection. We 
choose to study a more general case here as for $SU(2)$ the 
fundamental is pseudo-real, and certain features of the theory 
(such as presence of baryons) become more obscure.     

{}Let the scales $\Lambda_1$ and $\Lambda_2$ of the two $SU(N_c)$'s be 
very different: $\Lambda_1>>\Lambda_2$. Then the second $SU(N_c)$ can be 
regarded as dynamically inactive. We, therefore, have the $SU(N_c)_1$ 
theory with $N_f=N_c+1$ flavors $Q^i$, ${\tilde Q}_{\bar i}$ 
($i=1,\dots,N_f$, and similarly for ${\bar i}$). Here 
$Q^i=({\bf N}_c,{\overline {\bf N}}_c)$ for $i=1,\dots,N_c$, 
and $Q^{N_c+1}=({\bf N}_c,{{\bf 1}})$. The fields 
${\tilde Q}_{\bar i}$ are given by the complex conjugates of $Q^i$. 

{}For the $N_f=N_c+1$ theory the quantum moduli space is the same as 
the classical one \cite{Seib}. The latter is described by the mesons 
$M^i_{\bar j}=Q^i{\tilde Q}_{\bar j}$, and baryons 
$B_i=\epsilon_{i{j_1}\dots {j_{N_c}}}
Q^{j_1}\cdots Q^{j_{N_c}}$ and 
${\tilde B}^{\bar i}=\epsilon^{{\bar i}{{\bar j}_1}\dots {{\bar j}_{N_c}}} 
{\tilde Q}_{{\bar j}_1}\cdots {\tilde Q}_{{\bar j}_{N_c}}$. The classical 
constraints read:
\begin{eqnarray}\label{classconst}
 &&(M^{-1})^{\bar j}_{i} \det\,(M)-B_i {\tilde B}^{\bar j}=0~,\\
 && M^i_{\bar j}B_i=0~,~~~M^i_{\bar j} {\tilde B}^{\bar j}=0~.\nonumber
\end{eqnarray} 

{}Away from the origin $M=B={\tilde B}=0$, all the states in $M$, $B$ and 
${\tilde B}$ are physical and couple via the following superpotential 
\cite{Seib}: 
\begin{eqnarray}\label{supeff}
 {\cal W}_{eff}=-{1\over{\Lambda_1^{2N_c-1}}}(\det\,(M)-
 M^i_{\bar j}B_i {\tilde B}^{\bar j})~.
\end{eqnarray}
(The classical constraints (\ref{classconst}) arise from Eq (\ref{supeff}) as 
equations of motion for these fields.) Away from the origin the number of 
independent massless degrees of freedom is the same as in the classical 
theory since the components of $M$, $B$ and ${\tilde B}$ which are 
classically constrained acquire masses via the couplings in ${\cal W}_{eff}$ 
\cite{Seib}.

{}It is instructive to figure out which fields are generically massless, and 
what their charges are under the second $SU(N_c)$. We have (the r.h.s. is the 
representation of $SU(N_c)_2$):
\begin{eqnarray}
 &&M^{{N_c}+1}_{\overline{{N_c}+1}}={\bf 1}~,\\
 &&M^{i}_{\overline{{N_c}+1}}={\overline {\bf N}}_c~(i=1,\dots,N_c)~,\\
 &&M^{{N_c}+1}_{\bar j}={\bf N}_c~({\bar j}=1,\dots,N_c)~,\\
 &&M^i_{\bar j}={\bf 1}\oplus{\bf Adj}~(i,{\bar j}=1,\dots,N_c)~,\\
 &&B_{{N_c}+1}={\bf 1}~,\\
 &&{\tilde B}^{\overline{{N_c}+1}}={\bf 1}~,\\
 &&B_i={\bf N}_c~(i=1,\dots,N_c)~,\\
 &&{\tilde B}^{\bar j}={\overline {\bf N}}_c~({\bar j}=1,\dots,N_c)~.
\end{eqnarray}
Here ${\bf Adj}$ is the $N_c^2-1$ dimensional adjoint representation of 
$SU(N_c)$.

{}Let us see if we can recover the Abelian Coulomb branch discussed in the 
previous subsection. Consider the branch in the moduli space of mesons $M$ 
and baryons $B$ and ${\tilde B}$ where the singlet $B_{{N_c}+1}$ acquires a 
vev. Then, according to the superpotential (\ref{supeff}), the two singlets 
$M^{{N_c}+1}_{\overline{{N_c}+1}}$ and ${\tilde B}^{\overline{{N_c}+1}}$ 
pair up and acquire a mass of the order of the $B_{{N_c}+1}$ vev. Also, a 
linear combination of the fundamentals $M^{{N_c}+1}_{\bar j}$ and $B_i$ 
(note that in this case the singlet in $M^i_{\bar j}$ can also acquire a vev) 
pairs up with the antifundamental ${\tilde B}^{\bar j}$, and these fields 
acquire a mass. Thus, we are left with the gauge group $SU(N_c)_2$, two 
singlets, one adjoint and two ``flavors" (here we are taking into account 
the fields $({{\bf 1}},{\bf N}_c)$, $({\bf 1},{\overline {\bf N}}_c)$). 
Let us compare this directly to the classical result in the 
$SU(N_c)_1\otimes SU(N_c)_2$ model. Let 
us give a vev to the field $(
{\bf N}_c,{\overline {\bf N}}_c)$ in the direction that breaks the gauge 
group to the diagonal subgroup $SU(N_c)_{diag}$. Then the matter fields 
are two singlets, one adjoint and two ``flavors". If we have an 
appropriate tree-level superpotential as in the 
$SU(2)_1\otimes SU(2)_2$ case, the two ``flavors" will become heavy, 
and we will have $SU(N_c)$ theory with chiral matter supermultiplet 
transforming in the adjoint of the gauge group. (This is $N=2$ $SU(N_c)$ 
Yang-Mills with no matter.) One can explicitly check that provided such 
a superpotential is present at the tree-level, the two flavors in our above 
description (with $\Lambda_1>>\Lambda_2$) precisely decouple to match the 
classical result.       

{}Finally, we would like to return to the $SU(2)_1\otimes SU(2)_2$ case and 
give the meson and baryon fields in the notations of the previous subsection. 
Note that for $SU(2)$ there is no invariant distinction between the mesons 
and baryons (as the fundamental of $SU(2)$ is pseudo-real). Let us introduce 
the following meson fields of $SU(2)_1$ in the limit $\Lambda_1>>\Lambda_2$: 
\begin{eqnarray}
\label{meson1}
 {\cal O}_{ab\dot{\alpha}\dot{\beta}}\equiv{1\over \sqrt{2}}\epsilon^
 {\alpha \beta} Q_{a\alpha\dot{\alpha}} Q_{b\beta\dot{\beta}}~,~~~  
 V_{ai\dot{\alpha}}\equiv {1\over \sqrt{2}}\epsilon^
 {\alpha \beta}Q_{a\alpha\dot{\alpha}} L_{i\beta}~,~~~{\cal L}~.
\end{eqnarray}
Then we have the following correspondence between these meson fields and the 
mesons $M$ and baryons $B$ and ${\tilde B}$ (here we explicitly put $N_c=2$):
\begin{eqnarray}
 &&{\cal L}=M^3_{\bar 3}~\\
 &&V_{12\dot{\alpha}}=M^i_{\bar 3},~~~V_{21\dot{\alpha}}=M^3_{\bar j}~,~~~
 (i,{\bar j}=1,2)~,\\
 &&V_{11\dot{\alpha}}={\tilde B}^{\bar j},~~~V_{22\dot{\alpha}}=B_i~,~~~
 (i,{\bar j}=1,2)~,\\
 &&{\cal O}_{11\dot{\alpha}\dot{\beta}}=B_3~,~~~
 {\cal O}_{22\dot{\alpha}\dot{\beta}}={\tilde B}^{\bar 3}~,\\
 &&{\cal O}_{12\dot{\alpha}\dot{\beta}}=M^i_{\bar j}~,~~~(i,{\bar j}=1,2)~.
\end{eqnarray}
Note that in ${\cal O}_{11\dot{\alpha}\dot{\beta}}$ and 
${\cal O}_{22\dot{\alpha}\dot{\beta}}$ the indices 
$\dot{\alpha},\dot{\beta}$ must be antisymmetrized 
(this gives $SU(2)_2$ singlets), whereas in 
${\cal O}_{12\dot{\alpha}\dot{\beta}}$ both antisymmetric 
($SU(2)_2$ singlet) and symmetric ($SU(2)_2$ triplet) combinations 
are present.

\subsection{Comments}

{}We saw in the previous subsections that the classical moduli space 
is not modified in the hidden sector of the $S2$ model provided that 
the matter content is what we have been discussing so far. Therefore, supersymmetry 
with this matter content is {\em not} broken. One can ask 
a question whether at the string/anomalous $U(1)$ breaking scale by giving 
vevs to some of the fields $Q_1$, $Q_2$, $L_1$, $L_2$, $R_1$ and $R_2$ we 
can get a matter content (at the expense of breaking the gauge group) such 
that we have quantum modification in the moduli space of the resulting model 
once considered in the effective field theory. We already know that the 
fields $Q_1$ and $Q_2$ cannot do this. Since the spectrum is symmetric with 
respect to interchanging the two $SU(2)$'s, we are left to consider giving 
vevs to $R_1$ and $R_2$. Note that due to the $D$-flatness conditions if 
$R_1$ acquires a vev, then so must $R_2$. This breaks the gauge group from 
$SU(2)_1\otimes SU(2)_2$ to $SU(2)_1$, {\em i.e.}, the $SU(2)_2$ subgroup 
is completely Higgsed. Note that due
to the tree-level superpotential (\ref{exact}), some of the matter fields 
become heavy, and 
(below the anomalous $U(1)$ breaking scale $M_X$) we are in fact left with one 
``flavor" of $SU(2)_1$ (coming from $Q_1$ and $Q_2$; $L_1$ and $L_2$ pair up 
with the other components of $Q_1$ and $Q_2$ and become heavy) plus a 
singlet. Note that we can arrive at this model (with completely Higgsed 
$SU(2)_2$ subgroup in the $S2$ model) starting from the $\Sigma2$ model 
(see Table VI) and Higgsing the last two $U(1)$'s by giving vevs to the 
untwisted matter fields $({\bf 1},{\bf 1})(0,0,+6,0)$ and 
$({\bf 1},{\bf 1})(0,0,-3,\pm 3)$. In fact, the massless spectrum 
and the tree-level superpotential of the resulting model are the 
same as those of the $S2$ model after the Higgsing described above.

{}Thus, below the anomalous $U(1)$ breaking scale $M_X$ we have $SU(2)$ 
with one ``flavor'' (with ${\tilde b}_0=5$), whereas above the scale 
$M_X$ we have $SU(2)$ with three ``flavors''
in the above scenario. We have already discussed this case in 
Appendix \ref{SUSYBR} where we saw that the supersymmetry breaking 
scale in this scenario comes out way too low.

\section{Hidden Sector Matter Content of the $T5(1,0)$ Model}\label{T5(1,0)}

{}Following the rules for computing superpotentials given in Ref 
\cite{kst}, one can easily deduce the tree-level superpotential relevant 
for hidden sector dynamics (see first column of Table III for notation):
\begin{eqnarray}
W =\lambda \sum_{i,j=1}^{3} D_i (D_{i-} U_1+ D_{i+} U_2)~.
\end{eqnarray}
Upon the fields $U_1$ and $U_2$ acquiring vevs, $D_i$ and 3 linear combinations of $D_{i\pm}$ become heavy and decouple. Below the corresponding scale we therefore have $SU(2)\otimes SU(2)\otimes SU(2)$ with (effectively) 6 ``flavors'' in each $SU(2)$ (some of these 6 ``flavors'' come as bi- and tri-fundamentals). The one-loop $\beta$-function coefficient for this matter content vanishes. Although the gauge group is asymptotically free in the two-loop order, the strong scale comes out way below the electroweak scale. The reader can readily verify that Higgsing the gauge group by giving vevs to bi- and/or tri-fundamentals cannot remedy this situation. We thus conclude that this model does not seem to give rise to the non-perturbative dynamics required by supersymmetry breaking at the phenomenologically desired scale.   

\section{SUSY GUT Phenomenology of $SU(6)$ Models}\label{GUT6}

{}In this Appendix we provide some details on phenomenological properties of
the $S1(1,1)$, $\Sigma1$ and $S1$ models. In particular, we study the Yukawa
mass matrices. Our analysis of the corresponding 
superpotentials indicates that in the $S1(1,1)$ and $\Sigma1$ models there 
are two top-like generations of quarks, whereas in the $S1$ 
model generically there all three generations of quarks are 
top-like albeit with some fine-tuning one can achieve having 
only one top-like family.   

{}Let us start with the 
branching rules of the $SU(6)$ irreps ${\bf 35}$, ${\bf 15}$,  
${\overline {\bf 6}}$ and ${\bf 6}$ under the breaking 
$SU(6)\supset SU(3)_c\otimes SU(2)_w\otimes U(1)_Y$: 
\begin{eqnarray}
{\bf 35}=&&({\bf 1},{\bf 1})(0)+({\bf 1},{\bf 1})(0)+({\bf 8},
{\bf 1})(0) +({\bf 1}, {\bf 3})(0) +
 ({\bf 3},{\bf 2})(-5/3) +({\overline {\bf 3}},{\bf 2})(5/3)\nonumber\\
 \label{35-break}
&+&({\overline {\bf 3}},{\bf 1})(2/3) + ({\bf 1},{\bf 2})(1) + 
({\bf 3},{\bf 1})(-2/3) + ({\bf 1},{\bf 2})(-1)~,\\
 \label{15-break}
{\bf 15}=&&\stackrel{\textstyle E}{({\bf 1},{\bf 1})(2)} + 
\stackrel{\textstyle H_u}{({\bf 3},{\bf 1})(-2/3)} 
+\stackrel{\textstyle U}{({\overline {\bf 3}}, {\bf 1})(-4/3)}
 +\stackrel{\textstyle h_u}{({\bf 1},{\bf 2})(1)}+
\stackrel{\textstyle Q}{({\bf 3}, {\bf 2})(1/3)}~,\\
\label{6bar-break}
 {\overline {\bf 6}} = &&\stackrel{\textstyle L,\,h_d}{({\bf 1},{\bf
2})(-1)}
 +\stackrel{\textstyle D, H_d}{({\overline {\bf 3}}, {\bf 1})(+2/3)} +
 \stackrel{\textstyle\bf\tilde s}{({\bf 1},{\bf 1})(0)}~,\\
\label{6-break}
 {\bf 6} = &&\stackrel{\textstyle h_u}{({\bf 1},{\bf 2})(1)}
 +\stackrel{\textstyle H_u}{({\bf 3}, {\bf 1})(-2/3)} +
 \stackrel{\textstyle\bf s}{({\bf 1},{\bf 1})(0)}.
\end{eqnarray}
Here $h_u, h_d$ denote Higgs doublets, $H_u, H_d$ - Higgs triplets and
the usual notations for chiral fermions are used.  As discussed earlier, 
the possibility of $h_u$ coming from ${\bf 15}$ is
disfavored for phenomenological reasons (doublet-triplet splitting in
this case would require fine-tuning). Note that 
${\bf 6}$'s and ${\bf\overline 6}$'s contain singlets ${\bf s}$ and ${\bf \tilde s}$. 
Two linear combinations (one coming from ${\bf 6}$'s, and the other one coming from ${\bf\overline 6}$'s) of these singlets vevs at a high scale. (The rest of them either become 
heavy via the couplings ${\bf 6}\cdot{\bf 6}\cdot{\bf\overline 6}\cdot{\bf\overline 6}$ or describe 
right-handed neutrinos and anti-neutrinos.) From the above equations one can see that 
the up-quark (meaning $u,c,t$)Yukawa matrix comes from the couplings ${\bf 15\cdot 15\cdot 6\cdot 6}$ (times some power of the adjoint in certain cases), while the down-quark  (meaning $d,s,b$) and lepton Yukawa matrices come from the couplings ${\bf 15\cdot\overline {6}\cdot\overline {6}}$ in the superpotential. (The latter type of couplings are also responsible for decoupling of extra states in ${\bf 15}$'s and ${\bf\overline 6}$'s once a linear combination of ${\bf \tilde s}$ fields acquires a vev. This would, among other states, remove the extra $h_u$ fields in ${\bf 15}$'s.)

\subsection{The $S1(1,1)$ Model}

{}This model has five ${\bf 15}$'s $Q_i$ and $Q_0$ ($i=++,--,+-,-+$), two ${\overline {\bf 15}}$'s
$\tilde{Q}_\pm$, ten ${\overline {\bf 6}}$'s $H^\pm _0$ and $H^\pm _i$, and four ${\bf
6}$'s $\tilde{H}^+_\pm$ and $\tilde{H}^-_\pm$ (see Table IV).

{}The ${\bf 15}$-$\overline {\bf 15}$ pairing takes place via the
$Q\cdot {\tilde Q}\cdot H\cdot {\widetilde H}$ couplings. 
The corresponding $4\times 2$ matrix ($Q_0$ does not couple to 
$\tilde{Q}_\pm$) generically has rank two. (This is to be contrasted with the $T1(1,1)$
model, where the $F$-flatness constraints allow only one 
``generation-antigeneration'' pair to decouple. In the $S1(1,1)$ model the $F$-flatness constraints relevant for the ${\bf 15}$-$\overline {\bf 15}$ pairing can be
read-off from the $H\cdot H\cdot \tilde {H}\cdot \tilde {H}$ term in the 
superpotential.) Thus two linear combinations of the fields $Q_i$ become heavy. Writing
down the up-quark Yukawa matrix for the remaining three ${\bf 15}$'s, namely, $Q_0$ and the two linear combinations of the fields $Q_i$ that remain light, one
obtains a rank-2 matrix, {\em i.e.}, two top-like heavy quarks and one light quark
(the latter coming coming from $Q_0$). The corresponding 
term in the superpotential (see Appendix \ref{SupSU(6)}) 
reads $(\tilde{H}^-_-\tilde{H}^+_+ +
\tilde{H}^-_+\tilde{H}^+_-)(Q_{++}Q_{--} + Q_{+-}Q_{-+})$.     

\subsection{The $\Sigma1$ Model}

{}This model has five ${\bf 15}$'s $F_i$ and $F_0$ ($i=++,--,+-,-+$), 
two ${\overline {\bf 15}}$'s
$\tilde{F}_\pm$, eight ${\overline {\bf 6}}$'s $\tilde{S}_i$ and 
$\tilde{S}'_i$, 
and two ${\bf 6}$'s $S_\pm$ (see Table VI). This model is very 
similar to the $S1(1,1)$
model. Thus, the ${\bf 15}$-$\overline {\bf 15}$ pairing takes 
place at a high scale 
($\lambda_{19}$ term in the superpotential - see 
Appendix \ref{SupSU(6)}) leaving 
us with three ${\bf 15}$'s. These have a rank-2 up-quark Yukawa
matrix via the coupling $\lambda_{23}S_+S_-(F_{++}F_{--} + F_{+-}F_{-+})$.
$\lambda_{21}$ and $\lambda_{22}$ terms do not contribute to the 
Yukawa matrix because of the $F$-flatness condition 
$\partial W/\partial U_0=U_{++}U_{--}+U_{+-}U_{-+}=0$. 
Thus, just as in the $S1(1,1)$ model, we have two top-like 
quark families in the $\Sigma1$ model.

\subsection{The $S1$ Model}

{}This model has six ${\bf 15}$'s $F^A_\pm$, ($A=1,2,3$), 
three ${\overline {\bf 15}}$'s $\tilde{F}^A$, 
nine ${\overline {\bf 6}}$'s $\tilde{S}^A _0$ and 
$\tilde{S}^A _\pm$, and three ${\bf 6}$'s $S^A$ (see Table V).
 
{}Three ${\bf 15}$-$\overline {\bf 15}$ pairs generically decouple at 
a high scale (via the couplings $\lambda_{15}$ through
$\lambda_{24}$). The rest of the ${\bf 15}$'s have a rank-3 
up-quark Yukawa matrix (via the couplings  
$\lambda_{25}$ through $\lambda_{36}$). Some fine-tuning is therefore 
required to have only one top-like generation of quarks.

\section{Superpotentials for the $E_6$ and $SO(10)$ Models}\label{SupE_6}

{}In this Appendix we give superpotentials for the $E1$ and $T1(1,1)$ models. They were derived in Ref \cite{kst}.

\subsection{The $E_6$ Model}

{}The superpotential for the $E1$ model (see Table I for notation) reads:
\begin{eqnarray}
 W&=&\lambda_1 
      \chi_{0} (\chi_{++}\chi_{--} + \chi_{+-}\chi_{-+}) 
      \Phi + \lambda_2 \tilde{\chi}_{+}\tilde{\chi}_{-} 
               (\chi_{++}\chi_{--} + \chi_{+-}\chi_{-+}) \nonumber\\
 &+& \lambda_{3}  U_0
               ( U_{++}U_{--} + U_{+-}U_{-+}) + \lambda_{4}  
       [ \chi_{++}^3 U_{--} + \chi_{+-}^3 U_{-+} + 
         \chi_{-+}^3 U_{+-} + \chi_{--}^3 U_{++} ] \Phi^2 \nonumber\\
 &+& \lambda_{5}  (\chi_{0})^3 U_{0}  D_{+}D_{-} \Phi^2
 + \lambda_{6}   
   [ (\tilde{\chi}_{+})^3 \tilde{U}_{-} + \tilde{\chi}_{-}^3 \tilde{U}_{+} ]
       D_{+}D_{-} \Phi +...~,
\end{eqnarray}
where traces over the irreps of the 
gauge group are implicit. Here $\lambda_{k}\equiv
\lambda_k(\Phi^3,D_{+}D_{-})$ are 
certain polynomials of their  
arguments
\begin{equation}
\lambda_{k}(\Phi^{3},D_{+}D_{-})= \sum_{m,n} \lambda_{kmn} \Phi^{3m} 
(D_{+}D_{-})^{n}
\end{equation}
such that $\lambda_{km0},\lambda_{k0n}\not=0$.

\subsection{The $T1(1,1)$ Model}

{}The superpotential for the $T1(1,1)$ model can be deduced from that of the $E_6$ model by adjoint breaking $E_6\supset SO(10)\otimes U(1)$. It reads 
(see Table II for notation):
\begin{eqnarray}
 W&=& \lambda_1 
   \left[ Q_{0} (Q_{++}H_{--} + Q_{+-}H_{-+} + Q_{-+}H_{+-} + Q_{--}H_{++})
\nonumber \right. \\
      &+& H_{0} (Q_{++}Q_{--}+Q_{+-}Q_{-+}) 
      + H_{0} (H_{++}S_{--}+H_{+-}S_{-+}+H_{-+}S_{+-}+H_{--}S_{++})
\nonumber \\
&+&\left.  S_{0} (H_{++}H_{--}+H_{+-}H_{-+}) \right]  
      (\Phi + \phi^{\prime}) \nonumber\\
&+&\lambda_2 
   \left[ \tilde{Q}_{+}\tilde{Q}_{-} (Q_{++}Q_{--}+Q_{+-}Q_{-+})\right.
\nonumber\\ 
&+& \tilde{Q}_{+}\tilde{Q}_{-} ( H_{++}S_{--}+H_{+-}S_{-+}
                               + H_{-+}S_{+-}+H_{--}S_{++} )
\nonumber \\
&+&(\tilde{Q}_{+}\tilde{S}_{-}+\tilde{Q}_{-}\tilde{S}_{+})
          (Q_{++}S_{--}+Q_{+-}S_{-+} + Q_{-+}S_{+-} + Q_{--}S_{++}) 
               \nonumber\\
&+&(\tilde{H}_{+} \tilde{S}_{-} +\tilde{H}_{-} \tilde{S}_{+})
   (H_{++}S_{--}+H_{+-}S_{-+}+H_{-+}S_{+-}+H_{--}S_{++}) \nonumber \\
&+&(\tilde{H}_{+} \tilde{S}_{-} +\tilde{H}_{-} \tilde{S}_{+})
   (Q_{++}Q_{--}+Q_{+-}Q_{-+}) \nonumber \\
&+&(\tilde{Q}_{+}\tilde{H}_{-}+\tilde{Q}_{-}\tilde{H}_{+})
   (Q_{++}H_{--} + Q_{+-}H_{-+} + Q_{-+}H_{+-} + Q_{--}H_{++})
\nonumber \\
&+& \left. \tilde{H}_+ \tilde{H}_- (H_{++}H_{--}+H_{+-}H_{-+})
   + \tilde{S}_+ \tilde{S}_- (S_{++}S_{--}+S_{+-}S_{-+}) \right] 
\nonumber\\
 &+& \lambda_{3}  U_0
               ( U_{++}U_{--} + U_{+-}U_{-+}) \nonumber\\
 &+& \lambda_{4}  
     \left[ (Q_{++}^2 H_{++} + H_{++}^2 S_{++}) U_{--} + 
         (Q_{+-}^2 H_{+-} + H_{+-}^2 S_{+-}) U_{-+} \right. \nonumber \\ 
 &+& \left.  (Q_{-+}^2 H_{-+} + H_{-+}^2 S_{-+}) U_{+-} 
        +(Q_{--}^2 H_{--} + H_{--}^2 S_{--}) U_{++} \right] 
(\Phi + \phi^{\prime} )^2 +...~,
\end{eqnarray}
where the Clebsch-Gordan coefficients are
suppressed. 
The field $\phi^{\prime}$ is defined to be
\begin{equation}
\phi^{\prime}= \phi + \langle \phi \rangle~.
\end{equation}
Here $\lambda_{k}\equiv\lambda_{k}((\Phi + \phi^{\prime} )^3,D_{+}D_{-})$ are 
defined in the same way as in the $E_6$ model.

\section{Superpotentials for $SU(6)$ Models}\label{SupSU(6)}

{}In this Appendix we give superpotentials for the $S1(1,1)$, $S1$, $S2$,
$\Sigma1$ and $\Sigma2$ models. The superpotential for the 
$S1(1,1)$ can be deduced from that of the $E_6$ model 
(see Appendix \ref{SupE_6}) by adjoint breaking 
$E_6\supset SU(6)\otimes U(1)$. The superpotentials for the 
$S1$ and $S2$ models were derived in Ref \cite{kst}. Those for 
the $\Sigma1$ and $\Sigma2$ models can be obtained following the 
rules of Ref \cite{kst}. 

\subsection{The $S1(1,1)$ Model}

{}The superpotential for the $S1(1,1)$ model (see Table IV for notation) reads:
\begin{eqnarray}
 W&=& \lambda_1 
   [ Q_{0} (Q_{++}Q_{--} + Q_{+-}Q_{-+}) 
         +Q_{0} \sum_{i,j} z_{ij}
          (H^i_{++}H^j_{--}+H^i_{+-}H^j_{-+}) 
\nonumber  \\
      &+&  \sum_{i,j} z_{ij} 
      H^i_{0} (Q_{++}H^j_{--}+Q_{+-}H^j_{-+}+Q_{-+}H^j_{+-}+Q_{--}H^j_{++})]  
      (\Phi + \phi^{\prime}) \nonumber\\
&+&\lambda_2 
   [ \tilde{Q}_{+} \tilde{Q}_{-}
   (Q_{++}Q_{--}+Q_{+-}Q_{-+})
\nonumber \\
&+& \tilde{Q}_{+}\tilde{Q}_{-} \sum_{i,j} z_{ij}
          (H_{++}^i H_{--}^j + H_{+-}^i H_{-+}^j)
 + \sum_{i,j} z_{ij} \tilde{H}_{+}^i \tilde{H}_{-}^j 
            ( Q_{++}Q_{--}+Q_{+-}Q_{-+} )
               \nonumber\\
&+&  
\sum_{i,j,k,l} y_{ijkl} \tilde{H}_{+}^i\tilde{H}_{-}^j 
          (H_{++}^k H_{--}^l+H_{+-}^k H_{-+}^l)\nonumber\\
 &+& \sum_{i,j} z_{ij}
 (\tilde{Q}_{+}\tilde{H}_{-}^i+\tilde{Q}_{-}\tilde{H}_{+}^i)
 (Q_{++}H^j_{--}+Q_{+-}H^j_{-+}+Q_{-+}H^j_{+-}+Q_{--}H^j_{++})]
\nonumber \\
 &+& \lambda_{3}  U_0
               ( U_{++}U_{--} + U_{+-}U_{-+}) \nonumber\\
 &+& \lambda_{4}  
     \left[ (Q_{++}^3 + Q_{++} H_{++}^{+} H_{++}^{-} ) U_{--} + 
         (Q_{+-}^3 + Q_{+-} H_{+-}^{+} H_{+-}^{-}) U_{-+} 
\right. \nonumber \\ 
 &+& \left.  (Q_{-+}^3 + Q_{-+} H_{-+}^{+} H_{-+}^{-}) U_{+-} 
            +(Q_{--}^3 + Q_{--} H_{--}^{+} H_{--}^{-}) U_{++} \right] 
(\Phi + \phi^{\prime} )^2 +...~,
\end{eqnarray}
where the Clebsch-Gordan coefficients are
suppressed, and the coefficients $z_{ij}$ and $y_{ijkl}$ are defined as follows:
$z_{ij}=1-\delta_{ij}$, $y_{ijkl}=1$ if $(i,j,k,l)=(+,+,-,-)$ or
its permutations, and $y_{ijkl}=0$ otherwise. Here $i,j=+,-$. The couplings $\lambda_k$ are defined in the same way as in the $T1(1,1)$ model.

\subsection{The $S1$ Model}

{}The superpotential for the $S1$ model reads (see Table V for notation):
\begin{eqnarray}
 W &= &\lambda_{1}  U_0
               ( U_{++}U_{--} +U_{+-}U_{-+} )
      ~~+~  \lambda_{2}  T_0
               ( {\tilde T}_{+} U_{-+} +{\tilde T}_{-} U_{++})~~ +~ 
\lambda_{3}  {\tilde T}_0 T_- U_0~~ +~
           \lambda_{4}  {\tilde T}_0 T_0 {\tilde U}_- \nonumber\\[2mm]
     &+& \lambda_{5}  \sum_{A,B,C} y_{ABC}
 [ U_{++} ( F^A_{+}F^B_{+}F^C_{-} + {1 \over 3} F^A_{-}F^B_{-}F^C_{-} )
 + U_{-+} ( F^A_{+}F^B_{-}F^C_{-} + {1 \over 3} F^A_{+}F^B_{+}F^C_{+} )] 
\nonumber\\
     &+& [\lambda_{6}   \Phi^2 
         +\lambda_{7}  \Phi \phi
         +\lambda_{8}   \phi^2]~\sum_A \{ U_{++} 
[(F^A_{-})^3 + F^A_{-}(F^A_{+})^2] 
                        +U_{-+} [(F^A_{+})^3 + F^A_{+}(F^A_{-})^2] \}
\nonumber\\
     &+& \lambda_{9}  \sum_A 
            [F^A_{+} {\tilde S}^A_{-} + F^A_{-} {\tilde S}^A_{+}] 
      {\tilde S}^A_0 ~~+~~ [\lambda_{10}  \Phi + \lambda_{11}  \phi ]
      \sum_{A,B,C} y_{ABC} (F^A_{+} {\tilde S}^B_{-}+F^A_{-} {\tilde S}^B_{+})
      {\tilde S}^C_0  \nonumber\\
      &+& \lambda_{12}   {\tilde U}_+ T_+ {\tilde T}_0 
       \sum_{A,B,C} y_{ABC} {\tilde F}^A {\tilde F}^B 
        {\tilde F}^C ~~+~~ [\lambda_{13} \Phi + \lambda_{14} \phi ] 
{\tilde U}_+ T_+ {\tilde T}_0
      \sum_{A}  ({\tilde F}^A)^3   \nonumber\\
      &+& \lambda_{15} \sum_{A,B} z_{AB} {\tilde F}^A S^B
         ( F^A_+ {\tilde S}^B_{-} + F^A_- {\tilde S}^B_{+}  
          +F^B_+ {\tilde S}^A_{-} + F^B_- {\tilde S}^A_{+} )        
\nonumber\\
      &+& [\lambda_{16} \Phi + \lambda_{17}\phi]  
      \sum_{A,B,C} y_{ABC} {\tilde F}^A S^A 
        (F^B_{+} {\tilde S}^C_{-}+ F^B_{-} {\tilde S}^C_{+})  \nonumber\\
      &+& [\lambda_{18} \Phi^2 + 
     \lambda_{19} \Phi  \phi+
      \lambda_{20} \phi^2]
     \sum_{A,B,C} y_{ABC} [F^A_{+} {\tilde S}^A_{-} +F^A_{-} {\tilde S}^A_{+}]
       {\tilde F}^B S^C  
      \nonumber\\
     &+& [\lambda_{21} \Phi^3 + 
     \lambda_{22} \Phi^2 \phi+
      \lambda_{23} \Phi \phi^2+\lambda_{24} \phi^3]~
\sum_{A}[F^A_{+} {\tilde S}^A_{-}+F^A_{-}{\tilde S}^A_{+}]
        {\tilde F}^A S^A 
      \nonumber\\
     &+& \lambda_{25} \sum_{A,B,C} y_{ABC} 
                      [ (F^{A}_{+})^2+(F^{A}_{-})^2 ] S^B S^C 
                      ( U_{++}U_{+-} + U_{-+}U_{--} ) 
      \nonumber\\
     &+& \lambda_{26} \sum_{A,B,C} y_{ABC} 
                F^A_{+}F^A_{-}S^B S^C ( U_{++}U_{--}+U_{+-}U_{-+} )
      \nonumber\\
     &+& [ \lambda_{27} \Phi + \lambda_{28} \phi ]
                 \sum_{A,B} z_{AB}
                F^A_{+}F^B_{+}S^A S^B ( U_{++}U_{+-} + U_{-+}U_{--} ) 
      \nonumber\\
     &+& [ \lambda_{29} \Phi^2 
          +\lambda_{30} \Phi \phi 
          +\lambda_{31} \phi^2 ]~\sum_{A,B,C} y_{ABC} 
                F^A_{+}F^B_{+} (S^C)^2 ( U_{++}U_{+-} + U_{-+}U_{--} ) 
      \nonumber\\
     &+& [ \lambda_{32} \Phi + \lambda_{33} \phi ]~
\sum_{A,B} z_{AB} F^A_{+}F^B_{-}S^A S^B 
            ( U_{++}U_{--}+U_{+-}U_{-+} )
      \nonumber\\
     &+& [ \lambda_{34} \Phi^2 + \lambda_{35} \Phi \phi
          +\lambda_{36} \phi^2 ]~
           \sum_{A,B,C} y_{ABC} F^A_{+}F^B_{-}(S^C)^2
            ( U_{++}U_{--}+U_{+-}U_{-+} )   
      +...~,
\end{eqnarray}
where $\lambda_{k}\equiv
\lambda_k(\Phi,\phi)$ are 
certain polynomials of their  
arguments
\begin{equation}
\lambda_{k}(\Phi,\phi)= \sum_{m,n} \lambda_{kmn} \Phi^{3m-n} 
\phi^{n}
\end{equation}
such that $\lambda_{km0},\lambda_{k0n}\not=0$.
The coefficients $y_{ABC}$ and $z_{AB}$ are defined as follows: 
$y_{ABC}=\epsilon_{ABC}$, and $z_{AB}=1-\delta_{AB}$.

\subsection{The $S2$ Model}

{}The superpotential for the $S2$ model reads (see Table V for notation):
\begin{eqnarray}
 W&=&\lambda_{1}  U_0 ( d_{++}d_{--} +d_{+-}d_{-+})
      ~~+~~\lambda_{2} 
        (D_+ {\tilde d}_- \Delta_- + D_- {\tilde d}_+ \Delta_+) 
\nonumber\\[2mm]
&+& \lambda_{3}  U_0 \Delta_+ \Delta_- \sum_{A,B,C} y_{ABC}
       F^A F^B F^C 
~~+~~ [\lambda_{4}  \Phi^2 +\lambda_{5} \Phi \phi +
      \lambda_{6}  \phi^2 ]  U_0 \Delta_+ \Delta_- \sum_{A}
       (F^A)^3 \nonumber\\
    &+& \lambda_{7}  \sum_A F^A 
       [ {\tilde S}^A_{++}{\tilde S}^A_{--} 
        +{\tilde S}^A_{+-}{\tilde S}^A_{-+} ] 
~~+~~ [\lambda_{8} \Phi + \lambda_{9} \phi ]
      \sum_{A,B,C} y_{ABC} F^A ({\tilde S}^B_{++}{\tilde S}^C_{--} 
                              +{\tilde S}^B_{-+}{\tilde S}^C_{+-}) \nonumber\\ 
    &+& \lambda_{10}  \sum_{A,B} z_{AB} S^A_{+}S^B_{-}
     ({\tilde S}^A_{++} {\tilde S}^B_{--}+{\tilde S}^A_{-+} {\tilde S}^B_{+-}
     +{\tilde S}^A_{+-} {\tilde S}^B_{-+}+{\tilde S}^A_{--} {\tilde S}^B_{++})
       \nonumber \\ 
    &+& [\lambda_{11} \Phi^3 + 
         \lambda_{12} \Phi^2 \phi+
         \lambda_{13} \Phi \phi^2+\lambda_{14} \phi^3]~\sum_{A} S^A_+ S^A_- 
    ({\tilde S}^A_{++} {\tilde S}^A_{--} + {\tilde S}^A_{+-}{\tilde S}^A_{-+})
   \nonumber\\
    &+& [\lambda_{15} \Phi + \lambda_{16}  \phi]
     \sum_{A,B,C} y_{ABC} S^A_+ S^A_- 
      ({\tilde S}^B_{++}{\tilde S}^C_{--}+{\tilde S}^B_{-+}{\tilde S}^C_{+-}) 
\nonumber\\
  &+&[\lambda_{17} \Phi^2 + \lambda_{18} \Phi \phi+
   \lambda_{19} \phi^2 ]~\sum_{A,B,C} y_{ABC} S^B_{+}S^C_{-}
    ({\tilde S}^A_{++}{\tilde S}^A_{--}+{\tilde S}^A_{+-}{\tilde S}^A_{-+}) 
    +...~.
\end{eqnarray}
The couplings $\lambda_k$ and the coefficients $y_{ABC}$ and $z_{AB}$ are
defined in the same way as in the $S1$ model. 

\subsection{The $\Sigma1$ Model}

{}The superpotential for the $\Sigma 1$ model reads (see Table VI 
for notation):
\begin{eqnarray}
W &=& \lambda_1 U_0 (U_{++}U_{--}+U_{+-}U_{-+})~~+~~ 
\left( \lambda_2 \Phi + \lambda_3 \phi \right) 
        F_0 (F_{++}F_{--}+F_{+-}F_{-+}) \nonumber \\
   &+& \left( \lambda_4 \Phi^2 + \lambda_5 \Phi \phi + \lambda_6 \phi^2 \right)
   \left( 
    U_{++} F_{--}^3 + U_{+-} F_{-+}^3 + U_{-+} F_{+-}^3  + U_{--} F_{++}^3 
   \right) \nonumber \\
   &+& \left( \lambda_7 \Phi^2 + \lambda_8 \Phi \phi + \lambda_9 \phi^2 \right)
     U_0^2 \left( U_{++}U_{--}+U_{+-}U_{-+} \right) F_0^3 \nonumber\\
   &+& \left( \lambda_{10} \Phi + \lambda_{11} \phi \right)
   U_{0} \left( U_{++}U_{--}+U_{+-}U_{-+} \right)    
   \left( \tilde{U}_{+} \tilde{F}_{-}^3 + \tilde{U}_{-} \tilde{F}_{+}^3 \right)
   \nonumber \\
   &+& \left( \lambda_{12} \Phi^2 + \lambda_{13} \Phi \phi 
              + \lambda_{14} \phi^2 \right)
      U_0 F_0 \left( \tilde{S}_{++} \tilde{S}_{--} 
                   + \tilde{S}_{+-} \tilde{S}_{-+} \right) \nonumber \\
   &+& \left( \lambda_{15} \Phi + \lambda_{16} \phi \right)
      U_0 \left( U_{++}U_{--} + U_{+-}U_{-+} \right)
      F_0 \left( \tilde{S}^{\prime}_{++}\tilde{S}^{\prime}_{--}
                +\tilde{S}^{\prime}_{+-}\tilde{S}^{\prime}_{-+} \right)
\nonumber\\
   &+& \left( \lambda_{17} \Phi + \lambda_{18} \phi \right)[
      F_{++} ( \tilde{S}_{-+} \tilde{S}^{\prime}_{--}
                   +\tilde{S}_{--} \tilde{S}^{\prime}_{-+} )
   + F_{+-} ( \tilde{S}_{-+} \tilde{S}^{\prime}_{+-}
                   +\tilde{S}_{--} \tilde{S}^{\prime}_{++} )
\nonumber \\
   &&~~~~~~~+ 
      F_{-+} ( \tilde{S}_{++} \tilde{S}^{\prime}_{--}
                   +\tilde{S}_{+-} \tilde{S}^{\prime}_{-+} )+ 
      F_{--} ( \tilde{S}_{++} \tilde{S}^{\prime}_{+-}
                   +\tilde{S}_{+-} \tilde{S}^{\prime}_{++} )]
\nonumber \\
   &+& \lambda_{19} [ ( F_{++} S_{-}+ F_{+-} S_{+})
            ( \tilde{S}_{--} \tilde{F}_{+}
                                     +\tilde{S}_{-+} \tilde{F}_{-} )
+ ( F_{-+} S_{-}+F_{--} S_{+}) ( \tilde{S}_{+-} \tilde{F}_{+}
                                     +\tilde{S}_{++} \tilde{F}_{-} )]
\nonumber \\ 
   &+& \lambda_{20} U_0 \left( U_{++}U_{--}+U_{+-}U_{-+} \right) F_0
     \{ S_{+} \left( \tilde{S}^{\prime}_{++} \tilde{F}_{-}
                    +\tilde{S}^{\prime}_{+-} \tilde{F}_{+}
              \right)
       +S_{-} \left( \tilde{S}^{\prime}_{-+} \tilde{F}_{-}
                    +\tilde{S}^{\prime}_{--} \tilde{F}_{+}
              \right) \}
\nonumber \\
   &+& \left( \lambda_{21} \Phi + \lambda_{22} \phi \right)
       \left( U_{++}U_{--} + U_{+-} U_{-+} \right)
       \left( F_{+-}F_{--}S_{+}^2 +
       F_{++}F_{-+}S_{-}^2\right)
\nonumber \\
   &+& \lambda_{23} \left( F_{++}F_{--} + F_{+-}F_{-+} \right) S_+ S_-
\nonumber \\
   &+& \left( \lambda_{24} \Phi^2 + \lambda_{25} \Phi \phi 
             + \lambda_{26} \phi^2 \right)
       \left( U_0 + \tilde{U}_{+} \tilde{U}_{-} \right) 
       \left(
            \tilde{F}_{+}^2 \tilde{S}_{+-} \tilde{S}_{--}
          + \tilde{F}_{-}^2 \tilde{S}_{++} \tilde{S}_{-+}
       \right)
\nonumber \\
   &+& \left(  \lambda_{27} \Phi + \lambda_{28} \phi \right)
        U_0 \left( U_{++}U_{--}+U_{+-}U_{-+} \right) 
        \left(
             \tilde{F}_{+}^2 \tilde{S}^{\prime}_{+-} \tilde{S}^{\prime}_{--}
            +\tilde{F}_{-}^2 \tilde{S}^{\prime}_{++} \tilde{S}^{\prime}_{-+}
        \right)
\nonumber \\
   &+& \left( \lambda_{29} \Phi^2 + \lambda_{30} \Phi \phi 
             + \lambda_{31} \phi^2 \right)
        \left( U_0 + \tilde{U}_+ \tilde{U}_{-} \right)
        \tilde{F}_+ \tilde{F}_-
        \left( \tilde{S}_{++} \tilde{S}_{--} + \tilde{S}_{+-}\tilde{S}_{-+}
        \right)
\nonumber \\
   &+& \left(  \lambda_{32} \Phi + \lambda_{33} \phi \right)
       U_0 \left( U_{++}U_{--}+U_{+-}U_{-+} \right) 
        \tilde{F}_+ \tilde{F}_-
        \left( \tilde{S}^{\prime}_{++} \tilde{S}^{\prime}_{--} 
             + \tilde{S}^{\prime}_{+-} \tilde{S}^{\prime}_{-+}
        \right)
\nonumber \\
   &+& \left(  \lambda_{34} \Phi + \lambda_{35} \phi \right)
           \left( U_{++}U_{--}+U_{+-}U_{-+} \right) 
       \left( S_{+}^2 \tilde{S}^{\prime}_{-+} \tilde{S}^{\prime}_{--}+  
       S_{-}^2 \tilde{S}^{\prime}_{++} \tilde{S}^{\prime}_{+-}\right)
\nonumber \\
   &+& \left(  \lambda_{36} \Phi + \lambda_{37} \phi \right)
       S_{+}S_{-} \left( \tilde{S}^{\prime}_{++} \tilde{S}^{\prime}_{--}
                        +\tilde{S}^{\prime}_{+-} \tilde{S}^{\prime}_{-+} 
                  \right) + \dots ~.
\end{eqnarray}
The couplings $\lambda_k$ are
defined in the same way as in the $S1$ model.

\subsection{The $\Sigma2$ Model}

{}The superpotential for the $\Sigma 2$ model reads (see Table VI for 
notation):
\begin{eqnarray}
W &=& \left( \lambda_1 \Phi + \lambda_2 \phi \right) \Delta_{+} \Delta_{-}
               F^1 F^2 F^3 + 
\left(  \lambda_{3} \Phi^2 + \lambda_{4} \Phi \phi 
             + \lambda_{5} \phi^2 \right)  \Delta_{+} \Delta_{-}
              \sum_{A} U_A ( F^A )^3 
\nonumber \\
   &+& \left(  \lambda_{6} \Phi^2 + \lambda_{7} \Phi \phi 
              + \lambda_{8} \phi^2 \right) 
              \sum_{A} U_A F^A \left( \tilde{S}^{A}_{++} \tilde{S}^{A}_{--}
                                     +\tilde{S}^{A}_{+-} \tilde{S}^{A}_{-+}
                               \right)
\nonumber \\
   &+& \left( \lambda_{9} \Phi + \lambda_{10} \phi \right)
              \sum_{A,B,C} y_{ABC} F^A 
                              \left( \tilde{S}^{B}_{++} \tilde{S}^{C}_{--}
                                     +\tilde{S}^{B}_{+-} \tilde{S}^{C}_{-+}
                               \right)     
\nonumber \\
   &+& \left(  \lambda_{11} \Phi^2 + \lambda_{12} \Phi \phi 
             + \lambda_{13} \phi^2 \right)  
              \sum_{A} S^A_{+}S^A_{-}
                  \left( \tilde{S}^{A}_{++} \tilde{S}^{A}_{--}
                                     +\tilde{S}^{A}_{+-} \tilde{S}^{A}_{-+}
                               \right)
\nonumber \\
   &+& \left( \lambda_{14} \Phi + \lambda_{15} \phi \right)
               \sum_{A,B} z_{AB} S^A_{+}S^B_{-}
                   \left( \tilde{S}^{A}_{++} \tilde{S}^{B}_{--}
                         +\tilde{S}^{A}_{+-} \tilde{S}^{B}_{-+}
                         +\tilde{S}^{A}_{-+} \tilde{S}^{B}_{+-}
                         +\tilde{S}^{A}_{--} \tilde{S}^{B}_{++}
                  \right)
\nonumber \\
    &+& \left( \lambda_{16} \Phi + \lambda_{17} \phi \right)
               \sum_{A,B,C} y_{ABC} S^A_{+}S^A_{-} U_B U_C
                     \left( \tilde{S}^{B}_{++} \tilde{S}^{C}_{--}
                                     +\tilde{S}^{B}_{+-} \tilde{S}^{C}_{-+}
                               \right)     
\nonumber \\
    &+& \left(  \lambda_{18} \Phi^2 + \lambda_{19} \Phi \phi 
              + \lambda_{20} \phi^2 \right)  
              \sum_{A,B,C} y_{ABC} U_A S^B_{+} S^C_{-}
                \left( \tilde{S}^{A}_{++} \tilde{S}^{A}_{--}
                                     +\tilde{S}^{A}_{+-} \tilde{S}^{A}_{-+}
                               \right) + \dots ~.
\end{eqnarray}
The couplings $\lambda_k$ and the coefficients $y_{ABC}$ and $z_{AB}$ are
defined in the same way as in the $S1$ model.

\section{Construction of Some models}\label{models}

{}In this appendix we give the construction of the models 
$\Sigma1$, $\Sigma2$ and $\Sigma3$ discussed in section \ref{survey}. 
As we proceed, we will for completeness review the construction of the 
$E_6$ model, one of the $SO(10)$ models (namely, $T1(1,1)$), and two other 
$SU(6)$ models (namely, $S1$ and $S2$), which were previously constructed 
in \cite{class10,class5}.

\subsection{The Narain Model}

{}Consider the Narain model
with the momenta of the internal bosons spanning an even self-dual 
Lorentzian lattice $\Gamma^{6,22} =\Gamma^{6,6} \otimes
\Gamma^{16}$. Here $\Gamma^{16}$ is the  
${\mbox{Spin}}(32)/{\bf Z}_2$ lattice. The $\Gamma^{6,6}$ is the 
momentum lattice corresponding to the compactification on the $E_6$ six-torus 
defined by $X_I= X_I +E_I$. The dot product of the vectors $E_I$ 
defines the constant background metric $G_{IJ}=E_I \cdot E_J$ which is the 
same as the $E_6$ Cartan matrix. 
There is also the antisymmetric background field $B_{IJ}$ given by
$2B_{IJ} ={1\over 2} G_{IJ}$ for $I<J$ and $2B_{IJ} =-{1\over 2} G_{IJ}$ 
for $I>J$. 
The vectors $E_I$ (and also their duals ${\tilde E}^I$
defined so that $E_I \cdot {\tilde E}^J ={\delta_I}^J$) can be expressed 
in terms of the $SU(3)$ root and weight vectors
${e}_i$ and ${\tilde e}^i$ ($i=1,2$):
\begin{eqnarray}\label{}
 &&E_1=(e_1,0,0),~~~E_2=(e_2,0,0),\nonumber\\
 &&E_3=(0,e_1,0),~~~E_4=(0,e_2,0),\nonumber\\
 &&E_5=({\tilde e}^2,- {\tilde e}^2, {\tilde e}^2),~~~E_6=({\tilde e}^1, 
 -{\tilde e}^1, {\tilde e}^1).\nonumber 
\end{eqnarray}
Following the notations of Ref \cite{class10} we will refer to this Narain 
model as $N(1,1)$. This model has $N=4$ space-time supersymmetry and 
$E_6 \otimes SO(32)$ gauge group.
 
\subsection{Wilson Lines}

{}{}Next, we discuss the  Wilson lines that will give the Narain models 
with $SO(10)^3$ and $(E_6)^3$ gauge subgroups.   
Here we are writing the Wilson lines as shift vectors in the $\Gamma^{6,22}$
lattice. The shift vectors $U_1$ and $U_2$ to be introduced 
are order-$2$ shifts that break $SO(32)$ to $SO(10)^3 \otimes SO(2)$. 
For certain Wilson lines $U_1$ and $U_2$, however, 
the new (shifted) sectors introduce additional gauge bosons to 
enhance $SO(10)^3$ to $(E_6)^3$.  

{}Thus, starting from the Narain lattice $N(1,1)$ we have
two inequivalent choices:\\
$\bullet$ The $N1(1,1)$ model generated by the Wilson lines
\begin{eqnarray}
 &&U_1 =(0,0,0 \vert\vert e_1/2,0,0)({\bf s}\vert {\bf 0}\vert {\bf 0}\vert 
 {\overline    S})~,\nonumber\\
 &&U_2 =(0,0,0 \vert\vert e_2/2,0,0)({\bf 0}\vert {\bf s}\vert {\bf 0}
\vert {\overline S})~.\nonumber
\end{eqnarray} 
This model has $SU(3)^2 \otimes (E_6)^3$ gauge symmetry.
The $U_1$ and $U_2$ are order-$2$ (${\bf Z}_2$) shifts. 
The first three entries correspond to the right-moving complex world-sheet
bosons. 
The next three entries correspond to the left-moving complex world-sheet
bosons. Together they form the six-torus. The remaining 16 left-moving 
world-sheet bosons generate the ${\mbox{Spin}}(32)/{\bf Z}_2$ lattice. 
The $SO(32)$ shifts are given in the $SO(10)^3
\otimes SO(2)$ basis. In this basis, ${\bf 0}$(0) stands for the null vector, 
${\bf v}$($V$) is the vector weight, whereas
${\bf s}$($S$) and ${\overline {\bf s}}$(${\overline S}$) are the
spinor and anti-spinor weights of $SO(10)$($SO(2)$). (For $SO(2)$, $V=1$,
$S=1/2$ and ${\overline S}=-1/2$.) 
The unshifted sector provides gauge bosons of
$SU(3)^{2} \otimes U(1)^{2} \otimes SO(10)^{3} \otimes SO(2)$. 
The permutation symmetry of the three $SO(10)$'s is explicit here. 
There are additional gauge bosons from the new sectors. Recall that under
$E_6 \supset SO(10) \otimes U(1)$, 
\begin{equation}
{\bf 78}={\bf 1}(0)+{\bf 45}(0)+{\bf 16}(3)+\overline{\bf 16}(-3)~.
\end{equation}
It is easy to see that the $U_1$, $U_2$ and $U_1+U_2$ sectors provide
the necessary ${\bf 16}(3)$ and $\overline{\bf 16}(-3)$ gauge
bosons to the three $SO(10)$'s respectively. Consistency and the 
permutation symmetry of the three $SO(10)$'s implies the permutation 
symmetry of the three $E_6$'s.
The resulting Narain model $N1(1,1)$ has $N=4$ SUSY and gauge group
$SU(3)^2 \otimes (E_6)^3$.\\ 
$\bullet$ The $N2(1,1)$ model generated by the Wilson lines
\begin{eqnarray}
 &&U_1 =(0,e_1/2,e_1/2 \vert\vert e_1/2,0,0)({\bf s}\vert {\bf 0}
\vert {\bf 0}\vert 
 {\overline S})~,\nonumber\\
 &&U_2 =(0,e_2/2,e_2/2\vert\vert e_2/2,0,0)({\bf 0}\vert {\bf s}
\vert {\bf 0}\vert 
 {\overline S})~.\nonumber
\end{eqnarray}  
This model has  
$SU(3)^2 \otimes U(1)^2\otimes SO(10)^3 \otimes SO(2)$ gauge symmetry. 

\subsection{The $E_6$ and $SO(10)$ Models}

{}Before we describe the  ${\bf Z}_6$ asymmetric orbifolds that lead to the 
$E_6$ model and the $SO(10)$ model $T1(1,1)$, 
we will introduce some notation. By $\theta$ we will denote a 
$2\pi/3$ rotation of the corresponding two real chiral world-sheet bosons. 
Thus, $\theta$ is a ${\bf Z}_3$ twist. 
Similarly, by $\sigma$ we will denote a $\pi$ rotation of the 
corresponding two chiral world-sheet bosons. 
Thus, $\sigma$ is a ${\bf Z}_2$ twist. By ${\cal P}$ we will denote
the outer-automorphism of the three $SO(10)$'s that arise in the 
breaking $SO(32)\supset SO(10)^3 \otimes SO(2)$. 
Note that ${\cal P}$ is a ${\bf Z}_3$ twist. Finally, by $(p_1,p_2)$ 
we will denote the outer-automorphism of the corresponding two complex 
chiral world-sheet bosons.
Note that $(p_1,p_2)$ is a ${\bf Z}_2$ twist. The spin structures
of the world-sheet fermions in the right-moving sector are fixed by the 
world-sheet supersymmetry consistency. (Again, more details can be found in 
Ref \cite{kt}.)

{}Finally, we are ready to give the corresponding ${\bf Z}_3 \otimes {\bf Z}_2$ 
twists.\\
$\bullet$ The $E1$ model. Start from the $N1(1,1)$ model and perform 
the following twists:
\begin{eqnarray}
 &&T_3 =(\theta,\theta,\theta\vert\vert \theta,e_1/3,0)
 ({\cal P} \vert  2/3)~,\nonumber\\
 &&T_2=(\sigma, p_1,p_2\vert\vert  0,e_1/2,e_1/2)
 (0^{15} \vert 0)~.\nonumber
\end{eqnarray}  
This model has $SU(2)_1 \otimes (E_6)_3 \otimes U(1)^3$ gauge symmetry. 
The massless spectrum of the $E1$ model is given in Table I. The states are 
grouped according to where they come from, namely, the untwisted sector U,
the ${\bf Z}_3$ twisted ({\em i.e.}, $T_3$ and $2T_3$) sector T3, 
the ${\bf Z}_6$ twisted ({\em i.e.}, $T_3+T_2$ and $2T_3+T_2$) sector T6,
and ${\bf Z}_2$ twisted ({\em i.e.}, $T_2$) sector T2.\\
$\bullet$ The $E2$ model. Start from the $N1(1,1)$ model and perform the 
following twists:
\begin{eqnarray}
 &&T_3 =(0,\theta,\theta\vert\vert \theta,e_1/3,0)
 ({\cal P} \vert  2/3)~,\nonumber\\
 &&T_2=(\sigma, p_1,p_2\vert\vert  0,e_1/2,e_1/2)
 (0^{15} \vert 0)~.\nonumber
\end{eqnarray}  
This model has $SU(2)_1 \otimes (E_6)_3 \otimes U(1)^3$ gauge symmetry. 
The massless spectrum of the $E2$ model is given in Table I.\\
Here we note that the $E1$ and $E2$ models are the same, in particular, 
they have the same tree-level massless spectra and interactions. The $E1=E2$ 
model is what we refer to as the $E_6$ model.\\
$\bullet$ The $T1(1,1)$ model. Start from the $N2(1,1)$ model and perform 
the same twists as in the $E1$ model.
This model has $SU(2)_1 \otimes SO(10)_3 \otimes U(1)^4$ gauge symmetry. 
The massless spectrum of the $T1(1,1)$ model is given in Table II.\\
$\bullet$ The $T2(1,1)$ model. Start from the $N2(1,1)$ model and perform the 
same twists as in the $E2$ model. 
This model has $SU(2)_1 \otimes SO(10)_3 \otimes U(1)^4$ gauge symmetry. 
The massless spectrum of the $T2(1,1)$ model is given in Table II.\\
Here we note that the $T1(1,1)$ and $T2(1,1)$ models are the same, in 
particular, they have the same tree-level massless spectra and interactions.\\
$\bullet$ The $T5(1,0)$ and $T6(1,0)$ models.\\
These models are the same as they have the same tree-level massless spectra 
and interactions. They can be constructed in a way similar to the 
$T1(1,1)=T2(1,1)$
model starting from a different Narain model. Here we simply give the 
massless spectra of these models in Table III and refer the reader to 
Ref \cite{class10} for details.

\subsection{The $SU(6)$ Models}

{}We can start from the $E1$ or $E2$ model to generate some of the $SU(6)$ 
models by adding Wilson lines. Let us first consider the following Wilson line
\begin{eqnarray}\label{A3}
 &&A_3=(0,0,0 \vert\vert 0,{\tilde e}^2,{\tilde e}^2)(({1\over 3}{1\over 3}
{1\over 3} {1\over  3}{2\over 3} )^3 \vert 0   )~.
\end{eqnarray}  
$\bullet$ The $S1$ model. Start from the $E1$ model and add 
the $A_3$ Wilson line. 
Choose the relative phase between the $T_3$ and $A_3$ sectors 
to be $\phi(T_3, A_3)=0$ or $1/3$ (both choices give the same model). 
This model has $SU(3)_1 \otimes SU(6)_3 \otimes U(1)^3$ gauge symmetry. 
The massless spectrum of the $S1$ model is given in Table V.\\
$\bullet$ The $S2$ model. Start from the $E1$ model and add 
the $A_3$ Wilson line. Choose the relative phase between the $T_3$ and 
$A_3$  twisted sectors to be $\phi(T_3, A_3)=2/3$. 
This model has $SU(2)_1 \otimes SU(2)_1 \otimes SU(6)_3 \otimes U(1)^3$ 
gauge symmetry. The massless spectrum of the $S2$ model is given in Table V.\\
$\bullet$ The $S3$ model. Start from the $E2$ model and add  
the $A_3$ Wilson line. 
Choose the relative phase between the $T_3$ and $A_3$  twisted sectors to 
be $\phi(T_3, A_3)=0$ or $1/3$ (both choices give the same model). 
This model has $SU(3)_1 \otimes SU(6)_3 \otimes U(1)^3$ gauge symmetry. 
The massless spectrum of the $S3$ model was given in \cite{class5}. 
Note that $S3 = S1$. \\
$\bullet$ The $S4$ model. Start from the $E2$ model and add 
the $A_3$ Wilson line. Choose the relative phase between the $T_3$ and $A_3$ 
twisted sectors to be $\phi(T_3, A_3)=2/3$. This model has 
$SU(2)_1 \otimes SU(2)_1 \otimes SU(6)_3 \otimes U(1)^3$ gauge symmetry. 
The massless spectrum of the $S4$ model was given in \cite{class5}. 
Note that $S4 = S2$.

{}By modifying some of the relative phases between sectors of these models, 
and also the Narain lattice that we started with, one can build other $SU(6)$ 
(and $SU(5)$) models. In Table IV we give the massless spectrum of the 
$S1(1,1)$ model. For illustrative purposes we also present one of the 
$SU(5)$ models (which in Ref \cite{class5} was named $F11(1,0)$). 

{}Next we turn to construction of additional $SU(6)$ models by modifying 
the ${\bf Z}_2$ twist.\\
$\bullet$ The $\Sigma1A$ model. Start from the $N1(1,1)$ model and perform 
the following twists:
\begin{eqnarray}
 &&T_3 =(\theta,\theta,\theta\vert\vert \theta,e_1/3,0)
 ({\cal P} \vert  2/3)~,\nonumber\\
 &&T_2=(\sigma, p_1,p_2\vert\vert  0,e_1/2,0)
 (({1\over 4})^{15} \vert -{3\over 4})~.\nonumber
\end{eqnarray}  
This model has $SU(3)_1 \otimes SU(2)_3 \otimes SU(6)_3 \otimes U(1)^2$ gauge 
symmetry. 
The massless spectrum of the $\Sigma1A$ model is given in Table VIII.\\
$\bullet$ The $\Sigma1B$ model. Start from the $N1(1,1)$ model and perform the 
following twists:
\begin{eqnarray}
 &&T_3 =(0,\theta,\theta\vert\vert \theta,e_1/3,0)
 ({\cal P} \vert  2/3)~,\nonumber\\
 &&T_2=(\sigma, p_1,p_2\vert\vert  0,e_1/2,0)
 (({1\over 4})^{15} \vert -{3\over 4})~.\nonumber
\end{eqnarray}  
This model has $SU(3)_1 \otimes SU(2)_3 \otimes SU(6)_3 \otimes U(1)^2$
gauge symmetry. 
The massless spectrum of the $\Sigma1B$ model is given in Table VIII.\\
$\bullet$ The $\Sigma2A$ model. Start from the $N1(1,1)$ model and perform 
the following twists:
\begin{eqnarray}
 &&T_3 =(\theta,\theta,\theta\vert\vert \theta,e_1/3,0)
 ({\cal P} \vert  2/3)~,\nonumber\\
 &&T_2=(\sigma, p_1,p_2\vert\vert  0,0,e_1/2)
 (({1\over 4})^{15} \vert -{3\over 4})~.\nonumber
\end{eqnarray}  
This model has $SU(2)_1 \otimes SU(2)_3 \otimes SU(6)_3 \otimes U(1)^3$ 
gauge symmetry. 
The massless spectrum of the $\Sigma2A$ model is given in Table IX.\\
$\bullet$ The $\Sigma2B$ model. Start from the $N1(1,1)$ model and perform the 
following twists:
\begin{eqnarray}
 &&T_3 =(0,\theta,\theta\vert\vert \theta,e_1/3,0)
 ({\cal P} \vert  2/3)~,\nonumber\\
 &&T_2=(\sigma, p_1,p_2\vert\vert  0,0,e_1/2)
 (({1\over 4})^{15} \vert -{3\over 4})~.\nonumber
\end{eqnarray}  
This model has $SU(2)_1 \otimes SU(2)_3 \otimes SU(6)_3 \otimes U(1)^3$
gauge symmetry. 
The massless spectrum of the $\Sigma2B$ model is given in Table IX.\\
Here we note that if we give vev to the triplet (adjoint) of $SU(2)_3$ (which 
is part of the horizontal symmetry) in both $\Sigma1A$ and $\Sigma1B$, we 
obtain the same model, which we will refer to as $\Sigma1$. Similarly, if 
we give vev to the triplet of $SU(2)_3$ in both $\Sigma2A$ and $\Sigma2B$, 
we obtain the same model, which we will refer to as $\Sigma2$. Turning on 
vev for the triplet of $SU(2)_3$ in these models is equivalent to adding 
the following Wilson line:  
\begin{eqnarray}
 A^\prime_3 =(0,0,0\vert\vert 0,0,e_1/3)
 ( ({1\over 3}{1\over 3}{1\over 3}{1\over 3}{2\over 3})^{3}\vert  0)~.\nonumber
\end{eqnarray}  
The massless spectra of the models $\Sigma1$ and $\Sigma2$ are given in Table 
VI.\\
There is one other Wilson line we can add. This is precisely the $A_3$ Wilson 
line given by Eq (\ref{A3}). This way we get one more model, which we will 
refer to as $\Sigma3$.\\
$\bullet$ The $\Sigma3$ model. Start from the $\Sigma1A$ model and add the 
$A_3$ Wilson line. Choose the phase $\phi(T_3,A_3)=2/3$ (other choices lead 
to models already constructed). The resulting model has 
$SU(4)_1 \otimes SU(6)_3 \otimes U(1)^2$ gauge symmetry. 
Its massless spectrum is given in the first column of Table VII.\\
$\bullet$ There is another way of obtaining the $\Sigma3$ model. Start from 
the $\Sigma1B$ model and add the $A_3$ Wilson line. Choose the phase 
$\phi(T_3,A_3)=2/3$ (other choices lead to models already constructed). 
The resulting model has $SU(4)_1 \otimes SU(6)_3 \otimes U(1)^2$ gauge 
symmetry. Its massless spectrum is given in the second column of Table 
VII. Note that the two models in Table VII are the same, and we refer 
to this model as $\Sigma 3$.\\
We note that if we add the Wilson line $A_3$ to the $\Sigma2A$ model, we 
get the $S2$ model. Similarly, if we add the Wilson line $A_3$ to the 
$\Sigma2B$ model, we get the $S4$ model.

\begin{table}[t]
\begin{tabular}{|c|c|l||l|}
 &&$E1$ &  $E2$\\
 M &Field& $SU(2) \otimes E_6 \otimes U(1)^3$ & 
$SU(2) \otimes E_6 \otimes U(1)^3$ 
      \\ \hline
 & & &\\
  & $\Phi$& $ ({\bf 1},{\bf 78})(0,0,0)_L$ & $ ({\bf 1},{\bf 78})(0,0,0)_L$ \\
  $U$&$U_{+\pm},\,U_{-\pm}$ & $2 ({\bf 1},{\bf 1})(0,-{3},\pm 3)_L$ 
& $ ({\bf 1},{\bf 1})(\pm 3,+{3},0)_L$ \\
   &$U_0$& $ ({\bf 1},{\bf 1})(0,+6,0)_L$ & $ ({\bf 1},{\bf 1})(0,+6,0)_L$ \\
    & && $({\bf 2}, {\bf 1})(0,0,\pm 3)_L$ \\
  & & &\\
  \hline
 & & & \\
   &$\chi_0$&  $ ({\bf 1},{\bf 27})(0,-{2},0)_L$ &  
$ ({\bf 1},{\bf 27})(0,-{2},0)_L$ \\
    $T3$  &$\chi_{+\pm},\,\chi_{-\pm}$& $2({\bf  1}, {\bf 27})(0,+1,\pm1)_L$ 
& $ ({\bf 1},{\overline {\bf 27}}) (\pm 1,-1,0)_L$ \\
 & && \\
 \hline
  & && \\
  $T6$  &${\tilde \chi_\pm}$&  $ ({\bf 1},{\overline {\bf 27}}) 
(\pm 1,-1,0)_L$ & $2({\bf  1}, {\bf 27})(0,+1,\pm1)_L$  \\
    & &&  \\
 \hline
 & &&  \\
   $T2$ &$D_\pm$& $({\bf 2},{\bf 1})(0,0,{\pm 3})_L$ 
& $2 ({\bf 1},{\bf 1})(0,-{3},\pm 3)_L$ \\
    &${\widetilde U_\pm}$&  $ ({\bf 1},{\bf 1})(\pm {3},+{3},0)_L$ & \\
  & &&\\
   \hline
 & &&  \\
 $U(1)$ && $(1/ \sqrt{6}, ~1/{3\sqrt{2}}, ~1/\sqrt{6})$ & $(1/ \sqrt{6}, 
~1/{3\sqrt{2}}, ~1/\sqrt{6})$ 
\end{tabular}
\caption{The massless spectra of the two $E_6$  models $E1$ and $E2$ both 
with gauge symmetry $SU(2)_1 \otimes (E_6)_3\otimes U(1)^3$.
The $U(1)$ normalization
radii are given at the bottom of the Table.
The gravity, dilaton and gauge supermultiplets are not shown.}
\label{E1}
\end{table}

\begin{table}[t]
\begin{tabular}{|c|c|l||l|}
& &$T1(1,1)$ &  $T2(1,1)$\\
 M & Field & $SU(2) \otimes SO(10)\otimes U(1)^4$ & $SU(2) \otimes SO(10) 
\otimes U(1)^4$ 
      \\ \hline
 & & &\\
   &$\Phi$& $ ({\bf 1},{\bf 45})(0,0,0,0)_L$ &
              $ ({\bf 1},{\bf  45})(0,0,0,0)_L$ \\
   &$\phi$& $ ({\bf 1},{\bf 1})(0,0,0,0)_L$ & 
$ ({\bf 1},{\bf  1})(0,0,0,0)_L$ \\ 
  $U$ &$U_{+\pm},\,U_{-\pm}$& $2 ({\bf 1},{\bf 1})(0,-{3},\pm 3,0)_L$ & 
$ ({\bf 1},{\bf 1})(\pm 3,+{3},0,0)_L$ \\
   &$U_0$& $ ({\bf 1},{\bf 1})(0,+6,0,0)_L$ & $ ({\bf 1},{\bf 1})(0,+6,0,0)_L$
 \\
    & && $({\bf 2}, {\bf 1})(0,0,\pm 3,0)_L$ \\
  & & &\\
  \hline
 & &  &\\
   &$Q_0$&  $ ({\bf 1},{\bf  16})(0,-{2},0,-1)_L$ &  $ ({\bf 1},{\bf  16})
(0,-{2},0, -1)_L$ \\
   &$H_0$&   $ ({\bf 1},{\bf  10})(0,-{2},0,+2)_L$ &  $ ({\bf 1},{\bf  10})
(0,-{2},0,+2)_L$ \\
   &$S_0$&   $ ({\bf 1},{\bf  1})(0,-{2},0,-4)_L$ &  $ ({\bf 1},{\bf 1})
(0,-{2},0,-4)_L$ \\
 $T3$  
   &$Q_{+\pm},\,Q_{-\pm}$& $2({\bf  1}, {\bf  16})(0,+1,\pm1,-1)_L$ &  $ 
({\bf 1},{\overline {\bf 16}}) (\pm 1,-1,0,+1)_L$ \\
   &$H_{+\pm},\,H_{-\pm}$& $2({\bf  1}, {\bf 10})(0,+1,\pm1,+2)_L$ &  $ 
({\bf 1},{ {\bf 10}}) (\pm 1,-1,0,-2)_L$ \\
   &$S_{+\pm},\,S_{-\pm}$& $2({\bf  1}, {\bf 1})(0,+1,\pm1,-4)_L$ &  $ 
({\bf 1},{ {\bf 1}}) (\pm 1,-1,0,+4)_L$ \\
 & & &\\
 \hline
  & & &\\
  $T6$ &${\widetilde Q_\pm}$ &  $ ({\bf 1},{\overline {\bf 16}}) 
(\pm 1,-1,0,+1)_L$ & 2({\bf  1}, 
 ${\bf 16})(0,+1,\pm1,-1)_L$  \\
      &${\widetilde H_\pm}$&  $ ({\bf 1},{{\bf 10}}) (\pm 1,-1,0,-2)_L$ & 
2({\bf  1}, 
 ${\bf 10})(0,+1,\pm1,+2)_L$  \\
     &${\widetilde S_\pm}$&  $ ({\bf 1},{{\bf 1}}) (\pm 1,-1,0,+4)_L$ & 
2({\bf  1}, 
 ${\bf 1})(0,+1,\pm1,-4)_L$  \\

    & &&  \\
 \hline
 & &&  \\
   $T2$ &$D_\pm$& $({\bf 2},{\bf 1})(0,0,{\pm 3},0)_L$ & $2 ({\bf 1},{\bf 1})(0,-{3},\pm 3,0)_L$ \\
    &${\widetilde U}_\pm$&  $ ({\bf 1},{\bf 1})(\pm {3},+{3},0,0)_L$ & \\
  & &&\\
   \hline
 & &  &\\
 $U(1)$& & $(1/ \sqrt{6}, ~1/3\sqrt{2}, ~1/\sqrt{6},~1/6)$ & $(1/ \sqrt{6}, 
  ~1/{3\sqrt{2}},    ~1/\sqrt{6},~1/6)$\\
\end{tabular}
\caption{The massless spectra of the two $SO(10)$ models $T1(1,1)$ and $T2(1,1)$ both with gauge symmetry $SU(2)_1 \otimes SO(10)_3\otimes U(1)^4$.
The $U(1)$ normalization
radii are given at the bottom of the Table.
The gravity, dilaton and gauge supermultiplets are not shown.}
\label{SO(10)}
\end{table}


\begin{table}[t]
\begin{tabular}{|c|l|l||l|} 
 M& $T5(1,0)$&$T5(1,0)$ &  $T6(1,0)$\\ 
 & Field &$SU(2)^3 \otimes SO(10)\otimes U(1)^4$ &
  $SU(2)^3 \otimes SO(10) \otimes U(1)^4$ \\  \hline
   && & \\
   && $({\bf 1},{\bf 1},{\bf 1},{\bf 45})(0,0,0,0)_L$ & $({\bf 1},{\bf 1},{\bf 1},{\bf 45})(0,0,0,0)_L$ \\
   & & $ ({\bf 1},{\bf 1},{\bf 1},{\bf 1})(\pm 3,+3,0,0)_L$  
   &  $({\bf 2},{\bf 1},{\bf 1},{\bf 1})(\pm 3,0,0,0)_L$ \\
      $U$ & $U_1$ &  $({\bf 1},{\bf 1},{\bf 1},{\bf 1})(0,+6,0,0)_L$ &
            $ ({\bf 1},{\bf 1},{\bf 1},{\bf 1}) (0,+6,0,0)_L$  \\
   & $U_2$&  $({\bf 1},{\bf 1},{\bf 1},{\bf 1})(0,0,+6,0)_L$ & 
            $({\bf 1},{\bf 1},{\bf 1},{\bf 1})(0,0,+6,0)_L$ \\
   &&  $ ({\bf 2},{\bf 2},{\bf 2},{\bf 1})(0,0,+3,0)_L$ & $({\bf 1},{\bf 2},{\bf 2},{\bf 1})(0,-3,-3,0)_L$ \\
   &&  & $({\bf 1},{\bf 2},{\bf 2},{\bf 1})(0,\pm 3,\mp 3,0)_L$ \\
   &&  &  \\ \hline
     && &  \\
     && $ ({\bf 1},{\bf 1},{\bf 1},{\overline {\bf 16}}) ( 0,+2,+2,+1)_L$
     & $ ({\bf 1},{\bf 1},{\bf 1},{\overline {\bf 16}}) ( 0,+2,+2,+1)_L$ \\
 $T3$  
     && $ ({\bf 1},{\bf 1}, {\bf 1},{\bf 10})( 0,+2,+{2},-2)_L$
     & $ ({\bf 1},{\bf 1}, {\bf 1},{\bf 10})( 0,+2,+{2},-2)_L$ \\
     && $ ({\bf 1},{\bf 1},{\bf 1},{\bf 1})( 0,+2,+{2},+4)_L$
     & $ ({\bf 1},{\bf 1},{\bf 1},{\bf 1})( 0,+2,+{2},+4)_L$ \\
       && $({\bf 1},{\bf 1},{\bf 1},{\bf 16})({\pm 1},+1,-{2},-1)_L$  & \\
       && $({\bf 1},{\bf 1},{\bf 1},{\bf 10})({\pm 1},+1,-2,+2)_L$  & \\
       && $ ({\bf 1},{\bf 1},{\bf 1},{\bf 1})({\pm 1},+1,-{2},-4)_L$ & \\
      &&  & \\  \hline
 & & &\\
     &&  & $({\bf 1},{\bf 1},{\bf 1},{\bf 16})({\pm 1},+1,-{2},-1)_L$ \\
   $T6$ 
     & & & $({\bf 1},{\bf 1},{\bf 1},{\bf 10})({\pm 1},+1,-2,+2)_L$ \\
     &&  & $ ({\bf 1},{\bf 1},{\bf 1},{\bf 1})({\pm 1},+1,-{2},-4)_L$ \\
    & & $({\bf 1},{\bf 1},{\bf 1},{\bf 16})({\pm 1},-{2},+1,-1)_L$
    & $({\bf 1},{\bf 1},{\bf 1},{\bf 16})({\pm 1},-{2},+1,-1)_L$ \\
     & &$({\bf 1},{\bf 1},{\bf 1},{\bf 10})({\pm 1},-2,+1+2)_L$
     & $({\bf 1},{\bf 1},{\bf 1},{\bf 10})({\pm 1},-2,+1+2)_L$ \\
     & & $ ({\bf 1},{\bf 1},{\bf 1},{\bf 1})({\pm 1},-{2},+1,-4)_L$ 
     & $ ({\bf 1},{\bf 1},{\bf 1},{\bf 1})({\pm 1},-{2},+1,-4)_L$ \\
     & & &  \\  \hline 
 & & & \\
  $T2$   &  & $ ({\bf 1},{\bf 2},{\bf 1},{\bf 1})(\pm 3,0,0,0)_L$
                & $ ({\bf 1},{\bf 2},{\bf 1},{\bf 1})(\pm 3,0,0,0)_L$ \\
              &&  $ ({\bf 1},{\bf 1},{\bf 2},{\bf 1})(\pm 3,0,0,0)_L$
              & $ ({\bf 1},{\bf 1},{\bf 2},{\bf 1})(\pm 3,0,0,0)_L$ \\
               && $({\bf 2},{\bf 1},{\bf 1},{\bf 1})(\pm 3,0,0,0)_L$ 
               & $ ({\bf 1},{\bf 1},{\bf 1},{\bf 1})(\pm 3,+3,0,0)_L$ \\
               &&  $ ({\bf 1},{\bf 1},{\bf 1},{\bf 1})(\pm 3,0,+3,0)_L$ 
               & $ ({\bf 1},{\bf 1},{\bf 1},{\bf 1})(\pm 3,0,+3,0)_L$ \\
            &$D_2$& $ ({\bf 2},{\bf 1},{\bf 2},{\bf 1})(0,-3,-3,0)_L$ 
            & $ ({\bf 2},{\bf 1},{\bf 2},{\bf 1})(0,-3,-3,0)_L$ \\
            & $D_{2\pm}$& $ ({\bf 2},{\bf 1},{\bf 2},{\bf 1})(0,\pm 3,\mp 3,0)_L$
            & $ ({\bf 2},{\bf 1},{\bf 2},{\bf 1})(0,\pm 3,\mp 3,0)_L$ \\
            &   $D_{3}$ &$ ({\bf 2},{\bf 2},{\bf 1},{\bf 1})(0,-3,-3,0)_L$ 
            & $ ({\bf 2},{\bf 2},{\bf 1},{\bf 1})(0,-3,-3,0)_L$ \\
            &  $D_{3\pm}$&$ ({\bf 2},{\bf 2},{\bf 1},{\bf 1})(0,\pm 3,\mp 3,0)_L$
            & $ ({\bf 2},{\bf 2},{\bf 1},{\bf 1})(0,\pm 3,\mp 3,0)_L$ \\
            &&  $ ({\bf 2},{\bf 2},{\bf 2},{\bf 1})(0,+3,0,0)_L$
            & $ ({\bf 2},{\bf 2},{\bf 2},{\bf 1})(0,+3,0,0)_L$ \\
              & $D_{1}$&  $({\bf 1},{\bf 2},{\bf 2},{\bf 1})(0,-3,-3,0)_L$ 
              & $ ({\bf 2},{\bf 2},{\bf 2},{\bf 1})(0,0,+3,0)_L$ \\
              & $D_{1\pm}$&$({\bf 1},{\bf 2},{\bf 2},{\bf 1})(0,\pm 3,\mp 3,0)_L$ & \\
    & & &\\ \hline
 $U(1)$ &&  $(1/ \sqrt{6},~1/{3\sqrt{2}},~1/{3\sqrt{2}},~1/6)$ 
              & $(1/ \sqrt{6},~1/{3\sqrt{2}},~1/{3\sqrt{2}},~1/6)$ \\
\end{tabular}

\caption{The massless spectra of the $T5(1,0)$ and $T6(1,0)$ models 
both with gauge symmetry $SU(2)^3_1\otimes SO(10)_3\otimes U(1)^4$. The $U(1)$ normalization radii are given at the bottom of the Table.
The graviton, dilaton and gauge supermultiplets are not shown.}
\end{table}

\begin{table}[t]
\begin{tabular}{|c|c|l||l|}
  &&  $S1(1,1)$ & $F11(1,0)$ \\
 M  &Fields  &$SU(2) \otimes SU(6) \otimes U(1)^4$   
 &  $SU(2)^2 \otimes SU(5) \otimes U(1)^4$ \\[1mm] \hline
  && &\\
     &$\Phi$ & $({\bf 1},{\bf 35})(0,0,0,0)_L$& $ ({\bf 1},{\bf 1},{\bf 24})(0,0,0,0)_L$ \\
  $U$ &$U_{+\pm},\,U_{-\pm}$& $2 ({\bf 1},{\bf 1},{\bf 1})(0,-{3},\pm 3,0)_L$
    & $2 ({\bf 1},{\bf 1},{\bf 1})(0,-{3},\pm 3,0)_L$ \\
    &$U_0$& $ ({\bf 1},{\bf 1},{\bf 1})(0,+6,0,0)_L$ & $ 
({\bf 1},{\bf 1},{\bf 1})(0,+6,0,0)_L$ \\
    &$\phi$& $ ({\bf 1},{\bf 1},{\bf 1})(0,0,0,0)_L$ & $ ({\bf 1},{\bf 1},{\bf 1})(0,0,0,0)_L$ \\
     & && $ ({\bf 1},{\bf 3},{\bf 1})(0,0,0,0)_L$ \\
 & &&\\
 \hline
  & &&\\
                & $Q_0$& $ ({\bf 1},{\bf 15})(0,-{2},0 ,0)_L$    
                & $ ({\bf 1},{\bf 2},{\overline {\bf 5}})(0,-{2},0 ,+1)_L$  \\ 
                  &$H_0^+$& $ ({\bf 1},{\overline {\bf 6}})(0,-2,0,+1)_L$     
                       & $ ({\bf 1},{\bf 2},{\bf 1})(0,-2,0,-5)_L$  \\ 
                 &$H_0^-$& $ ({\bf 1},{\overline {\bf 6}})(0,-2,0,-1)_L$
                 & $ ({\bf 1},{\bf 1},{\bf 10})(0,-2,0,-2)_L$\\ 
     && &  $ ({\bf 1},{\bf 1},{\bf 5})(0,-2,0,+4)_L$ \\ 
 $T3$        
           &$Q_{+\pm},\,Q_{-\pm}$& $2({\bf 1},{\bf 15})(0,+1,\pm 1 ,0)_L$
         & $2({\bf 1},{\bf 2},{\overline {\bf 5}})(0,+1,\pm 1 ,+1)_L$  \\ 
 &$H_{+\pm}^+,\,H_{-\pm}^+$  & 
$2 ({\bf 1},{\overline {\bf 6}})(0,+1, \pm 1,+1)_L$ 
                 & $2 ({\bf 1},{\bf 2},{\bf 1})(0,+1, \pm 1,-5)_L$  \\ 
       &$H_{+\pm}^+,\,H_{-\pm}^+$& $2 ({\bf 1},{\overline {\bf 6}})
(0,+1, \pm 1,-1)_L$
                        & $ 2({\bf 1},{\bf 1},{\bf 10})(0,+1, \pm 1,-2)_L$\\ 
               &  & & $2 ({\bf 1},{\bf 1},{\bf 5})(0,+1,\pm 1,+4)_L$ \\ 
   & & &\\
 \hline
  & & &\\
   & ${\widetilde Q}_\pm$  & $ ({\bf 1},{\overline {\bf 15}})(\pm 1,-1,0,0)_L$
                    & $ ({\bf 1},{\bf 2},{{\bf 5}})(\pm 1,-{1},0 ,-1)_L$  \\ 
          $T6$   &${\widetilde H}_\pm^+$& $ ({\bf 1},{\bf 6})(\pm 1,-1,0,+1)_L$
                & $ ({\bf 1},{\bf 2},{\bf 1})(\pm 1,-1,0,+5)_L$ \\ 
                 &${\widetilde H}_\pm^-$& $ ({\bf 1},{\bf 6})(\pm 1,-1,0,-1)_L$
            & $ ({\bf 1},{\bf 1},{\overline {\bf 10}})(\pm 1,-1,0,+2)_L$\\ 
       &  & & $ ({\bf 1},{\bf 1},{\overline {\bf 5}})(\pm 1,-1,0,-4)_L$ \\ 
 & & &\\
 \hline
&  & &\\
   $T2$  & $D_\pm$    & $({\bf 2},{\bf 1},{\bf 1})(0,0,{\pm 3},0)_L$
   & $({\bf 2},{\bf 1},{\bf 1})(0,0,{\pm 3},0)_L$ \\
  &  ${\widetilde U}_\pm$& $({\bf 1},{\bf 1},{\bf 1})(\pm {3},+{3},0,0)_L$ 
   & $({\bf 1},{\bf 1},{\bf 1})(\pm {3},+{3},0,0)_L$ \\
& & &\\
\hline
 & & &\\
 $U(1)$   &   & $(1/\sqrt{6}, 1/{3\sqrt{2}},1/\sqrt{6},1/\sqrt{6})$
      & $(1/\sqrt{6}, 1/{3\sqrt{2}},1/\sqrt{6},1/{3\sqrt{10}})$ \\
\end{tabular}
\caption{The massless spectra of the $S1(1,1)$ and 
$F11(1,0)$ models with gauge groups $SU(2)_1 \otimes SU(6)_3 \otimes U(1)^4$ 
and 
$SU(2)_1 \otimes SU(2)_3 \otimes SU(5)_3 \otimes U(1)^4$, respectively.
The $U(1)$ normalization radii are given at the bottom of the Table.
The gravity, dilaton and gauge supermultiplets are not shown.}
\label{S11}
\end{table}

\begin{table}[t]
\begin{tabular}{|c|c|l||c|l|} 
M    &&  $S1$ & & $S2$ \\
       &Field& $SU(3)   \otimes SU(6) \otimes U(1)^3$ &
    Field&$SU(2)^2 \otimes SU(6) \otimes U(1)^3$ \\
   \hline
 & & &&\\  
 &$\Phi$& $({\bf 1},{\bf 35})(0,0,0)_L$
   &$\Phi$& $ ({\bf 1},{\bf 1},{\bf 35})(0,0,0)_L$ \\
 &$\phi$&  $({\bf 1},{\bf 1})(0,0,0)_L$
   &$\phi$& $ ({\bf 1},{\bf 1},{\bf 1})(0,0,0)_L$ \\
 $U$    &$U_0$&  $ ({\bf 1},{\bf 1}) (+6,0,0)_L$ 
   & $U_0$&$ ({\bf 1},{\bf 1},{\bf 1})(0,0,-6)_L$ \\
    &$U_{+\pm},\,U_{-\pm}$& $2 ({\bf 1},{\bf 1})(-3, {\pm 3},{\pm 3})_L$ &
    $d_{+\pm},\,d_{-\pm}$& $2 ({\bf 1},{\bf 2},{\bf 1})
({\pm 1},{\mp 3},+3)_L$ \\
    &${\widetilde T}_\pm$& $2 ({\overline {\bf 3}},{\bf 1})(+3,-3,+1)_L$ & 
     $D_\pm$&$  ({\bf 2},{\bf 1}, {\bf 1})({\pm 2},0,+3)_L$\\
   &$T_0$& $({\bf 3},{\bf 1})(0,0,-4)_L$ & \\
  & & &&\\
 \hline
   & & &&\\
    & $F_\pm^A$&$6 ({\bf 1},{\bf 15})(+1,-1,-{1})_L$ &
     $F^A$&$3 ({\bf 1},{\bf 1},{\bf 15})(0,0,+2)_L$ \\
  $T3$ & ${\widetilde S}^A_\pm$&
$6 ({\bf 1},{\overline {\bf 6}})(-2,+1,-{1})_L$ & 
  ${\widetilde S}^A_{+\pm},\,{\widetilde S}^A_{-\pm}$&  
 $6 ({\bf 1},{\bf 1},{\overline {\bf 6}})(\pm 1,\mp {1},-1)_L$ \\
    &${\widetilde S}^A_0$&  $3 ({\bf 1},{\overline {\bf 6}})(+1,0,+2)_L$ & \\ 
&&  & & \\
   \hline
 &&& & \\ 
 $T6$ & $S^A$&      $3({\bf 1},{\bf 6})(+2,+1,+1)_L$ &
     $S^A_\pm$& $ 3({\bf 1},{\bf 1}, {\bf 6})(\pm 1,\pm {1},+1)_L$ \\
  &${\widetilde F}^A$& $ 3({\bf 1},{\overline {\bf 15}}) 
( -1,-{1},+{1})_L$ & \\
  & &&& \\
   \hline 
 & & &&\\   
    &  $T_\pm$&$({\bf 3},{\bf 1})(\pm 3,-3,-1)_L$ &
  $\Delta_\pm$&$({\bf 2},{\bf 2},{\bf 1})(\pm 1,\mp 3,0)_L$ \\
 $T2$   &${\widetilde T}_0$&  $({\overline {\bf 3}},{\bf 1})(-3,+3,+1)_L$ &
 ${\tilde d}_\pm$&$({\bf 1},{\bf 2},{\bf 1})(\pm 1,\pm 3, -3)_L$ \\
 &${\widetilde U}_\pm$& $ ({\bf 1},{\bf 1})(+3,\pm 3,\mp 3)_L$ & \\
 & &&& \\
 \hline
 $U(1)$&  &
    $({1\over{3\sqrt{2}}},~{1\over{2\sqrt{3}}},~{1\over 
    {2\sqrt{3}}})$ &&
  $({1\over 2},~{1\over {2\sqrt{3}}},~{1\over {3\sqrt{2}}})$~ \\
\end{tabular}
\caption{The massless spectra of the two $SU(6)$ models $S1$ and 
$S2$ with gauge symmetries 
$SU(3)_1 \otimes SU(6)_3 \otimes U(1)^3$ and 
$SU(2)_1 \otimes SU(2)_1 \otimes SU(6)_3 \otimes U(1)^3$, respectively. 
Note that double signs (as in $({\bf 1},{\bf 2},{\bf 1})(\pm 1,\pm 3, -3)_L$)
are correlated. The $U(1)$ normalization 
radii are given at the bottom of the table.
The graviton, dilaton and gauge supermultiplets are not shown.}
\label{S1}
\end{table}

\begin{table}[t]
\begin{tabular}{|c|c|l|c|l|}
 &&$\Sigma1$ (from $\Sigma 1A$ with $A^\prime_3$)& 
 &$\Sigma2$ (from $\Sigma 2A$ with $A^\prime_3$)\\
 M &Field& $SU(3) \otimes SU(6) \otimes U(1)^3$ & 
Field& $SU(2)\otimes SU(6) \otimes U(1)^4$ 
      \\ \hline
 & & &&\\
   & $\Phi$&$({\bf 1},{\bf 35})(0,0,0)_L$ &
          $\Phi$&      $({\bf 1},{\bf 35})(0,0,0,0)_L$  \\
   & $\phi$& $({\bf 1},{\bf 1})(0,0,0)_L$ &
          $\phi$&      $({\bf 1},{\bf 1})(0,0,0,0)_L$  \\
   $U$ & $U_0$& $({\bf 1},{\bf 1})(0,+6,0)_L$ &
             $U_1$&   $({\bf 1},{\bf 1})(0,0,+6,0)_L$  \\
   &$U_{+\pm},\,U_{-\pm}$&  $2({\bf 1},{\bf 1})(0,-3,\pm 3)_L$
              &$U_2$& $({\bf 1},{\bf 1})(0,0,-3,-3)_L$  \\    
&&&$U_3$&$({\bf 1},{\bf 1})(0,0,-3,+3)_L$\\
  & & &&\\
  \hline
 & &  &&\\
   & $F_0$& $({\bf 1},{\bf 15})(0,-2,0)_L$ &
         $F_1$&        $({\bf 1},{\bf 15})(0,0,-2,0)_L$\\
    & $F_{+\pm},\,F_{-\pm}$& $2({\bf 1},{\bf 15})(0,+1,\pm 1)_L$ &
                 $F_2$&$({\bf 1},{\bf 15})(0,0,+1,+1)_L$\\
&&&$F_3$&$({\bf 1},{\bf 15})(0,0,+1,-1)_L$\\  
  $T3$ 
   &  ${\widetilde S}'_{+\pm}$&$({\bf 1},{\overline {\bf 6}})(\pm1,+1,+ 1)_L$ &
      ${\widetilde S}_{+\pm}^2,\,{\widetilde S}_{-\pm}^2$& 
$2({\bf 1},{\overline {\bf 6}})(\pm1,0,+1,+1)_L$\\
 & ${\widetilde S}'_{-\pm}$&$({\bf 1},{\overline {\bf 6}})(\pm1,+1,- 1)_L$ &
     ${\widetilde S}_{+\pm}^3,\,{\widetilde S}_{-\pm}^3$& 
            $2({\bf 1},{\overline {\bf 6}})(\pm1,0,+1,-1)_L$\\
 &${\widetilde S}_{+\pm},\,{\widetilde S}_{-\pm}$
& $2({\bf 1},{\overline {\bf 6}})(\pm 1,-2,0)_L$ &
     ${\widetilde S}_{+\pm}^1,\,{\widetilde S}_{-\pm}^1$& 
                 $2({\bf 1},{\overline {\bf 6}})(\pm 1,0,-2,0)_L$\\
 & & &&\\
 \hline
  & & &&\\
     &  ${\widetilde F}_\pm$&$({\bf 1},{\overline {\bf 15}})(\pm1,-1,0)_L$ &
                 $S_\pm^1$&$({\bf 1},{{\bf 6}})(0,\pm1,+2,0)_L$\\ 
   $T6$ & $S_\pm$& $({\bf 1},{ {\bf 6}})(0,+2,\pm1)_L$ &
                 $S_\pm^2$&$({\bf 1},{{\bf 6}})(0,\pm1,-1,-1)_L$\\ 
            &  &&$S_\pm^3$&
                 $({\bf 1},{{\bf 6}})(0,\pm1,-1,+1)_L$\\
    & &&&  \\
 \hline
 & &  &&\\
   $T2$ &${\widetilde U}_\pm$  
              & $({\bf 1},{\bf 1})(\pm3,+ 3,0)_L$ &
 $\Delta_\pm$&$({\bf 2},{\bf 1})(\pm3,0,0,0)_L$ \\
   & & &&\\
   \hline
 & &  &&\\
 $U(1)$ && $(1/\sqrt{6},~1/{3\sqrt{2}}, ~1/\sqrt{6})$ & 
& $(1/\sqrt{6},~1/\sqrt{6},~1/{3\sqrt{2}},  ~1/\sqrt{6})$\\
\end{tabular}
\caption{The massless spectra of the two $SU(6)$ models $\Sigma1$ and 
$\Sigma2$ (obtained by adding $A^\prime_3$ Wilson line to the model 
$\Sigma1A$ and to the model $\Sigma1B$, respectively). The model 
$\Sigma1$ has gauge symmetry $SU(3)_1 \otimes SU(6)_3\otimes U(1)^3$. 
The model $\Sigma2$ has gauge symmetry $SU(2)_1 \otimes SU(6)_3\otimes U(1)^4$.
The $U(1)$ normalization
radii are given at the bottom of the Table.
The gravity, dilaton and gauge supermultiplets are not shown.}
\label{Sigma1}
\end{table}

\begin{table}[t]
\begin{tabular}{|c|l|l|}
 &$\Sigma3$ (from $\Sigma 1A$ with $A_3$)& 
 $\Sigma3$ (from $\Sigma 1B$ with $A_3$)\\
 M & $SU(4)\otimes SU(6) \otimes U(1)^2$ & 
 $SU(4)\otimes SU(6) \otimes U(1)^2$ 
      \\ \hline
 & & \\
   & $({\bf 1},{\bf 35})(0,0)_L$ &
                $({\bf 1},{\bf 35})(0,0)_L$  \\
   &  $({\bf 1},{\bf 1})(0,0)_L$ &
                $({\bf 1},{\bf 1})(0,0)_L$  \\
   $U$ &  $({\bf 1},{\bf 1})(+6,0)_L$ &
                $({\bf 1},{\bf 1})(+6,0)_L$  \\
   &  $2({\bf 4},{\bf 1})(-3,- 3)_L$
              & $({\bf 4},{\bf 1})(+3,+ 3)_L$  \\
        &$2({\overline {\bf 4}},{\bf 1})(-3,+3)_L$
              & $({\overline {\bf 4}},{\bf 1})(+3,-3)_L$  \\      
  & & \\
  \hline
 & &  \\
   &  $3({\bf 1},{\bf 15})(-2,0)_L$ &
                 $3({\bf 1},{\bf 15})(-2,0)_L$\\
    $T3$ 
   &  $3({\bf 1},{\overline {\bf 6}})(+1,\pm 1)_L$ &
                 $3({\bf 1},{\overline {\bf 6}})(+1,\pm 1)_L$\\
 & & \\
 \hline
  & & \\
  $T6$     &  ------------& ------------\\
    & &  \\
 \hline
 & &  \\
   $T2$   
              & $({\bf 4},{\bf 1})(+3,+ 3)_L$ 
                   &  $2({\bf 4},{\bf 1})(-3,- 3)_L$ \\
        
              & $({\overline {\bf 4}},{\bf 1})(+3,-3)_L$ 
                   &$2({\overline {\bf 4}},{\bf 1})(-3,+3)_L$ \\
& &\\
   \hline
 & &  \\
 $U(1)$ & $(1/{3\sqrt{2}}, ~1/\sqrt{12})$ & $(1/{3\sqrt{2}},  ~1/\sqrt{12})$\\
\end{tabular}
\caption{The massless spectrum of the $SU(6)$ model $\Sigma3$ constructed by adding $A_3$ Wilson line to the model $\Sigma1A$ (first column) and to the model  $\Sigma1B$ (second column). The model $\Sigma3$ has gauge symmetry $SU(4)_1  \otimes SU(6)_3\otimes U(1)^2$.
The $U(1)$ normalization
radii are given at the bottom of the Table.
The gravity, dilaton and gauge supermultiplets are not shown.}

\end{table}

\begin{table}[t]
\begin{tabular}{|c|l|l|}
 &$\Sigma 1A$ &  $\Sigma 1B$\\
 M & $SU(3) \otimes SU(2)\otimes SU(6) \otimes U(1)^2$ & 
 $SU(3) \otimes  SU(2)\otimes SU(6) \otimes U(1)^2$ 
      \\ \hline
 & & \\
   & $({\bf 1},{\bf 1},{\bf 35})(0,0)_L$ &
                $({\bf 1},{\bf 1},{\bf 35})(0,0)_L$  \\
   &  $({\bf 1},{\bf 3},{\bf 1})(0,0)_L$ &
                $({\bf 1},{\bf 3},{\bf 1})(0,0)_L$  \\
   $U$ &  $({\bf 1},{\bf 1},{\bf 1})(+6,0)_L$ &
                $({\bf 1},{\bf 1},{\bf 1})(+6,0)_L$  \\
   &  $2({\bf 1},{\bf 1},{\bf 1})(-3,\pm 3)_L$
              & $({\bf 1},{\bf 1},{\bf 1})(+3,\pm 3)_L$  \\
   &  $2({\bf 1},{\bf 2},{\bf 20})(0,0)_L$
              &  $2({\bf 1},{\bf 2},{\bf 20})(0,0)_L$ \\   
  & & \\
  \hline
 & &  \\
   &  $({\bf 1},{\bf 1},{\bf 15})(-2,0)_L$ &
                 $({\bf 1},{\bf 1},{\bf 15})(-2,0)_L$\\
    $T3$ &  $2({\bf 1},{\bf 1},{\bf 15})(+1,\pm 1)_L$ &
                 $({\bf 1},{\bf 1},{\overline {\bf 15}})(-1,\pm 1)_L$ \\
   &  $({\bf 1},{\bf 2},{\overline {\bf 6}})(+1,\pm 1)_L$ &
                 $({\bf 1},{\bf 2},{\overline {\bf 6}})(+1,\pm 1)_L$\\
   &  $2({\bf 1},{\bf 2},{\overline {\bf 6}})(-2,0)_L$ &
                 $({\bf 1},{\bf 2},{ {\bf 6}})(+2,0)_L$\\
 & & \\
 \hline
  & & \\
  $T6$     &  $({\bf 1},{\bf 2},{\overline {\bf 15}})(-1,0)_L$ &
                 $2({\bf 1},{\bf 2},{\bf 15})(+1,0)_L$\\
   &  $({\bf 1},{\bf 1},{\bf 6})(+2,\pm 1)_L$ &
                 $2({\bf 1},{\bf 1},{\overline {\bf 6}})(-2,\pm 1)_L$ \\
    & &  \\
 \hline
 & &  \\
   & $2({\bf 1},{\bf 2},{\bf 1})(-3,0)_L$ & 
 $2({\bf 1},{\bf 2},{\bf 1})(+3,0)_L$ \\
  $T2$  & $ ({\bf 1},{\bf 4},{\bf 1})(+{3},0)_L$  &  
              $ 2({\bf 1},{\bf 4},{\bf 1})(-{3},0)_L$  \\
 & $({\bf 1},{\bf 2},{\bf 1})(+3,0)_L$ & \\
  & &\\
   \hline
 & &  \\
 $U(1)$ & $(1/{3\sqrt{2}}, ~1/\sqrt{6})$ & $(1/{3\sqrt{2}},  ~1/\sqrt{6})$\\
\end{tabular}
\caption{The massless spectra of the two $SU(6)$  models $\Sigma1A$ and $\Sigma1B$ both with gauge symmetry $SU(3)_1 \otimes SU(2)_3 \otimes SU(6)_3\otimes U(1)^2$.
The $U(1)$ normalization
radii are given at the bottom of the Table.
The gravity, dilaton and gauge supermultiplets are not shown.}

\end{table}

\begin{table}[t]
\begin{tabular}{|c|l|l|}
 &$\Sigma 2A$ &  $\Sigma 2B$\\
 M & $SU(2) \otimes SU(2)\otimes SU(6) \otimes U(1)^3$ & 
 $SU(2) \otimes  SU(2)\otimes SU(6) \otimes U(1)^3$ 
      \\ \hline
 & & \\
   & $({\bf 1},{\bf 1},{\bf 35})(0,0,0)_L$ &
                $({\bf 1},{\bf 1},{\bf 35})(0,0,0)_L$  \\
   &  $({\bf 1},{\bf 3},{\bf 1})(0,0,0)_L$ &
                $({\bf 1},{\bf 3},{\bf 1})(0,0,0)_L$  \\
   $U$ &  $({\bf 1},{\bf 1},{\bf 1})(0,+6,0)_L$ &
                $({\bf 1},{\bf 1},{\bf 1})(0,+6,0)_L$  \\
   &  $({\bf 1},{\bf 1},{\bf 1})(0,-3,\pm 3)_L$
              & $({\bf 1},{\bf 1},{\bf 1})(0,-3,\pm 3)_L$  \\
   &  $2({\bf 1},{\bf 2},{\bf 20})(0,0,0)_L$
              &  $2({\bf 1},{\bf 2},{\bf 20})(0,0,0)_L$ \\
   &   & $({\bf 2},{\bf 1},{\bf 1})(\pm 3,0,0)_L$
                \\    
  & & \\
  \hline
 & &  \\
   &  $({\bf 1},{\bf 1},{\bf 15})(0,-2,0)_L$ &
                 $({\bf 1},{\bf 1},{\bf 15})(0,-2,0)_L$\\
    $T3$ &  $({\bf 1},{\bf 1},{\bf 15})(0,+1,\pm 1)_L$ &
                 $({\bf 1},{\bf 1},{ {\bf 15}})(0,+1,\pm 1)_L$ \\
   &  $2({\bf 1},{\bf 2},{\overline {\bf 6}})(0,-2,0)_L$ &
                 $({\bf 1},{\bf 2},{{\bf 6}})(0,+2,0)_L$\\
   &  $2({\bf 1},{\bf 2},{\overline {\bf 6}})(0,+1,\pm1)_L$ &
                 $({\bf 1},{\bf 2},{ {\bf 6}})(0,-1,\pm1)_L$\\
 & & \\
 \hline
  & & \\
         &  $({\bf 1},{\bf 1},{{\bf 6}})(\pm1,+2,0)_L$ &
                 $2({\bf 1},{\bf 1},{\overline {\bf 6}})(\pm1,-2,0)_L$\\
  $T6$     &  $({\bf 1},{\bf 1},{{\bf 6}})(\pm1,-1,-1)_L$ &
                 $2({\bf 1},{\bf 1},{\overline {\bf 6}})(\pm1,+1,+1)_L$\\
           &  $({\bf 1},{\bf 1},{{\bf 6}})(\pm1,-1,+1)_L$ &
                 $2({\bf 1},{\bf 1},{\overline {\bf 6}})(\pm1,+1,-1)_L$\\
       & &  \\
 \hline
 & &  \\
   $T2$ & $({\bf 2},{\bf 2},{\bf 1})(0,0,0)_L$ & 
 $2({\bf 2},{\bf 2},{\bf 1})(0,0,0)_L$ \\
  & $ ({\bf 2},{\bf 4},{\bf 1})(0,0,0)_L$  &  
                \\
  & &\\
   \hline
 & &  \\
 $U(1)$ & $(1/\sqrt{6},~1/{3\sqrt{2}}, ~1/\sqrt{6})$ &
 $(1/\sqrt{6},~1/{3\sqrt{2}},  ~1/\sqrt{6})$\\
\end{tabular}
\caption{The massless spectra of the two $SU(6)$  models $\Sigma2A$ and $\Sigma2B$ both with gauge symmetry $SU(2)_1 \otimes SU(2)_3 \otimes SU(6)_3\otimes U(1)^3$.
The $U(1)$ normalization
radii are given at the bottom of the Table.
The gravity, dilaton and gauge supermultiplets are not shown.}

\end{table}


\begin{references}
\bigskip

\bibitem{rev} For recent reviews, see, {\em e.g.},\\
A. Faraggi, hep-ph/9707311;\\
F. Quevedo, hep-ph/9707434;\\
P. Nath, hep-ph/9708221;\\
Z. Kakushadze, hep-th/9707249;\\
G. Cleaver, hep-th/9708023.

\bibitem{FFC} H. Kawai, D.C. Lewellen and S.-H.H. Tye, Phys. Rev. Lett. {\bf 57} (1986) 1832; (E) Phys. Rev. Lett. {\bf 58} (1987) 492; Nucl. Phys. {\bf B288} (1987) 1;\\
I. Antoniadis, C. Bachas and C. Kounnas, Nucl. Phys. {\bf B289} (1987) 87;\\
I. Antoniadis and C.P. Bachas, Nucl. Phys. {\bf B298} (1988) 586;\\
H. Kawai, D.C. Lewellen, J.A. Schwarz and S.-H.H. Tye, Nucl. Phys. {\bf B299} (1988) 431.

\bibitem{Orb} L. Dixon, J. Harvey, C. Vafa and E. Witten, Nucl. Phys. {\bf B261} (1985) 678; Nucl. Phys. {\bf B274} (1986) 285;\\
L.E. Ib{\'a}{\~n}ez, H.-P. Nilles and F. Quevedo, Phys. Lett. {\bf B187} (1987) 25;\\
K.S. Narain, M.H. Sarmadi and C. Vafa, Nucl. Phys. {\bf  B288} (1987) 55;\\ 
Z. Kakushadze and S.-H.H. Tye, Phys. Rev. {\bf D54} (1996) 7520.

\bibitem{NonAbe} D.S. Freed and C. Vafa, Comm. Math. Phys. {\bf 110} (1987) 349;\\
P. Ginsparg, Nucl. Phys. {\bf B295} (1988) 153;\\
R. Dijkgraaf, C. Vafa, E. Verlinde and H. Verlinde, Comm. Math. Phys. {\bf 123}
(1989) 485;\\
Z.S. Li and C.S. Lam, Int. J. Mod. Phys. {\bf A7} (1992) 5739;\\
Z. Kakushadze, G. Shiu and S.-H. H. Tye, Phys. Rev. {\bf D54} (1996) 7545.

\bibitem{scatt}D. Friedan, E. Martinec and S. Shenker, Phys. Lett. {\bf B160} (1985) 55; Nucl. Phys. {\bf B271} (1986) 93;\\
S. Hamidi and C. Vafa, Nucl. Phys. {\bf B279} (1987) 465;\\
L. Dixon, D. Friedan, E. Martinec and S. Shenker, Nucl. Phys. {\bf B282} (1987) 13;\\
Z. Kakushadze, G. Shiu and S.-H.H. Tye, Nucl. Phys. {\bf B501} (1997) 547, hep-th/9704113.

\bibitem{standard} For a partial list, see, {\em  e.g.},\\
L.E. Ib\'a\~nez, J.E. Kim, H.-P. Nilles and F. Quevedo,
Phys. Lett. {\bf B191} (1987) 282; \\
D. Bailin, A. Love and S. Thomas, Phys. Lett. {\bf B194} (1987) 385;\\
B.R. Greene, K.H. Kirklin, P.J. Miron and G.G. Ross, Nucl. Phys. 
{\bf B292} (1987) 606;\\
L.E. Ib\'a\~nez, H.-P. Nilles and F. Quevedo, Nucl. Phys. {\bf B307}
(1988) 109; \\
R. Arnowitt and P. Nath, Phys. Rev. {\bf D40} (1989) 191;\\
A. Font, L.E. Ib\'a\~nez, F. Quevedo and A. Sierra,
Nucl. Phys. {\bf B331} (1990) 421;\\
A. Font, L.E. Ib\'a\~nez, and F. Quevedo,
Nucl. Phys. {\bf B345} (1990) 389;\\  
L.E. Ib\'a\~nez, D. L\"{u}st and G.G. Ross, Phys. Lett. {\bf B272} 
(1991) 251;\\
A.E. Faraggi, Phys. Lett. {\bf B278} 
(1992) 131; Phys. Rev. {\bf D47} (1993) 5021;\\  
S. Kachru, Phys. Lett. {\bf B349} (1995) 76;\\
S. Chaudhuri, G. Hockney and J.D. Lykken, Nucl. Phys. {\bf B469} (1996) 357.

\bibitem{semi} For a partial list, see, {\em  e.g.},\\
I. Antoniadis, J. Ellis, J. Hagelin and D.V. Nanopoulos,
Phys. Lett. {\bf B194} (1987) 231; {\bf B208} (1988) 209; 
{\bf B231} (1989) 65;\\
J. Lopez, D.V. Nanopoulos and K. Yuan, Nucl. Phys. {\bf B399} (1993) 654;\\
J. Lopez and D.V. Nanopoulos,  Nucl. Phys. {\bf B338} (1989) 73; Phys. Rev.
Lett. {\bf 76} (1996) 1566;\\
I. Antoniadis, G.K. Leontaris and J. Rizos, Phys. Lett. {\bf B245} 
(1990) 161;\\
G.K. Leontaris, Phys. Lett. {\bf B372} (1996) 212;\\
D. Finnell, Phys. Rev. {\bf D53} (1996) 5781;\\
R. Barbieri, G. Dvali and A. Strumia, Phys. Lett. {\bf B333} (1994) 79;\\
A. Maslikov, S. Sergeev and G. Volkov, Phys. Rev. {\bf D50} (1994) 7440;\\
A. Maslikov, I. Naumov and G. Volkov, Int. J. Mod. Phys. {\bf A11} (1996) 
1117.

\bibitem{Lew} D.C. Lewellen, Nucl. Phys. {\bf B337} (1990) 61.

\bibitem{dienu} For a recent review, see, {\em e.g.},\\
K.R. Dienes, Phys. Rep. {\bf 287} (1997) 447.

\bibitem{coupling}
C. Giunti, C.W. Kim and U.W. Lee, Mod. Phys. Lett. {\bf A6} (1991) 1745;\\
J. Ellis, S. Kelly and D.V. Nanopoulos, Phys. Lett. {\bf B260} (1991) 131;\\
U. Amaldi, W. de Boer and H. Furstenau, Phys. Lett. {\bf B260} (1991) 447;\\
P. Langacker and M. Luo, Phys. Rev. {\bf D44} (1991) 817.

\bibitem{dienf} K.R. Dienes and A. Faraggi, Phys. Rev. Lett. {\bf 75} (1995) 2646;
Nucl. Phys. {\bf B457} (1995) 409.

\bibitem{witten} E. Witten, Nucl. Phys. {\bf B471} (1996) 135.

\bibitem{Sch} J.A. Schwartz, Phys. Rev. {\bf D42} (1990) 1777.

\bibitem{CCHL} S. Chaudhuri, S.-W. Chung and J.D. Lykken, hep-ph/9405374;\\
S. Chaudhuri, S.-W. Chung, G. Hockney and J.D. Lykken, hep-th/9409151; Nucl. Phys. {\bf B456} (1995) 89;\\
G.B. Cleaver, hep-th/9409096; Nucl. Phys. {\bf B456} (1995) 219; hep-th/9506006; hep-th/9604183.

\bibitem{AFIU} G. Aldazabal, A. Font, L.E. Ib\'a\~nez and A.M. Uranga,
Nucl. Phys. {\bf B452} (1995) 3.

\bibitem{Erler} J. Erler, Nucl. Phys. {\bf B475} (1996) 597.

\bibitem{kst0} Z. Kakushadze, G. Shiu and S.-H. H. Tye (Ref \cite{NonAbe}).

\bibitem{kt} Z. Kakushadze and S.-H.H. Tye (Ref \cite{Orb}).

\bibitem{vafa} K.S. Narain, M.H. Sarmadi and C. Vafa (Ref \cite{Orb}). 

\bibitem{kt10} Z. Kakushadze and S.-H.H. Tye, Phys. Rev. Lett. {\bf 77} 
(1996) 2612.

\bibitem{kt5} Z. Kakushadze and S.-H.H. Tye, Phys. Lett. {\bf B392} (1997) 335.

\bibitem{class10} Z. Kakushadze and S.-H.H. Tye, Phys. Rev. {\bf D55} (1997) 7878.

\bibitem{class5} Z. Kakushadze and S.-H.H. Tye, Phys. Rev. {\bf D55} (1997) 7896.

\bibitem{GSDSW} M. Green and J.H. Schwarz, Phys. Lett. {\bf 149B} (1984) 117; \\
E. Witten,  Phys. Lett. {\bf 149B} (1984) 351; \\
M. Dine, N. Seiberg and E. Witten, Nucl. Phys. {\bf B289} (1987) 585; \\
J. Attick, L. Dixon and A. Sen, Nucl. Phys. {\bf B292} (1987) 109; \\
M. Dine, I. Ichinose and N. Seiberg, Nucl. Phys. {\bf B293} (1987)253.

\bibitem{kst} Z. Kakushadze, G. Shiu and S.-H.H. Tye (Ref \cite{scatt}).

\bibitem{Seiberg} See, {\em e.g.},\\
N. Seiberg, Nucl. Phys. {\bf B435} (1995) 129;\\
K. Intriligator and N. Seiberg, Nucl. Phys. {\bf B444} (1995) 125; Nucl. Phys. Proc. Suppl. {\bf 45BC} (1996) 1.

\bibitem{kstv} Z. Kakushadze, G. Shiu, S.-H.H. Tye and Y. Vtorov-Karevsky, Phys. Lett. 
{\bf B408} (1997) 173, hep-ph/9705202.

\bibitem{dien}K.R. Dienes and J. March-Russell, 
Nucl. Phys. {\bf B479} (1996) 113; \\
K.R. Dienes, Nucl. Phys. {\bf B488} (1997) 141.

\bibitem{Georgi} H. Georgi and S. Glashow, Phys. Rev. Lett. {\bf 32} (1974) 438.
\bibitem{pseudo}
Z. Berezhiani and G. Dvali, Sov. Phys. Lebedev Inst. Reports {\bf 5} (1989) 55;\\
R. Barbieri, G. Dvali and M. Moretti, Phys. Lett. {\bf B322} (1993) 173;\\
R. Barbieri, Z. Berezhiani, G. Dvali, L. Hall and A. Strumia, Nucl. Phys. {\bf B432} (1994) 49;\\ 
Z. Berezhiani, hep-ph/9412372; Phys. Lett. {\bf B355} (1995) 481; hep-ph/9602325; hep-ph/9703426;\\
Z. Berezhiani, C. Csaki and L. Randall, Nucl. Phys. {\bf B444} (1995) 61;\\
G. Dvali and S. Pokorski, Phys. Rev. Lett. {\bf 78} (1997) 807.

\bibitem{Kap} P. Ginsparg, Phys. Lett. {\bf B197} (1987) 139;\\
V. Kaplunovsky, Nucl. Phys. {\bf B307} (1988) 145; (E) Nucl. Phys. {\bf B382} (1992) 436.

\bibitem{Wu} For a recent discussion, see, {\em e.g.},\\
T. Banks and M. Dine, Phys. Rev. {\bf D50} (1994) 7454;\\
P. Bin{\'e}truy, M.K. Gaillard and Y.-Y. Wu, Nucl. Phys. {\bf B481} (1996) 109; 
Nucl. Phys. {\bf B493} (1997) 27; hep-th/9702105;\\  
J.A. Casas, Phys. Lett. {\bf B384} (1996) 103; Nucl. Phys. Proc. Suppl. {\bf 52A} (1997) 289.

\bibitem{SUGRA} 
R. Arnowitt, A.H. Chamseddine and P. Nath, Phys. Rev. Lett. {\bf 49} 
(1982) 970; Phys. Lett. {\bf B121} (1983) 33; Nucl. Phys. {\bf B227} (1983) 
121;\\
R. Barbieri, S. Ferrara and C.A. Savoy, Phys. Lett.
{\bf B119} (1982) 343;\\ 
L. Hall, J. Lykken and S. Weinberg, Phys. Rev. {\bf D27} (1983) 2359;\\
H.-P. Nilles, M. Srednicki and D. Wyler, Phys. Lett. {\bf B120} (1983) 346.

\bibitem{Gia} G. Dvali and Z. Kakushadze, hep-ph/9707287.

\bibitem{adj} C. Bachas, C. Fabre and T. Yanagida, Phys. Lett. {\bf B370} (1996) 49.

\bibitem{int} K. Intriligator, Phys. Lett. {\bf B336} (1994) 409.

\bibitem{SW} N. Seiberg and E. Witten, Nucl. Phys. {\bf B426} (1994)19. 

\bibitem{Coulomb} M. Gremm, hep-th/9707071.

\bibitem{poppitz} E. Poppitz, Y. Shadmi and S. Trivedi, Nucl. Phys. {\bf B480} (1996) 125.

\bibitem{Seib} N. Seiberg, Phys. Rev. {\bf D49} (1994) 6857.



 



\end{references}
\end{document}